\documentclass[11pt,a4paper]{article}
\pdfoutput=1 

\usepackage[utf8x]{inputenc}
\usepackage{epic}
\usepackage{graphicx}

\usepackage[T1]{fontenc}
\usepackage{amssymb,amsmath,amsfonts}
\usepackage{stackrel}
\usepackage{float}
\usepackage{dsfont}
\usepackage{jheppub} 

\usepackage{color}

\newcommand{\be}{\begin{equation}}
\newcommand{\ee}{\end{equation}}
\newcommand{\bse}{\begin{subequations}}
\newcommand{\ese}{\end{subequations}}
\newcommand{\ba}{\begin{eqnarray}}
\newcommand{\ea}{\end{eqnarray}}
\newcommand{\bea}{\begin{eqnarray}}
\newcommand{\eea}{\end{eqnarray}}

\newcommand{\g}{g} 
\newcommand{\mn}{{\mu\nu}}
\newcommand{\mns}{{(\mu\nu)}}
\newcommand{\gt}{\tilde{g}} 
\newcommand{\gB}{g^{\rm (B)}} 

\newcommand{\Go}{\Gamma}
\newcommand{\Gt}{\tilde\Gamma} 
 
\newcommand{\nablao}{\nabla}
\newcommand{\nablat}{\tilde{\nabla}}

\newcommand{\T}{T_1} 
\newcommand{\To}{\T}
\newcommand{\Tt}{T_2}
\newcommand{\TT}{\mathcal{T}}

\newcommand{\x}{a} 
\newcommand{\xt}{\tilde{a}} 
\newcommand{\y}{b} 
\newcommand{\yt}{\tilde{b}} 

\newcommand{\uo}{u} 
\newcommand{\ut}{\tilde{u}}
\newcommand{\vo}{v} 
\renewcommand{\v}{v} 
\newcommand{\vt}{\tilde{v}} 
\newcommand{\w}{\nu}
\newcommand{\wo}{\nu} 
\newcommand{\wt}{\tilde{\nu}}
\newcommand{\vi}{\upsilon} 

\newcommand{\EM}{energy-momentum}
\newcommand{\tit}{\tilde{t}} 
\newcommand{\ttot}{T} 
\newcommand{\tu}{\underline{t}}
\newcommand{\tut}{\underline{\tit}}

\newcommand{\e}{\epsilon_1} 
\newcommand{\et}{\epsilon_2} 
\newcommand{\p}{P_1}
\newcommand{\pt}{P_2} 
\newcommand{\s}{s_1}
\newcommand{\st}{s_2} 
\newcommand{\ET}{\mathcal{E}}
\newcommand{\PT}{\mathcal{P}}
\newcommand{\ST}{\mathcal{S}}
\newcommand{\CVT}{\mathcal{C}_V}

\newcommand{\ga}{\gamma}
\newcommand{\gap}{\gamma'}
\newcommand{\gTf}{\ga\TT^4}
\newcommand{\gqT}{\ga^{1/4}\TT}

\newcommand{\n}{n_1}
\newcommand{\no}{n_1}
\newcommand{\nt}{n_2}
\newcommand{\alf}{\no}
\newcommand{\alft}{\nt}
\newcommand{\kap}{\kappa_1}
\newcommand{\kapt}{\kappa_2}

\newcommand{\bo}{\beta}
\newcommand{\bt}{\tilde{\beta}}
\newcommand{\gammamet}{\gamma}
\newcommand{\gammamett}{\tilde{\gamma}}
\newcommand{\z}{\chi}
\newcommand{\zt}{\tilde\chi}

\newcommand{\pio}{\pi}
\newcommand{\pit}{\tilde{\pi}}
\newcommand{\tauo}{\tau}
\newcommand{\taut}{\tilde{\tau}}

\newcommand{\exppar}{\partial_t}

\newcommand{\tr}[1]{{\rm tr}\left\{#1\right\}}

\begin{document}

\title{
Hybrid 
Fluid Models from\\ Mutual Effective Metric Couplings}

\author[a,b]{Aleksi Kurkela,}
\author[c,d]{Ayan Mukhopadhyay,}
\author[c]{Florian Preis,}
\author[c]{Anton Rebhan}
\author[c]{and Alexander Soloviev}

\affiliation[a]{Theoretical Physics Department, CERN,\\ CH-1211 Geneva, Switzerland}
\affiliation[b]{Faculty of Science and Technology, University of Stavanger,\\ 4036 Stavanger, Norway}

\affiliation[c]{Institut f\"{u}r Theoretische Physik, Technische Universit\"{a}t Wien,\\
Wiedner Hauptstr.~8-10, A-1040 Vienna, Austria}

\affiliation[d]{Department of Physics, Indian Institute of Technology Madras,\\ Chennai 600036, India}

\emailAdd{aleksi.kurkela@cern.ch}
\emailAdd{ayan@physics.iitm.ac.in}
\emailAdd{fpreis@hep.itp.tuwien.ac.at}
\emailAdd{rebhana@hep.itp.tuwien.ac.at}
\emailAdd{alexander.soloviev@tuwien.ac.at}

\abstract{
Motivated by a semi-holographic approach to the dynamics of quark-gluon plasma  which combines holographic and perturbative descriptions
of a strongly coupled infrared and a more weakly coupled ultraviolet sector, we construct a hybrid two-fluid model where interactions between its two sectors are encoded by their effective metric backgrounds, which are determined mutually by their energy-momentum tensors.
We derive the most general consistent ultralocal interactions such that the full system has a total conserved energy-momentum tensor in flat Minkowski space and study its consequences in and near thermal equilibrium
by working out its phase structure and its hydrodynamic modes.
}

\maketitle


\section{Introduction}
The hydrodynamic analysis of heavy-ion collisions performed at RHIC and the LHC
suggests that a droplet of strongly interacting matter is generated in the collisions. The value of 
the specific viscosity  that  best describes these data is very low \cite{Heinz:2013th,Romatschke:2017ejr}, $\eta/s \ll 1$, suggesting that 
the plasma is strongly coupled and does not have a description in terms of weakly interacting quasi-particles.
This has encouraged much work in describing the plasma formed in terms of strongly coupled models, such as $\mathcal N = 4$ super Yang-Mills theory as described by AdS/CFT duality \cite{CasalderreySolana:2011us} (and references therein). 

While the low value of $\eta/s$ implies that the system is strongly coupled, the collisions exhibit also hallmarks of weak coupling dynamics. In particular, it is seen that the hard components of high-$p_T$ jets go largely unmodified and resemble those created in $p$-$p$ collisions {\cite{Connors:2017ptx}}. This suggests that the medium formed in heavy-ion collisions cannot be strongly coupled at all scales and even if some of the modes are strongly coupled, others are weakly coupled. Even more strikingly, the interpretation of the observed long range rapidity correlations in $p$-$A$ and high multiplicity $p$-$p$ collisions through final state interactions, combined with no signature of jet quenching in these systems {may be seen to suggest} the presence of both strongly and weakly coupled modes. 

The simultaneous presence of strongly and weakly coupled modes poses a theoretical challenge. In absence of any fully developed non-perturbative method to access real-time properties of QCD in the non-perturbative regime, we may attempt to model the non-perturbative modes using a theory that we can solve in the strong coupling limit, while discussing the perturbative sector in a weak coupling approximation. Corresponding attempts have been made in \cite{Casalderrey-Solana:2014bpa,Iancu:2014ava,Mukhopadhyay:2015smb}.
Such approaches in general pose the non-trivial question then, how the two sectors, described with different models describing different degrees of freedoms, be coupled.

A consistent coupling requires that the quantities that mediate the coupling should be well-defined in each theory and also gauge-invariant. In the context of jet-quenching, such couplings have been suggested for example in \cite{Liu:2006ug,Casalderrey-Solana:2014bpa}. 

In a {different} attempt to formulate a generic coupling between the two subsectors for the study of collective dynamics and equilibration, {a local coupling of all the marginal operators of the two subsectors was proposed \cite{Iancu:2014ava,Mukhopadhyay:2015smb},
following previous examples of a \emph{semi-holographic} framework where only
part of the dynamics is described by gauge/gravity duality \cite{Faulkner:2010tq,Nickel:2010pr,Jensen:2011af,Mukhopadhyay:2013dqa}.}
This includes in particular a coupling between the energy-momentum tensors of the two subsectors,
which can be induced by deforming the boundary metric of a holographic sector.

Specifically, as a semi-holographic model of the early stages of heavy-ion collisions, the perturbative sector was assumed
to be described by classical Yang-Mills equations as in the glasma effective theory \cite{Gelis:2010nm} that describes the color-glass condensate initial conditions \cite{Kovner:1995ts,Kovner:1995ja} of the deconfined gluonic matter liberated in the heavy-ion collisions, and the nonperturbative infrared sector by AdS/CFT, corresponding to strongly coupled $\mathcal{N}=4$ super-Yang-Mills theory. The toy model studied in \cite{Mukhopadhyay:2015smb} demonstrated that in this way a closed system with a conserved
energy-momentum tensor in Minkowski space can be obtained.\footnote{{However, by only considering a gravitational coupling in
a strictly homogeneous and isotropic situation which precludes propagating degrees of freedom in the bulk,
the far-from-equilibrium system did not show any thermalization. When
also a coupling between the gravitational dilaton field and the Yang-Mills Lagrangian density is turned on, the infrared sector turns out \cite{Ecker:2018ucc} to be heated up, thereby showing at least an onset of thermalization.}}

In this work, we explore the implications of the ``democratic'' couplings proposed in \cite{Banerjee:2017ozx} (and extensions thereof), where the effective metric of \emph{each} subsystem depends on the energy-momentum fluctuations of the complements. Instead of far-from-equilibrium systems studied previously we concentrate on systems that are in equilibrium and the equilibration of near-equilibrium systems. Furthermore, we again restrict the couplings to that between the respective energy-momentum tensors of the subsystems.

We first address the question of what is the equilibrium state of two coupled conformal systems. As the system is assumed
to be in thermal equilibrium, and the coupling depends only on the energy-momentum tensors of the subsystems, the microscopic features of the subsystems do not enter the discussion and therefore the results are generic
for conformal subsystems and depend only on the properties of the coupling between the subsystems. We observe that 
requiring causality and ultraviolet completeness restricts the range the model parameters describing the coupling can take. In addition we find that the composite system -- that breaks conformal symmetry due to dimensionful parameters of the coupling -- exhibits a rich phase structure with a phase transition that takes the system from a sum of two separate conformal subsystems at low temperatures to a new emergent conformal system at high temperatures. As a function of the model parameters, this transition is either a cross-over or a first-order transition, and the two are separated by second-order critical endpoint with specific heat critical exponent $\alpha=2/3$.

Next we study in detail the collective dynamics of near-equilibrium systems.
We first assume
that each subsystem can be separately described as a conformal fluid in terms of first-order hydrodynamics. This assumption is generally valid if the 
length scale of the deviation of global equilibrium is sufficiently long for a well behaved gradient expansion and if
no long-lived non-hydrodynamic modes are excited. Within this approximation we follow how linearized energy-momentum perturbations of the composite system approach global equilibrium and find a rich structure of two-fluid dynamics.
In the shear sector we find that the overall viscosity interpolates between those of the subsystems
and decreases with the coupling between the subsystems. In the sound sector we obtain two modes where
only one is propagating with the thermodynamic speed of sound at large coupling. However both
have attenuation vanishing with the square of momentum, implying that spatially homogeneous
density perturbations of the individual subsystems are not attenuated and 
therefore more dynamics is required for the thermal equilibrium to be established between the two sectors. 
Indeed, this is in line with the findings in the semi-holographic toy model of Ref.~\cite{Mukhopadhyay:2015smb},
where also interactions beyond the ones between the energy-momentum tensors are needed for thermalization \cite{Ecker:2018ucc}.

Finally, we study to what extent non-hydrodynamic modes in one subsystem are attenuated 
because of coupling to the other dissipative subsystem.

The organisation of the paper is as follows. In Section \ref{Sec:Setup}, we 
{describe the general setup, its motivation in the semi-holographic context, as well as the concrete
mutual metric coupling and how a total energy-momentum tensor that is conserved with respect to the (Minkowski) background
metric of the full system arises.}
In Section \ref{Sec:Thermodynamics}, we {discuss the requirements of causality and UV-completeness and}
study the consequences of our couplings for the thermodynamics {and phase structure} of the full system. 
In Section \ref{Sec:Bihydro}, we study the hydrodynamic limit of the full system, and in Section \ref{Sec:KinHydro} we further study the case when the weakly coupled system can be described by kinetic theory and the strongly coupled sector as a conformal fluid with appropriate transport coefficients. 

\section{General setup}\label{Sec:Setup}
\subsection{Semi-holography and democratic coupling}

We consider a dynamical system $\mathfrak{S}$ in a fixed background metric $\gB_\mn$ (to be
set to the Minkowski metric {$\eta_\mn$} eventually) which consists of two subsystems $\mathfrak{S}_1$
and $\mathfrak{S}_2$.

In the previous approach to semi-holography {\cite{Mukhopadhyay:2015smb}}, the full effective action $S$ {of system $\mathfrak{S}$} was constructed as \cite{Mukhopadhyay:2015smb}:
\begin{equation}
S = S^{\rm pert}[A_\mu^a, \cdots] + S^{\rm hol}[g^{\rm (b)}_{\mn} = {\gB_\mn} + \gamma t_\mn^{\rm pert}, \cdots],
\end{equation}
where $S^{\rm pert}$ is the effective perturbative action {for $\mathfrak{S}_1$}, and {$\mathfrak{S}_2$ is} represented by $S^{\rm hol}$, which is the holographic on-shell gravitational action in presence of sources{. These sources}, such as a non-trivial boundary metric $g^{\rm (b)}_{\mn}$, are functionals of the gauge-invariant operators of the perturbative sector, and $\gamma$ is a hard-soft coupling.
{T}he full conserved energy-momentum tensor {calculated by varying the action with respect to $\gB_\mn$ cannot be written in terms of} the {\textit{effective}} operators of each sector and therefore the low energy dynamics of the full system cannot be readily derived from the coarse-grained descriptions of the individual subsystems. {Furthermore, the way the two sectors are coupled is somewhat asymmetric. On the one hand, the coupling amounts to deforming the metric of $\mathfrak{S}_2$ by the energy-momentum tensor of $\mathfrak{S}_1$. On the other hand, the energy-momentum tensor of $\mathfrak{S}_2$ enters via the equations of motion of $\mathfrak{S}_1$. Nevertheless the main improvement of \cite{Iancu:2014ava} made in \cite{Mukhopadhyay:2015smb} was that the full energy-momentum tensor is conserved, provided that the effective operators of $\mathfrak{S}_2$ satisfy a separate Ward identity and the equation of motion for the fields in $\mathfrak{S}_1$ are in effect.

Motivated by the semi-holographic approach in the democratic formulation \cite{Banerjee:2017ozx}, the two subsystems
are assumed to have covariant dynamics with respect to individual \emph{effective} metrics $\g_\mn$ and $\gt_\mn$, respectively.%
\footnote{In the following, quantities relating to the subsystems $\mathfrak{S}_1$
and $\mathfrak{S}_2$ will be distinguished either by indices 1 and 2 or by
a tilde for those pertaining to $\mathfrak{S}_2$ (a tilde is used in particular when indices might be
confusing).} Interactions between the two subsystems are introduced 
by promoting each effective metric to functions that are locally determined by {the state of the complement system,}
\begin{equation}\label{eq:eff-metric}
\g_\mn =\g_\mn[\tit^{\alpha\beta},\ldots], \qquad \gt_\mn = \gt_\mn[t^{\alpha\beta},\ldots].
\end{equation}
The two subsystems are assumed to share the same topological space so that we can use the same
coordinates {for both of them} (and thus the total system)\footnote{Coordinate transformations
would thus affect the background metric of the complete system and the effective
metrics of the subsystems simultaneously.}. Furthermore, the
 subsystems appear as closed systems with respect to their individual effective metrics, but they can exchange
energy and momentum defined with respect to the actual physical background metric $\gB_\mn$. Thus the effective metric tensors encode the interactions between the two subsystems.

The diffeomorphism invariance of the respective theories describing the two subsystems imply the Ward identities
\begin{align}\label{eq:subsysWI}
\nablao_\mu t^{\mu \nu} = 0, \quad  \nablat_\mu \tilde t^{\mu \nu} = 0,
\end{align}
where $\nablao$ and $\nablat$ refer to the covariant derivatives with respect to the different effective metrics with the Levi-Civita connections
\begin{eqnarray}\label{eq:Christoffels}
&&\Go^\mu_{\nu\rho} = \frac{1}{2}\g^{\mu\sigma} (\partial_\nu {\g}_{\sigma\rho}+\partial_\rho {\g}_{\sigma\nu}-\partial_\sigma {\g}_{\nu\rho}) = \Gamma^{\mu\rm (B)}_{\nu\rho} + \frac{1}{2}\g^{\mu\sigma} (\nabla^{\rm (B)}_\nu {\g}_{\sigma\rho}+\nabla^{\rm (B)}_\rho {\g}_{\sigma\nu}-\nabla^{\rm (B)}_\sigma {\g}_{\nu\rho}) , \nonumber\\
&&\Gt^\mu_{\nu\rho}= \frac{1}{2}\gt^{\mu\sigma} (\partial_\nu {\gt}_{\sigma\rho}+\partial_\rho {\gt}_{\sigma\nu}-\partial_\sigma {\gt}_{\nu\rho}) = \Gamma^{\mu\rm (B)}_{\nu\rho} + \frac{1}{2}\gt^{\mu\sigma} (\nablat^{\rm (B)}_\nu {\gt}_{\sigma\rho}+\nablat^{\rm (B)}_\rho {\gt}_{\sigma\nu}-\nablat^{\rm (B)}_\sigma {\gt}_{\nu\rho}).\qquad
\end{eqnarray}
Above, $\nabla^{\rm (B)}$ is the covariant derivative with respect to $\gB_\mn$ and $\Gamma^{\mu\rm (B)}_{\nu\rho}$ is the corresponding Levi-Civita connection, and the second equalities in \eqref{eq:Christoffels} indicate that from the point of view of the actual physical background metric $\gB_\mn$ 
the identities \eqref{eq:subsysWI} actually imply that work is done on the respective subsystems by external forces.
In what follows, we restrict the forms \eqref{eq:eff-metric} of the effective metrics $\g$ and $\gt$ (in a generally covariant manner) such that there exists a $T^\mn$ for the full system that is locally conserved with respect to the physical background metric $\gB_\mn$, i.e., we can enforce the Ward identity for the total system:
\be
\nabla^{\rm (B)}_\mu T^{\mu\nu}=0.
\ee
It turns out that the full energy-momentum tensor {$T^\mn$} is a functional only of the effective operators $t^\mn$ and $\tit^\mn$ of the two sectors. Hence one can readily construct effective descriptions of the full dynamics from the effective description of each sector.\footnote{Additionally such couplings can generate expectation values of high-dimensional irrelevant operators without the need of introducing a non-trivial irrelevant deformation of the respective theory \cite{Banerjee:2017ozx}. This feature is needed for the cancellation of the Borel poles of the perturbative expansion.} The main advantage of our method in the context of phenomenology is that it works even when we cannot invoke action principles for the effective descriptions of one or both subsystems. The full dynamics is obtained by solving the subsystems in a mutually self-consistent way as has been illustrated in case of the vacuum state in a toy example \cite{Banerjee:2017ozx}.
 
In the present paper, we utilize this to construct the low energy phenomenology by considering appropriate effective description of each subsector.  First we assume that both sectors are described by fluids. Then we describe the perturbative sector by an effective kinetic theory and the non-perturbative sector by a strongly coupled fluid. We will be able to find consistent solutions for the full thermal equilibrium and also study its linear perturbations. 

As a general remark, the principle of democratic coupling can be extended to other couplings such as that between scalar operators $O$ and $\tilde{{O}}$ \cite{Banerjee:2017ozx}. Let the theory describing the non-perturbative sector be also a (strongly coupled holographic) Yang-Mills theory with the coupling $\tilde{g}_{\rm YM}$ whereas $g_{\rm YM}$ is the coupling of the perturbative sector. These mutual deformations by scalar operators lead to the modified Ward identities (we turn off other couplings including the effective metric couplings for purpose of illustration)
\begin{eqnarray}\label{simple-WIs}
\partial_\mu t^\mu_{{\,}\nu} = O\partial_\nu g_{\rm YM}, \quad  \partial_\mu \tit^\mu_{{\,}\nu} = \tilde{O}\partial_\nu \tilde{g}_{\rm YM}.
\end{eqnarray}
Then we may postulate a democratic coupling of the form:
\begin{eqnarray}\label{simple-coupling}
g_{\rm YM} =g_{\rm YM}^0 + \alpha \tilde{O}, \quad \tilde{g}_{\rm YM} =\tilde{g}_{\rm YM}^0 + \alpha O,
\end{eqnarray}
where $g_{\rm YM}^0$ and $\tilde{g}_{\rm YM}^0$ are constants.
It is clear then that the above Ward identities along with \eqref{simple-coupling} imply the existence of the conserved energy-momentum tensor $T^\mu_{{\,}\nu}$ given by
\begin{equation}
T^\mu_{{\,}\nu} = t^\mu_{{\,}\nu} + \tit^\mu_{{\,}\nu} - \alpha O\tilde{O}\delta^\mu_{{\,}\nu}
\end{equation}
satisfying $\partial_\mu T^\mu_{{\,}\nu} = 0$.

In \cite{Banerjee:2017ozx}, the most general scalar couplings of the form have been explored and a toy construction has been done to illustrate how these ``hard-soft'' couplings (such as $\alpha$) along with the parameters of the holographic classical gravity determining $S^{\rm hol}$ can be derived as functions of the perturbative couplings in $S^{\rm pert}$ via simple consistency rules. In the following subsection, we extend and correct the democratic effective metric type couplings
set up in \cite{Banerjee:2017ozx}.

\subsection{Consistent mutual effective metric couplings}\label{Sec:metriccouplings}

We start the construction of the coupling rules between the two subsystems by demanding that the total system
a conserved energy-momentum tensor $T^{\mu\nu}$ can be written for the total system $\mathfrak{S}$ in the flat 
background metric {(from now on we choose $\g^{\rm (B)}_\mn = \eta_\mn$ {unless explicitly mentioned otherwise})}
\begin{align}
\partial_\mu T^{\mu\nu} = 0,
\end{align}
while simultaneously satisfying the Ward identities of the two subsystems in their respective
curved metrics
\begin{align}
\nablao_\mu t^{\mu \nu} = 0, \quad  \nablat_\mu \tilde t^{\mu \nu} = 0,
\end{align}
where $\nablao$ and $\nablat$ refer to the covariant derivatives with respect to the different metrics of the subsystems, with the
corresponding Christoffel symbols \eqref{eq:Christoffels}.

For the rest of the paper, unless explicitly indicated otherwise, by $t^\mu_{{\,}\nu}$ we will mean $t^{\mu\rho} g_{\rho\nu}$ and by $t_{\mu\nu}$ we will mean $g_{\mu\rho}t^{\rho\sigma} g_{\sigma\nu}$, etc., with all lowering (and raising) of indices done by the effective metric (and its inverse) of the respective theory.
The Ward identity of subsystem $\mathfrak{S}_1$ implies that
\begin{equation}
\nabla_\mu t^{\mu}_{\phantom{\mu}\nu} = 0, 
\end{equation}
i.e.,
\begin{equation}\label{WI1int}
\partial_\mu t^{\mu}_{\phantom{\mu}\nu} + \Go^\mu_{\mu\rho}t^{\rho}_{\phantom{\mu}\nu} -\Go^\mu_{\nu\rho}t^{\rho}_{\phantom{\mu} \mu} = 0, 
\end{equation}
or
\begin{equation}\label{WI1}
\partial_\mu (t^{\mu}_{\phantom{\mu}\nu}\sqrt{-\g}) - \frac{1}{2}t^{\mu\sigma}\sqrt{-\g}\partial_\nu {\g}_{\mu\sigma} = 0,
\end{equation}
where we have used
\begin{equation}
\Go^\mu_{\mu\nu} = \partial_\nu (\ln \sqrt{-\g}), \quad \Go^\mu_{\nu\rho} t^{\rho}_{\phantom{\mu}\mu} = \frac{1}{2}t^{\mu\rho} \partial_\nu {\g}_{\mu\rho},
\end{equation}
and multiplied both sides of (\ref{WI1int}) with $\sqrt{-g}$ to obtain (\ref{WI1}). Similarly, the Ward identity for subsystem $\mathfrak{S}_2$ implies that
\begin{equation}\label{WI2}
\partial_\mu (\tit^{\mu}_{\phantom{\mu}\nu}\sqrt{-\gt}) - \frac{1}{2}\tit^{\mu\sigma}\sqrt{-\gt}\partial_\nu{\gt}_{\mu\sigma} = 0.
\end{equation}
We require 
that both $t^{\mu\nu}$ and $\tit^\mn$ are symmetric tensors. Using these Ward identities, it is straightforward to verify that the following
local relations for the effective metrics 
\ba\label{coupling-rule}
\g_{\mn} &=& \eta_{\mn} +\ga\, \eta_{\mu\alpha} \tit^{\alpha\beta} \eta_{\beta\nu}\sqrt{-\gt}
+\gap\, \eta_\mn \eta_{\alpha\beta} \tit^{\alpha\beta}\sqrt{-\gt}
, \nonumber\\ 
{\gt}_{\mu\nu} &=& \eta_{\mu\nu}+ \ga\, \eta_{\mu\alpha} {t}^{\alpha\beta} \eta_{\beta\nu}\sqrt{-\g}
+\gap\, \eta_\mn \eta_{\alpha\beta} t^{\alpha\beta}\sqrt{-\g},
\ea
where $\ga$ and $\gap$ are coupling constants (with mass dimension $-4$), allow us to construct
a symmetric conserved tensor for the full system in flat space.

From (\ref{WI1}) and (\ref{WI2}) it follows that
\be
K^\mu_{\phantom{\mu}\nu} = t^\mu_{\phantom{\mu}\nu}\sqrt{-\g} + \tit^\mu_{\phantom{\mu}\nu}\sqrt{-\gt} 
+ \Delta K \delta^\mu_{\phantom{\mu}\nu},
\ee
with
\be\label{DeltaK}
\Delta K = - \frac{1}{2}\left[\ga\,
(t^{\rho\alpha}\sqrt{-\g}) \eta_{\alpha\beta} (\tit^{\beta\sigma}\sqrt{-\gt}) \eta_{\sigma\rho} 
+\gap\, (t^{\alpha\beta}\sqrt{-\g}) \eta_{\alpha\beta} (\tit^{\sigma\rho}\sqrt{-\gt}) \eta_{\sigma\rho} \right]
\ee
satisfies
\begin{equation}
\partial_\mu K^\mu_{\phantom{\mu}\nu} = 0.
\end{equation}
Similarly it is easy to see that
\begin{equation}
L_\mu^{\phantom{\mu}\nu} = t_\mu^{\phantom{\mu}\nu}\sqrt{-\g} + \tit_\mu^{\phantom{\mu}\nu}\sqrt{-\gt} + \Delta K \delta_\mu^{\phantom{\mu}\nu}
\end{equation}
satisfies
\begin{equation}
\partial_\nu L_\mu^{\phantom{\mu}\nu} = 0.
\end{equation}
A symmetric and conserved total \EM\ tensor $T^{\mu\nu} = \eta^{\mu\rho}T_\rho^{\phantom{\mu}\nu}= T^\mu_{\phantom{\mu}\rho}\eta^{\rho\nu}$ 
with $\partial_\mu T^{\mu\nu} = 0$ (also $\partial_\mu T^{\mu}_{\phantom{\mu}\nu} = 0$)
can therefore be defined by
\begin{equation}\label{emt-full}
T^{\mu}_{\phantom{\mu}\nu} = \frac{1}{2}(K^\mu_{\phantom{\mu}\nu}+L_\nu^{\phantom{\nu}\mu}). 
\end{equation}

We can easily generalize the above construction for 
a curved background metric $\g^{\rm (B)}_{\mu\nu}$ instead of the Minkowski metric using {the second identities in \eqref{eq:Christoffels}} which imply
\begin{eqnarray}
&\Go^{\mu}_{\nu\mu} -  \Go^{{\rm (B)}\mu}_{\phantom{{\rm (B)}}\nu\mu} =\partial_\nu (\ln \sqrt{-\g}) -\partial_\nu (\ln \sqrt{-\g^{\rm(B)}}) = \partial_\nu\left(\ln \frac{\sqrt{-\g}}{\sqrt{-\g^{\rm (B)}}}\right)\nonumber\\
&= \frac{\sqrt{-\g^{\rm (B)}}}{\sqrt{-\g}}\partial_{\nu}\left(\frac{\sqrt{-\g}}{\sqrt{-\g^{\rm (B)}}}\right)= \frac{\sqrt{-\g^{\rm (B)}}}{\sqrt{-\g}}\nabla^{\rm (B)}_{\nu}\left(\frac{\sqrt{-\g}}{\sqrt{-\g^{\rm (B)}}}\right),
\end{eqnarray}
where we have used that $\sqrt{-\g}/\sqrt{-\g^{\rm (B)}}$ is a scalar under general coordinate transformations.

With the help of these relations, one can readily see that the consistent coupling rules {have the following general covariant forms}
\ba\label{coupling-rule-cov}
\g_{\mu\nu} &=& \g^{\rm (B)}_{\mu\nu} +\left(\ga\, \g^{\rm (B)}_{\mu\alpha} \tit^{\alpha\beta} \g^{\rm (B)}_{\beta\nu}
+\gap\, \g^{\rm (B)}_{\mu\nu} \tit^{\alpha\beta} \g^{\rm (B)}_{\alpha\beta}
\right)\frac{\sqrt{-\gt}}{\sqrt{-\g^{\rm (B)}}}, \nonumber\\ 
{\gt}_{\mu\nu} &=& \g^{\rm (B)}_{\mu\nu}+ \left(\ga\, \g^{\rm (B)}_{\mu\alpha} {t}^{\alpha\beta} \g^{\rm (B)}_{\beta\nu}
+\gap\, \g^{\rm (B)}_{\mu\nu} {t}^{\alpha\beta} \g^{\rm (B)}_{\alpha\beta}
\right)\frac{\sqrt{-{\g}}}{\sqrt{-\g^{\rm (B)}}}.
\ea
Then with
\bea
\Delta K&=&- \frac{\ga}{2}\left(t^{\rho\alpha}\frac{\sqrt{-\g}}{\sqrt{-\g^{\rm (B)}}}\right) \g^{\rm (B)}_{\alpha\beta} \left(\tit^{\beta\sigma}\frac{\sqrt{-\gt}}{\sqrt{-\g^{\rm (B)}}}\right) \g^{\rm (B)}_{\sigma\rho}\nonumber\\
&&- \frac{\gap}{2}\left(t^{\alpha\beta}\frac{\sqrt{-\g}}{\sqrt{-\g^{\rm (B)}}}\right) \g^{\rm (B)}_{\alpha\beta} \left(\tit^{\sigma\rho}\frac{\sqrt{-\gt}}{\sqrt{-\g^{\rm (B)}}}\right) \g^{\rm (B)}_{\sigma\rho},
\ea
the full conserved energy-momentum tensor is again
given by (\ref{emt-full}), and it satisfies $\nabla^{\rm (B)}_\mu T^\mu_{\phantom{\mu}\nu} = 0$ in the actual background where all degrees of freedom live. (Note $T^{\mu\nu} = T^\mu_{\phantom{\mu}\rho} g^{\rm (B)\rho\nu}$).

More general consistent couplings can be constructed if we permit higher powers of the \EM\ tensors $t^\mn$ and $\tit^\mn$
together with new coupling constants carrying correspondingly higher inverse mass dimension. This is done in
Appendix \ref{App:GeneralCouplings} (correcting and generalizing
Ref.~\cite{Banerjee:2017ozx} in this respect); in the following we will restrict ourselves to the above two coupling terms
with coupling constants $\ga$ and $\gap$.{\footnote{Appendix  \ref{App:GeneralCouplings} points out that in the most general set-up, where the total energy-momentum tensor satisfies thermodynamic consistency proven in Appendix \ref{App:ThdynConsistency}, each possible interaction term in the total energy-momentum tensor can be obtained via an appropriate coupling rule as a result of an interesting combinatoric identity. This is significant because a generic interaction term is not ruled out by any symmetry, and therefore it should indeed be reproduced by our way of introducing interactions via effective metrics.}

While the dimensionful coupling constants introduced here appear to be arbitrary at this stage, we shall see that
certain restrictions appear when further physical requirements are imposed. In particular, the complete
dynamics should be such that causality remains intact. This means that the effective lightcone speed in the subsystems
should not exceed the actual speed of light defined by $\gB_\mn$. At least in the following applications to
equilibrium and near-equilibrium situations, we can confirm that with just the two terms of in the consistent coupling
rules corresponding to $\ga$ and $\gap$ causality can be ensured -- at arbitrary energy scales -- by choosing $\ga>0$
and $r\equiv -\gap/\ga > 1$. Interestingly enough, a positive value of the tensorial coupling constant $\ga$ was also found
to be required in the semi-holographic study of Ref.~\cite{Mukhopadhyay:2015smb} in order that interactions lead to a positive
interaction measure, $\mathcal E-3P=-T^\mu_{\phantom{\mu}\mu}>0$, which is a feature
of (lattice) Yang-Mills theories at finite temperature \cite{Boyd:1996bx,Borsanyi:2010cj}.

\section{Thermodynamics}\label{Sec:Thermodynamics}

\subsection{General equilibrium solution}\label{Sect:EquilibriumSolution}

We now assume that the full system $\mathfrak{S}$, living in a flat Minkowski background metric 
$\gB_\mn=\eta_\mn$ and composed of two sectors $\mathfrak{S}_1$ and $\mathfrak{S}_2$
that interact through mutually determining their effective metrics, 
has reached a homogeneous and isotropic equilibrium state with temperature $\TT$ 
with total \EM\ tensor
\be
T^\mn=(\ET+\PT)U^\mu U^\nu +\PT \eta^\mn, \quad U^\mu=(1,0,0,0).
\ee

Assuming furthermore that the subsystems $\mathfrak{S}_1$ and $\mathfrak{S}_2$ have also thermalized due to their internal dynamics
taking place in their respective effective metrics, we expect that the latter will have a static, homogeneous and isotropic form
for which we introduce the ans\"atze
 \begin{equation}\label{equilibriummetrics}
\g_{\mu\nu} = {\rm diag}(- \x^2, \y^2, \y^2, \y^2 ), \quad \gt_{\mu\nu} = {\rm diag}(-\xt^2,\yt^2,\yt^2,\yt^2 ),
\end{equation}
with constants $\x,\y,\xt,\yt$ to be determined self-consistently.\footnote{If one of the systems is to be described by
gauge/gravity duality, the simple metric ansatz above is of course not pertaining to the bulk, but to the boundary of the gravity dual.} 

The \EM\ tensors of the subsystems are then of the form
\begin{eqnarray}
& t^{\mu\nu} = (\e(\To)+ \p(\To))  \uo^\mu \uo^\nu + \p(\To) \g^{\mu\nu}, \quad {\rm with} \quad \uo^\mu = (1/\x,0,0,0), \nonumber\\
& \tit^{\mu\nu} = (\et(\Tt)+ \pt(\Tt))  \ut^\mu \ut^\nu + \pt(\Tt) \gt^{\mu\nu}, \quad {\rm with} \quad \ut^\mu = (1/\xt,0,0,0),
\end{eqnarray}
i.e.,
\begin{eqnarray}
 t^{\mu\nu} &=& {\rm diag}\left(\frac{\e(\To)}{\x^2},\frac{\p(\To)}{\y^2},\frac{\p(\To)}{\y^2},\frac{\p(\To)}{\y^2}\right), \nonumber\\  \tit^{\mu\nu} &=& {\rm diag}\left(\frac{\et(\Tt)}{\xt^2},\frac{\pt(\Tt)}{\yt^2},\frac{\pt(\Tt)}{\yt^2},\frac{\pt(\Tt)}{\yt^2}\right),
\end{eqnarray}
with individual temperatures $\To$ and $\Tt$. 

The simplest coupling rules (\ref{coupling-rule}) now read
\begin{eqnarray}\label{xsandys}
1-\x^2 =\left(\ga \frac{\et(\Tt)}{\xt^2}-\gap\left(-\frac{\et(\Tt)}{\xt^2}+\frac{3\pt(\Tt)}{\yt^2}\right)\right)\xt \yt^3,\nonumber\\ \y^2-1 =\left(\ga \frac{\pt(\Tt)}{\yt^2}+\gap\left(-\frac{\et(\Tt)}{\xt^2}+\frac{3\pt(\Tt)}{\yt^2}\right)\right) \xt \yt^3, \nonumber\\ 1-\xt^2 = \left(\ga\frac{\e(\To)}{{\x}^2}-\gap\left(-\frac{\e(\To)}{{\x}^2}+\frac{3\p(\To)}{{\y}^2}\right)\right){\x}{\y}^3,\nonumber\\ \yt^2-1 =\left(\ga \frac{\p(\To)}{{\y}^2}+\gap\left(-\frac{\e(\To)}{{\x}^2}+\frac{3\p (\To )}{{y}^2}\right)\right) {\x} {\y}^3,
\end{eqnarray}
and these determine $\x$, $\y$, $\xt$ and $\yt$ as functions of $\To$, $\Tt$ and the coupling constants $\ga$ and $\gap$. 
The full energy-density and pressure following from (\ref{emt-full}) are
\begin{eqnarray}\label{EandP}
\ET
&=& \e(\To){\x} {\y}^3 + \et(\Tt) \xt \yt^3\nonumber\\&&+  \frac{\ga}{2}\left(\frac{\e(\To)\et(\Tt)}{\x^2\xt^2}+\frac{3\p(\To)\pt(\Tt)}{\y^2\yt^2}\right)\xt \yt^3 {\x} {\y}^3
\nonumber\\&&+  \frac{\gap}{2}\left(-\frac{\e(\To)}{{\x}^2}+\frac{3\p(\To)}{{\y}^2}\right)\left(-\frac{\et(\Tt)}{\xt^2}+\frac{3\pt(\Tt)}{\yt^2}\right)\xt \yt^3 {\x} {\y}^3, \nonumber\\
\PT
&=& \p (\To){\x} {\y}^3 + \pt(\Tt) \xt \yt^3\nonumber\\&& -\frac{\ga}{2}\left(\frac{\e(\To)\et(\Tt)}{\x^2\xt^2}+\frac{3\p(\To)\pt(\Tt)}{\y^2\yt^2}\right)\xt \yt^3 {\x} {\y}^3\nonumber\\&&-  \frac{\gap}{2}\left(-\frac{\e(\To)}{{\x}^2}+\frac{3\p(\To)}{{\y}^2}\right)\left(-\frac{\et(\Tt)}{\xt^2}+\frac{3\pt(\Tt)}{\yt^2}\right)\xt \yt^3 {\x} {\y}^3.
\end{eqnarray}

In a thermal equilibrium for the full system as well as for its individual subsystems, the physical temperature $\TT$ of $\mathfrak{S}$
living in Minkowski space is given by the inverse
of $\int_0^\beta d\tau$, where $\tau$ is imaginary time and $\beta$ its period. The temperature of the subsystem $\mathfrak{S}_1$, which effectively
lives in a metric with constant $\sqrt{-g_{00}}=\x$, is then given by 
\be
\To^{-1}=\int_0^\beta \sqrt{-g_{00}} \, d\tau=\x\beta=\x \TT^{-1};
\ee
by the same token we have $\Tt^{-1}=\xt \TT^{-1}$. Hence,
\begin{equation}\label{eq-condition}
\TT = \To \x = \Tt\xt.
\end{equation}
Thus $\TT$ alone parametrizes the space of equilibrium solutions.

Using the thermodynamic identities
\be\label{thermo-identity-1}
\epsilon_{1,2}+P_{1,2}=T_{1,2} s_{1,2},\quad
\ET + \PT = \TT \ST,
\ee
the result (\ref{EandP}) implies
\begin{equation}\label{Sdyn}
\TT \ST = \To  \s (\To ) \x \y^3 + \Tt  \st (\Tt ) \xt\yt^3
=\TT\left[\s(\To) \y^3 + \st(\Tt) \yt^3\right],
\end{equation}
showing that the total entropy density is the sum of the two entropy densities.
Therefore, we identify the total entropy current as
\begin{equation}
\ST^\mu = \sqrt{-\g}\s^\mu + \sqrt{-\gt}\st^\mu
\end{equation}
for $\s^\mu = \s(\To)\uo^\mu$, $\st^\mu = \st(\Tt)\ut^\mu$, and
$\ST^\mu=\ST \mathcal{U}^\mu$ and $\mathcal{U}_\mu = (-1, 0,0,0)$.

This indeed makes perfect sense in a general non-equilibrium situation. 
When each sector has an entropy current $s^\mu_{1,2}$ satisfying
 \begin{equation}
\nabla_\mu \s^\mu \geq 0, \quad \tilde{\nabla}_\mu \st^\mu \geq 0,
\end{equation}
this implies
\begin{equation}
\partial_\mu (\sqrt{-\g}\s^\mu) \geq 0, \quad \partial_\mu (\sqrt{-\gt}\st^\mu) \geq 0,
\end{equation}
such that
\begin{equation}
\partial_\mu (\sqrt{-\g}\s^\mu + \sqrt{-\gt}\st^\mu) = \partial_\mu \ST^\mu \geq 0.
\end{equation} 

In thermal equilibrium, we also need to have
\be\label{thermo-identity}
{\rm d} \ET =  \TT {\rm d} \ST
\ee
or, equivalently, 
${\rm d}\PT/{\rm d}\TT =  \ST$, for thermodynamic consistency. {In Appendix \ref{App:ThdynConsistency} we prove this relation and the consistency of \eqref{eq-condition}, \eqref{Sdyn} and \eqref{thermo-identity} for the coupling discussed here as well as for the coupling rules that generalize \eqref{coupling-rule}.} The mutual compatibility of the thermodynamic identities \eqref{thermo-identity-1} and \eqref{thermo-identity} of the full system with the global equilibrium condition \eqref{eq-condition} (along with the additivity of the total entropies that can be expected from the fact that each subsystem is closed in an effective point of view) provides a strong low-energy consistency check of our approach.

\subsection{Causal structure of equilibrium solution}

Since the causal structure of the dynamics taking place in the subsystems is
dictated by the respective effective metrics only, 
causality in the full system, which is living in Minkowski space, is not guaranteed.
For example, massless excitations from the point of view of subsystem $\mathfrak{S}_1$ with
metric $\g_{\mu\nu} = {\rm diag}(- \x^2, \y^2, \y^2, \y^2 )$ propagate
with velocity $\v=\x/\y$ with respect to the actual physical spacetime 
that the full system
is occupying. (Recall that the two subsystems and the full system share the same topological
space; the effective metrics of the subsystems just encode the effects of interactions between
the two components of the full system.)

At least for 
the above solution for the equilibrium configuration obtained in the case of the simplest
coupling rules (\ref{coupling-rule}) we can ensure the absence of superluminal propagation
by requiring that the tensorial coupling constant $\ga\ge0$ together with
$P_{1,2}\ge0$ and $\epsilon_{1,2}\ge0$. To see this, take the sum of
the first and second as well as of the third and fourth equation in (\ref{xsandys}).
This leads to
\bea
\y^2-\x^2&=&\ga\left( \frac{\et(\Tt)}{\xt^2}+\frac{\pt(\Tt)}{\yt^2}\right) \xt \yt^3 \ge 0, \nonumber\\
\yt^2-\xt^2&=&\ga\left( \frac{\e(\T)}{\x^2}+\frac{\p(\T)}{\y^2}\right) \x \y^3 \ge 0,
\eea
independent of $\gap$,
implying that the effective lightcones defined by the metrics $\g_\mn$ and $\gt_\mn$ are 
contained within the lightcone defined by the background Minkowski metric.

\subsection{Conformal subsystems}\label{sec:conformalsubsystems}

In the following we shall consider the case of two conformal subsystems. For example one may think of
a gas of nearly free massless particles for $\mathfrak{S}_1$ coupled to a strongly interacting quantum liquid
for $\mathfrak{S}_2$. The \EM\ tensors $t^\mn$ and $\tit^\mn$ are assumed to be traceless with respect to
the effective metrics $g_\mn$ and $\gt_\mn$, thus the equations of state of the two subsystems are
then simply
\begin{eqnarray}\label{bi-conformal}
\e (\To ) = 3 \p (\To ) = 3 \no \To ^4, \nonumber\\
\et (\Tt ) = 3 \pt (\Tt ) = 3 \nt  \Tt ^4,
\end{eqnarray}
with constant prefactors $\no$ and $\nt$. 

The entropy is a simple expression in terms of the effective lightcone velocities $\v,\vt$ {associated with the effective metrics $\g_\mn$ and $\gt_\mn$, respectively},
\be\label{ST}
\ST=4\TT^3\left( \frac{\n}{\v^3}+\frac{\nt}{\vt^3} \right),
\ee
where
\be
\v:=\frac{\x}{\y},\quad \vt:=\frac{\xt}{\yt}.
\ee
Similarly we obtain for (\ref{xsandys})
\bea\label{vb-equations}
1-\v^2 \y^2&=& 3 \gTf \nt \frac{1-r(1-\vt^2)}{\vt^5 \yt^2},\nonumber\\
\y^2-1 &=&\gTf \nt \frac{\vt^2+3r(1-\vt^2)}{\vt^5 \yt^2},\nonumber\\
1-\vt^2 \yt^2&=& 3 \gTf \n \frac{1-r(1-\v^2)}{\v^5 \y^2},\nonumber\\
\yt^2-1 &=&\gTf \n \frac{\v^2+3r(1-\v^2)}{\v^5 \y^2},
\eea
where
\be
r:=-\frac{\gap}{\ga}
\ee
is a dimensionless coupling constant that we shall use from now on in exchange for $\gap$.
Eliminating $\y$ and $\yt$ yields the two equations
\bea\label{gT4v-nonsymmetric1}
\n \gTf &=& \frac{\v^5(1-\vt^2)(3+\vt^2)}{[3+\v^2\vt^2-3r(1-\v^2)(1-\vt^2)]^2},\\
\nt \gTf &=& \frac{\vt^5(1-\v^2)(3+\v^2)}{[3+\v^2\vt^2-3r(1-\v^2)(1-\vt^2)]^2}.
\label{gT4v-nonsymmetric2}
\eea
Since causality implies $0<\v,\vt<1$, we see that solutions exist for arbitrarily high $\TT$ only when
the denominator on the right-hand side of (\ref{gT4v-nonsymmetric1}) is able to reach a zero {and is positive}.
This is the case when $r>1$, which thus turns out to be a necessary (as well as sufficient) condition
for ultraviolet completeness for
the simplest coupling rules (\ref{coupling-rule}); otherwise this model would exist only up to some finite value of $\TT$.

As shown in Appendix \ref{App:LowHigh}, the high-temperature behavior of the total system
is governed by the fact that the metric factors $\x,\xt,\y,\yt$ asymptote to linear functions
of the physical temperature $\TT$. Since 
the effective temperatures of the subsystems are given by $\To=\TT/\x$ and $\Tt=\TT/\xt$,
they stop growing together with $\TT$ and
instead saturate at finite values proportional to $\gamma^{-1/4}$.
For $r=2$ figure \ref{fig:effectiveTs} displays this behavior for equal and unequal subsystems, i.e., $\no=\nt$ and $\no\neq\nt$, respectively.

\begin{figure}[tbp]
\centering 
\includegraphics[width=.5\textwidth]{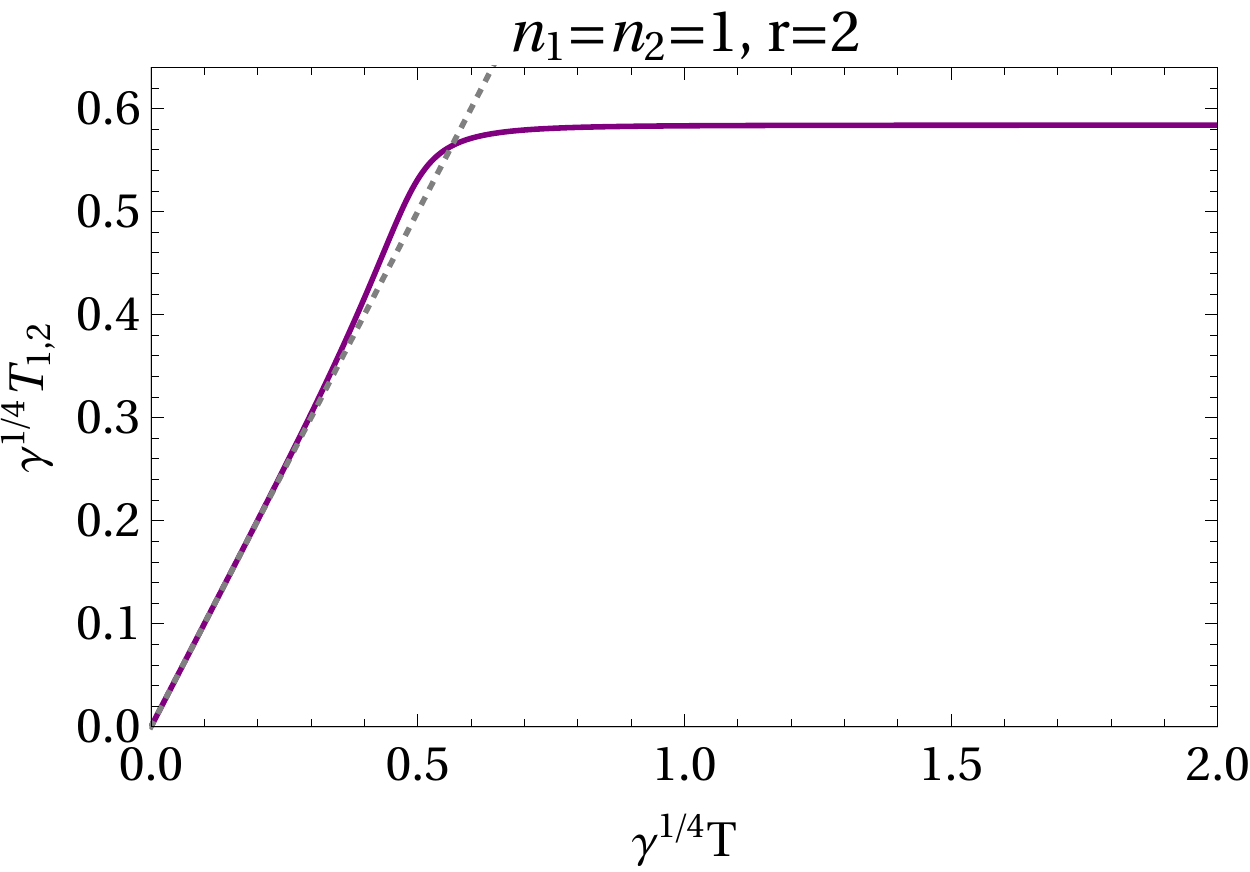}\includegraphics[width=.5\textwidth]{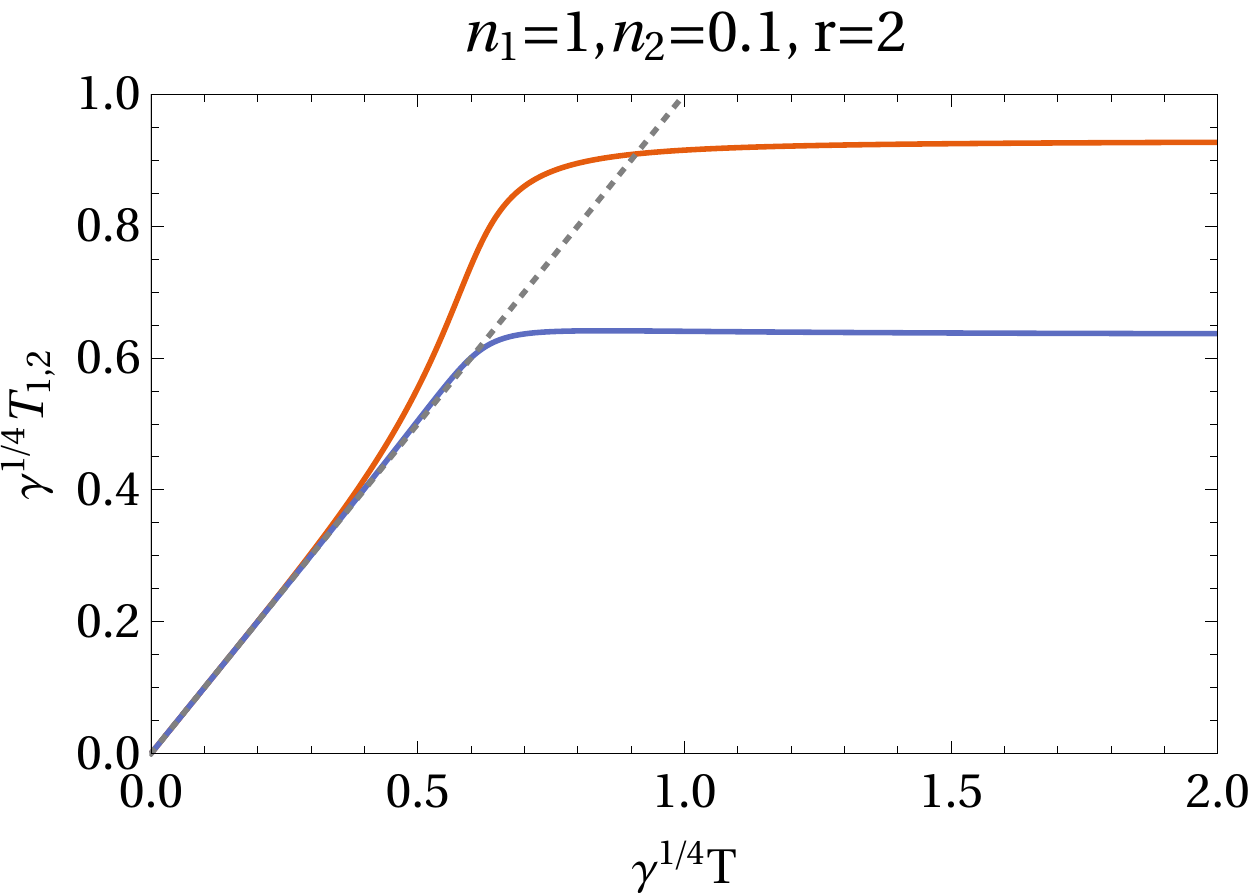}
\caption{\label{fig:effectiveTs} Effective temperatures of the subsystems as a function of the physical temperature
with $r=2$ for equal and nonequal ($\nt=\n/10$) subsystems (left and right panel, respectively). As the physical
temperature increases, the effective temperature of the subsystems first increases in line with the former
(the dotted line marks equality), but when $\cal T$ becomes larger than $\gamma^{-1/4}$, the effective temperatures
asymptote to a limiting value. This limiting value is larger for the subsystem with fewer degrees of freedom.}
\end{figure}

{Although the subsystems are conformal, when the two sectors interact, the full system in general
is no longer conformally invariant. With the simplest
coupling rules (\ref{coupling-rule}) and the resulting solution (\ref{EandP}) we find
\be
\ET-3\PT=\frac{6\ga\v^2\vt^2\no\nt\TT^8}{\x^2\xt^2}\left[
(3+\v^2\vt^2)-3r(1-\v^2)(1-\vt^2) \right].\label{eq:intmeasure_confsub}
\ee
Note that the term in square brackets in \eqref{eq:intmeasure_confsub} is the square root of the denominator in \eqref{gT4v-nonsymmetric2}; it is positive in the uncoupled case where $\v=\vt=1$, and it cannot change sign for any finite value of $\gTf$. Therefore, the conditions for causality $\ga>0$ and condition for ultraviolet completeness, $r>1$, imply that the interaction measure $\mathcal E-3P=-T^\mu_{\phantom{\mu}\mu}$ is positive (as is the case with lattice QCD results), and that the full system approaches conformality at large temperature $\TT$.

While in general we have to resort to numerical evaluations, one can also derive perturbative expansions
for all quantities (see Appendix \ref{App:LowHigh}).
For small couplings {or for small temperature,}\footnote{{When
writing down perturbative results, we shall assume that $\ga\TT^4$ and $\gap\TT^4$ are of the same order, i.e., that $r$ is of order 1.}} $\vert\ga\vert,\vert\gap\vert\ll \TT^{-4}$, the resulting $\x$, $\xt$, $\v$, and $\vt$ are all close to unity, and thus $\ET-3\PT \approx 24\gamma \no\nt \TT^8$, i.e., the full system approaches conformality at small temperature as expected.

This behavior can also be seen in the speed of sound (squared) of the full system, defined thermodynamically by
\be\label{csthdyndef}
c_s^2=\frac{d\PT}{d\ET}=\left(\frac{d\ln\ST}{d\ln\TT}\right)^{-1},
\ee
which expanded up to third order in $\gTf$ reads
\be
c_{s}(\TT)=\frac{1}{\sqrt{3}}-\frac{8}{\sqrt{3}}\ga\TT^{4}\frac{\alf\alft}{\alf+\alft}-32\sqrt{3}\ga^{2}\TT^{8}\frac{\alf\alft(\alf^{2}+\alft^{2})}{(\alf+\alft)^{2}}+\mathcal{O}((\ga\TT^4)^{3}).
\ee
With conformal subsystems the dependence on
$r=-\gap/\ga$ appears only at third order; in quantities which only depend on $\v$ and $\vt$, as is the case for the entropy,
also the third-order term is still independent of $r$.}

\subsubsection{Equal subsystems}

For the special case $\no=\nt$ where $\v=\vt$, the numerical solution of (\ref{gT4v-nonsymmetric1})
is displayed in
Fig.~\ref{fig:lightconespeeds-symmetric}
for various values of $r>1$.\footnote{Note that having equal equations of states does not imply that the subsystems
are identical. Later on, we shall consider hydrodynamical results with subsystems that have $\no=\nt$ but
different transport coefficients.}

\begin{figure}[tbp]
\centering 
\includegraphics[width=.75\textwidth]{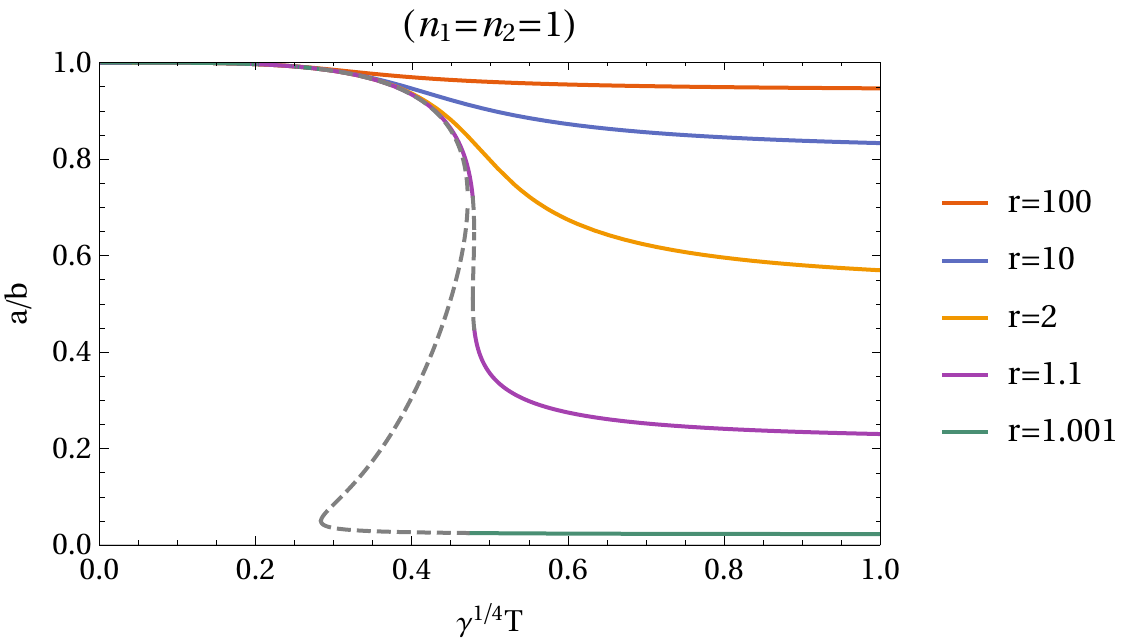}
\caption{\label{fig:lightconespeeds-symmetric} Effective light-cone speeds of the two subsystems with $\n=\nt=1$ and different values of $r=-\gap/\ga$.
Above $r=r_c\approx1.1145$ there is a unique solution for all values of $\gqT$ (full lines), while below $r_c$
there are ranges of $\gqT$ with three solutions (dashed lines).}
\end{figure}

It turns out that for $1<r<r_c \approx 1.1145$ more than one solution exists. This corresponds to a phase transition that will be discussed
in section \ref{subsubsec:phasetransition}. Concentrating first on the case $r>r_c$, the behavior of the pressure and the interaction measure (divided by $\TT^4$)
is shown in the left panel of Fig.~\ref{fig:Pcsr2equal} for a typical case ($r=2$).
Intriguingly, $\PT/\TT^4$ shows an increase somewhat reminiscent of the deconfinement crossover transition in QCD.

The speed of sound (squared) \eqref{csthdyndef} shown in the right panel of Fig.~\ref{fig:Pcsr2equal}
exhibits a pronounced dip, indicating a crossover as opposed to a phase transition as $\gqT$ is increased from the conformal situation at $\gqT=0$
to large values, where it asymptotes again to conformal value $1/3$.

Since $\ST/\TT^3\propto \v^{-3}$, the entropy increases from its interaction free value at $\TT=0$, where $\v=1$,
in parallel to the drop in $\v$ displayed in Fig.~\ref{fig:lightconespeeds-symmetric}

\begin{figure}[tbp]
\centering 
\includegraphics[width=.45\textwidth]{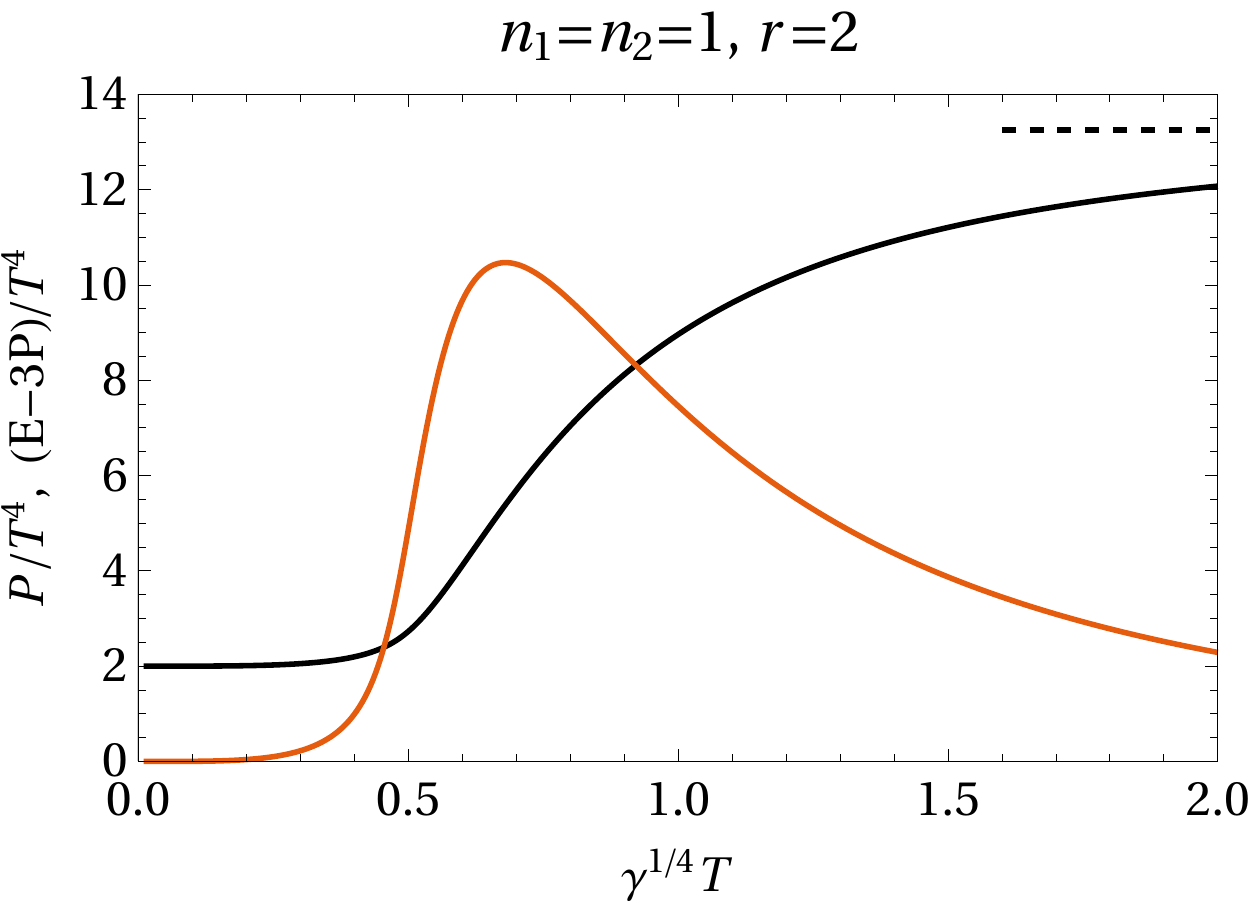}\qquad\includegraphics[width=.45\textwidth]{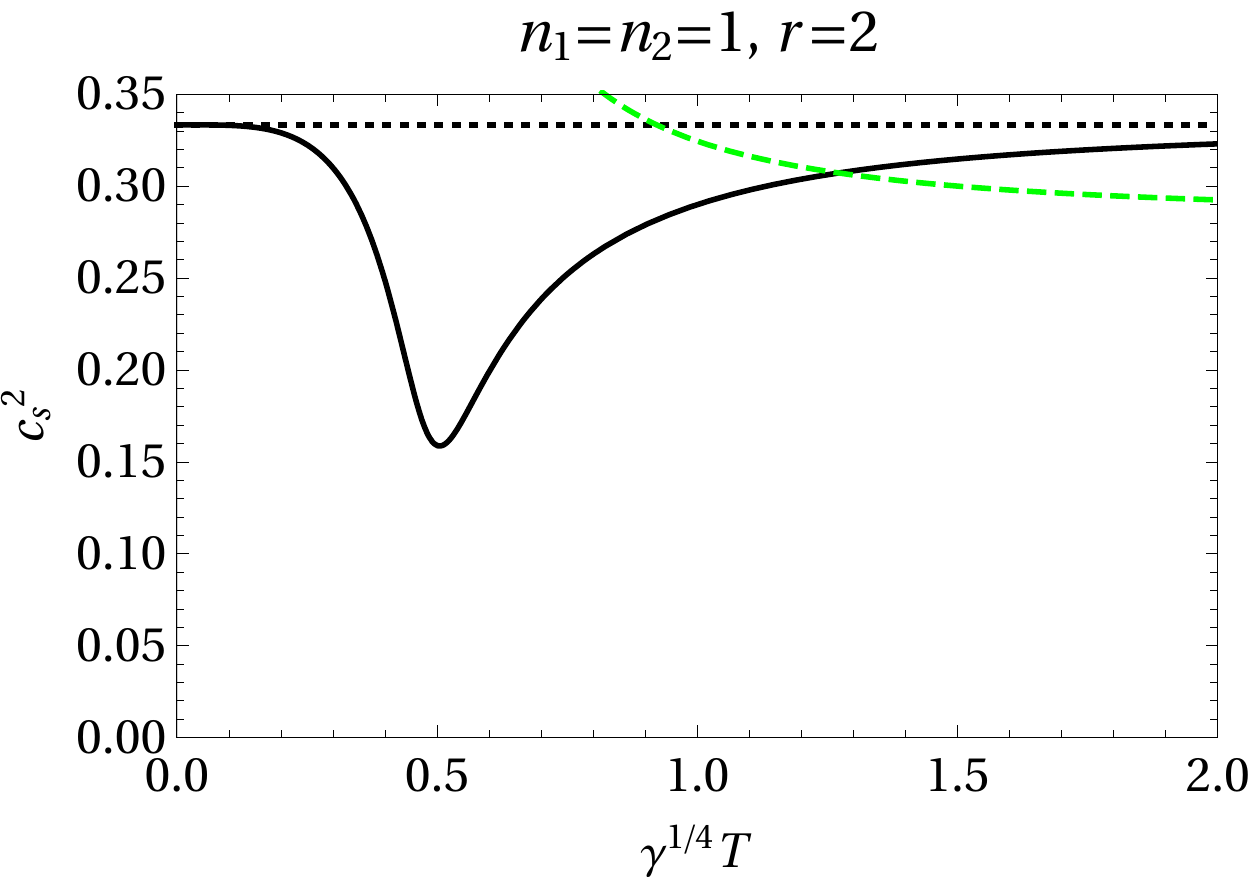}
\caption{\label{fig:Pcsr2equal} Left panel: Pressure (black line) and trace of the energy-momentum tensor (red) divided by $\TT^4$, with the asymptotic value of the pressure indicated by the short dashed line; 
right panel: speed of sound squared (full black line) -- both
for $\no=\nt=1$ and $r=2$. As $\gqT$ increases from small to large values, a crossover between regimes
with different values of $\PT/\TT^4$ takes place that is accompanied by a dip in the speed of sound which
takes on a conformal value in both asymptotic regimes. At large $\gqT$ and for sufficiently low values of $r$
(including the case $r=2$ at hand), the speed of sound
in the full system turns out to be larger than the effective lightcone speed $\v$ of the subsystems (green dashed line: $\v^2$). }
\end{figure}

In the case of two identical conformal subsystems, the relation between the effective
lightcone velocity $\v$ and $\gTf$ is given by the roots of a polynomial equation of 9th degree
(explicitly given in (\ref{gT4v-symmetric})), which in general can only be solved
numerically. The asymptotic value of $v$
is however determined by a simple quadratic equation which yields
\be\label{vasymp-symmetric}
v_\infty^2:=\lim_{\gTf\to\infty} v^2 = \frac{3 r-\sqrt{3} \sqrt{4 r-1}}{3 r-1}.
\ee
Evidently, the entire physical range $0<v_\infty<1$ is covered as $r$ varies between unity and infinity. 

Since the speed of sound $c_s$ approaches the conformal value $1/\sqrt3$ at large $\gqT$, for sufficiently small
values of $r$ (namely $r<7/3$), $c_s$ can be larger than the effective lightcone speed $\v$ of the subsystems.
This is no contradiction to causality, since besides dynamics within the subsystems, there is
also collective dynamics between them. (In Section \ref{Sec:BiHydroSound} this will be studied further.)

\subsubsection{Unequal subsystems}\label{subsubsec:uneqsys}

For unequal systems one can show (using formulae (\ref{gT4v-nonsymmetric1}) and (\ref{gT4v-nonsymmetric2})) that
there exist solutions for $\v$ and $\vt$ in the limit $\gTf\to\infty$ for any value of $\nt/\no$ and $r>1$. 
They are given
by the (sextic) equations
\be\label{vasymp-unequal}
\frac{3[r(1-\v_\infty^2)-1]^{5/2}}{(4r-1)\v_\infty^5[r(1-\v_\infty^2)+\v_\infty^2/3]^{1/2}}
=\frac{\nt}{\no}=
\frac{(4r-1)\vt_\infty^5[r(1-\vt_\infty^2)+\vt_\infty^2/3]^{1/2}}{3[r(1-\vt_\infty^2)-1]^{5/2}},
\ee
which have a unique solution in the domain $0<\v_\infty,\vt_\infty<\infty$ when $r>1$.
In the extreme limit that one of the systems completely dominates, say $\nt/\n\to0$,
the asymptotic effective lightcone velocity of the smaller system approaches zero, $\vt_\infty \sim O((\nt/\no)^{1/5})$, while
the dominant system has the limit $\v_\infty\to\sqrt{1-r^{-1}}$.

In Fig.~\ref{fig:Scsr2unequal}, the full numerical solution of the effective lightcone velocities is displayed for $\nt/\no=1/10$
and $r=2$ as well as the resulting entropies of the two subsystems. While the smaller subsystem has a much larger relative growth of 
$S/T^3$ than the larger subsystem, the latter remains dominant. (Considering again the extreme limit $\nt/\no\to0$,
$S_2/S_1$ changes from being of order $\nt/\no$ at low $\gTf$ to $(\nt/\no)^{2/5}$ at high $\gTf$.)

\begin{figure}[tbp]
\centering 
\includegraphics[width=.5\textwidth]{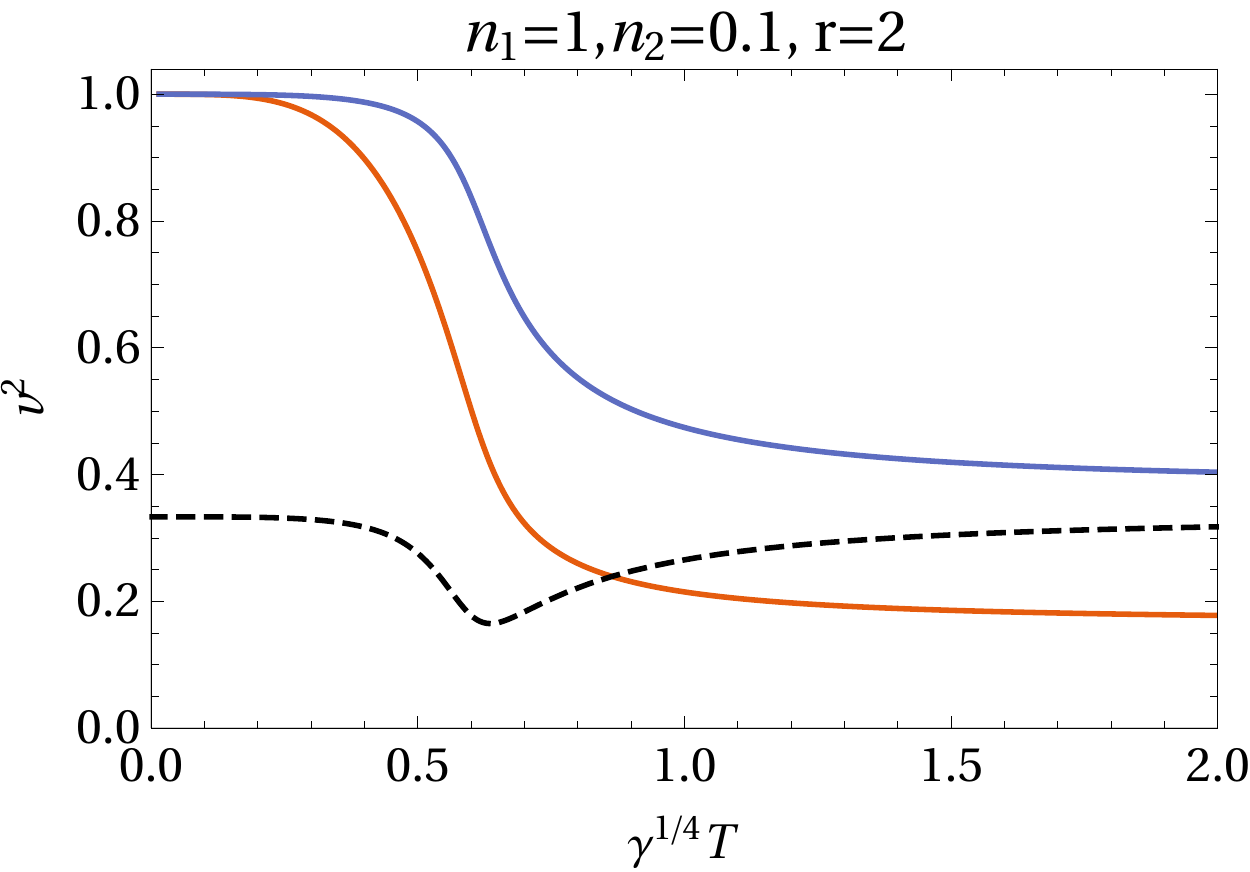}\includegraphics[width=.5\textwidth]{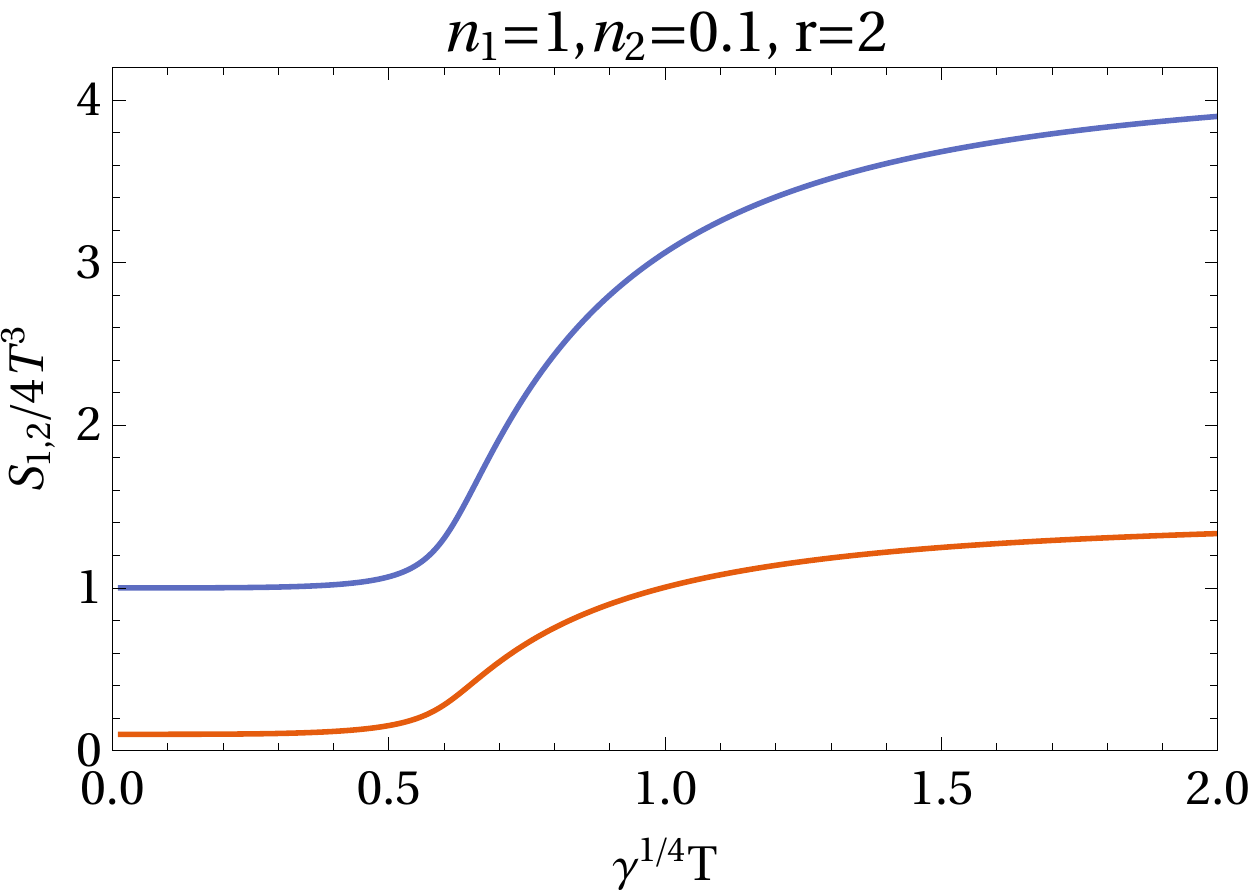}
\caption{\label{fig:Scsr2unequal} Unequal systems with $\no=1,\nt=0.1$ and $r=2$. Left panel: light-cone velocities 
squared
in the two subsystems ($\vo^2$: upper, blue line, $\vt^2$: lower, red line) compared to $c_s^2$ (dashed black line). 
Right panel: entropies of the two subsystems ($\ST_1$: upper, blue line, $\ST_2$: lower, red line).}
\end{figure}

At the value $r=2$ used in Fig.~\ref{fig:Scsr2unequal}, the behavior of the speed of sound is
similar to the case shown in Fig.~\ref{fig:Pcsr2equal}. Again, there is a dip at the crossover between
the regimes of small and large $\gqT$, where $c_s$ asymptotes to the conformal value $1/\sqrt{3}$.
In the case displayed in Fig.~\ref{fig:Scsr2unequal}, now only one of the effective
lightcone velocities, namely $\vt$, falls below the (conformal) value of the speed of sound at large $\gqT$.

\subsubsection{Phase transition}\label{subsubsec:phasetransition}

Perturbative expansions in the dimensionless parameter $\gTf$ turn out to work rather poorly and indeed have
to break down for $1<r<r_c$ where multiple solutions appear at finite values of $\gTf$, as shown in
Fig.~\ref{fig:lightconespeeds-symmetric} for $\no=\nt$. For $1<r<r_c\approx1.1145$
in the case $\no=\nt$ and $1<r<r_c\approx1.25$ for $\no\not=\nt$, this corresponds to a first-order phase
transition that turns into a second-order phase transition at $r_c$. 

In Fig.~\ref{fig:phasetransition}
pressure and entropy are plotted in the region around the first-order phase transition with $\no=\nt=1$ and $r=1.1$.%
\footnote{For $r\le 1$ and the simplest coupling rules (\ref{coupling-rule}),
two solutions for the pressure exist up to a maximal value of $\gTf$, where they merge
with different slopes and infinite second derivatives, after which there is (at least) no homogeneous and isotropic solution
to (\ref{coupling-rule}). One solution, whose beginning can be found
perturbatively, starts at zero pressure for $\gTf=0$; the other solution has smaller pressure (i.e., higher free energy)
and is thus thermodynamically disfavored.}
The range in $\gqT$ where the pressure has three solutions corresponds to the possibility of superheating or
supercooling (depending on whether the phase transition is approached from higher or lower temperatures).
This happens if one does not immediately switch to the thermodynamically preferred phase with higher pressure
(lower free energy). The third solution which directly connects the endpoints of superheating and supercooling
is always thermodynamically disfavored and cannot be accessed physically, because it comes with negative
specific heat (corresponding to the part of the curve for the entropy with negative slope).

\begin{figure}[tbp]
\centering 
\includegraphics[width=.45\textwidth]{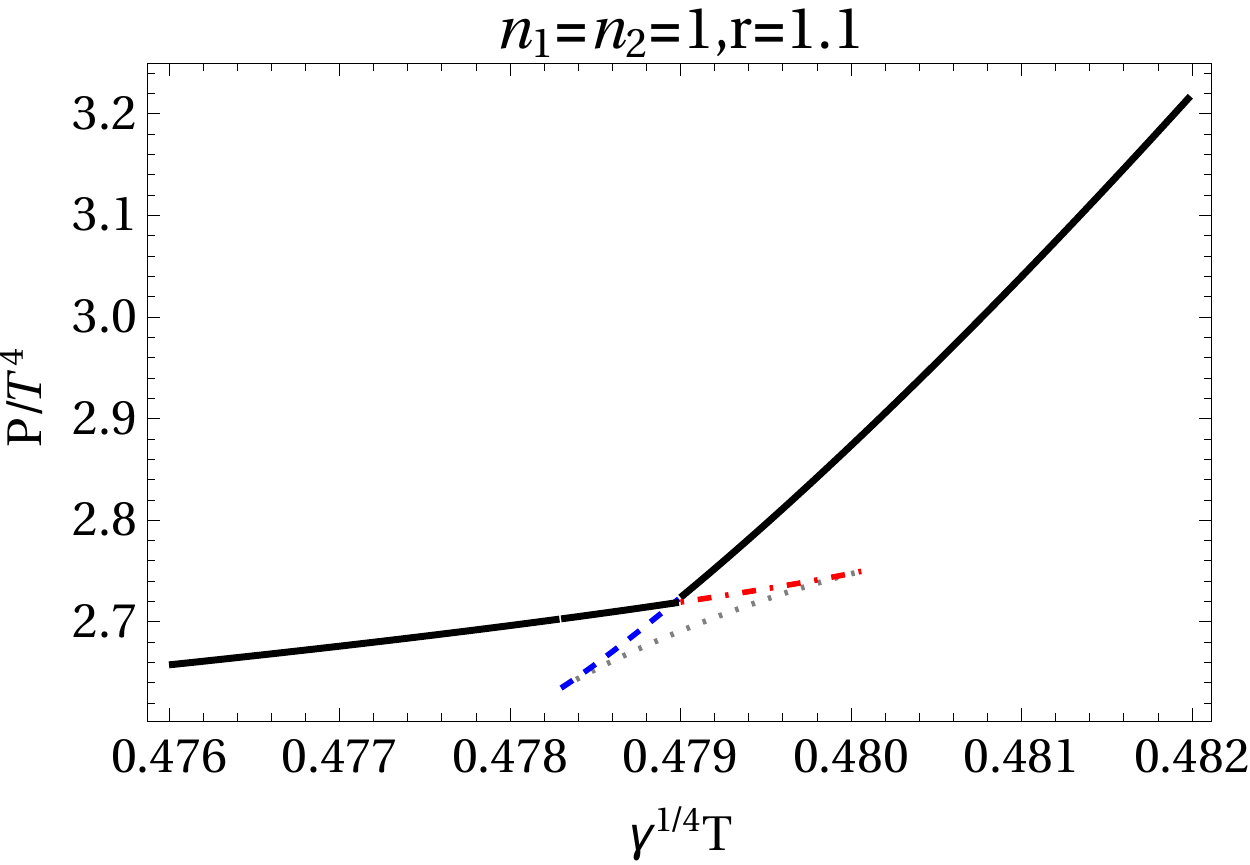}
\qquad\includegraphics[width=.45\textwidth]{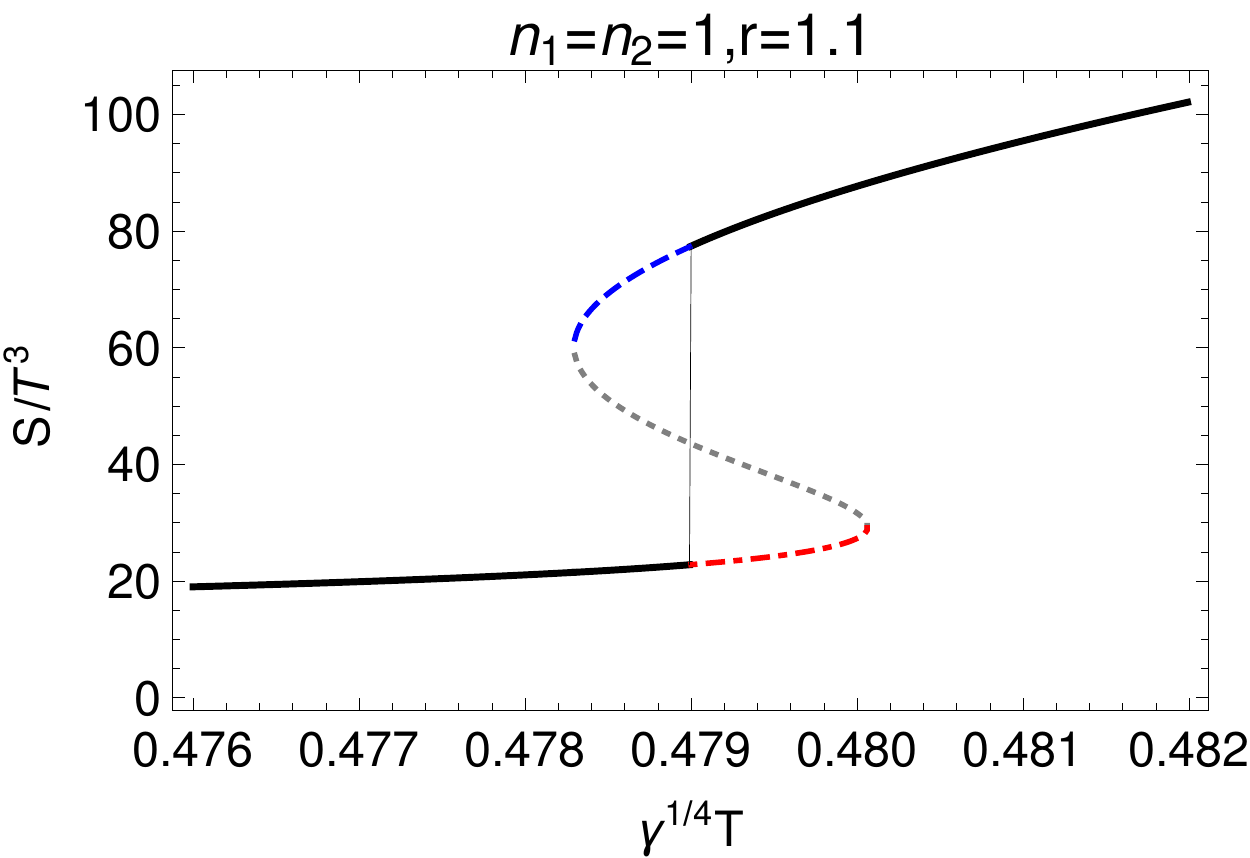}
\caption{\label{fig:phasetransition} Pressure (left panel) and entropy (right panel) around the first-order phase transition with $\no=\nt=1$ and $r=1.1$. The pressure 
of the ground state is given by the maximal value at each temperature. At the critical temperature
the slope changes discontinuously. The lines which extend smoothly 
beyond this point when coming from lower or higher temperatures correspond to superheating or supercooling phases,
respectively. (The lower line connecting the endpoints of supercooling and superheating corresponds to the 
entropy curve with negative slope and thus cannot be accessed physically.) The dotted line in the entropy curve
indicates the jump in the entropy that occurs when there is no supercooling or superheating.}
\end{figure}

In Fig.~\ref{fig:phasetransitionT1vsTt} the effective temperature of the subsystems is shown for the same
set of parameters. This shows a curious nonmonotonic behavior; at the phase transition the effective
temperature jumps and approaches the asymptotic value from above as the physical temperature goes to infinity.
In fact, although hardly perceptible in the left plot in Fig.~\ref{fig:effectiveTs}, the effective temperature 
also approaches the asymptotic value from above for $r=2$ in the crossover region; only for $r\gtrsim2.048$ (in the case
of $\no=\nt$) the effective temperature eventually shows monotonic behavior.

\begin{figure}[tbp]
\centering 
\includegraphics[width=.45\textwidth]{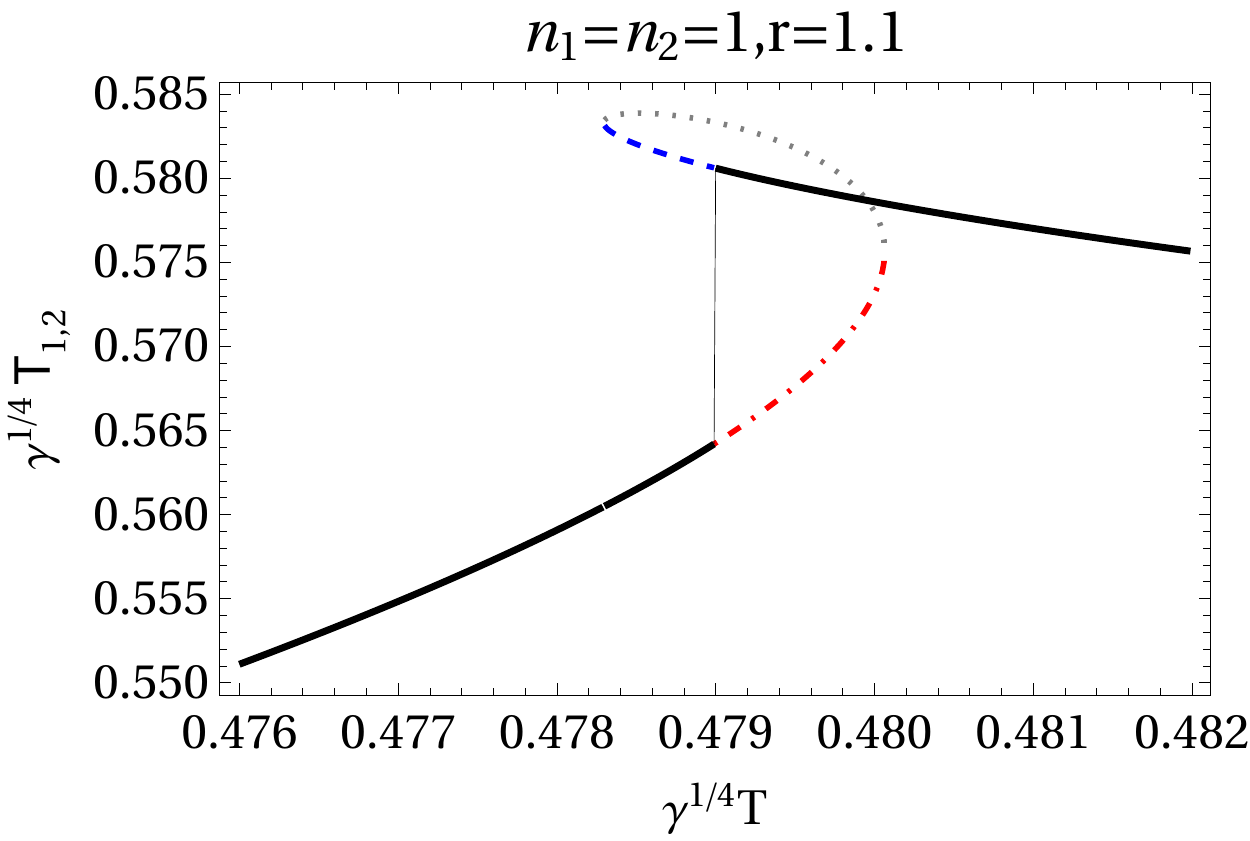}
\caption{\label{fig:phasetransitionT1vsTt} The behavior of the effective temperature of the subsystems 
during the first-order phase transition with $\no=\nt=1$ and $r=1.1$.
The dotted line in the entropy curve
indicates the jump in the effective temperature when there is no supercooling or superheating.}
\end{figure}

At $r=r_c$ the phase transition becomes second-order with continuous pressure and entropy.
In Appendix \ref{App:PhaseTransition} the parameters of the second-order phase transition are obtained in
closed form for $\no=\nt$. In particular the critical exponent $\alpha$ that characterizes the behavior
of specific heat, is obtained, with the result
\be\label{CVTalpha}
\CVT\sim |\TT-\TT_c|^{-\alpha}, \quad \alpha=\frac23,
\ee
which is independent of $\nt/\no$.

It is thus different from any mean-field result,
and it is also larger than the value in the Ising model ($\alpha\approx 0.11$) or in the polymer models ($\alpha
\approx 0.236$), which are the largest values occurring in $N$ vector models (for $N=1$ and $N=0$, respectively) \cite{Guida:1998bx}.
The comparatively large value of $\alpha$ in (\ref{CVTalpha}) is curiously close to that obtained
in the matrix model for deconfinement of Ref.~\cite{Pisarski:2012bj}, which yields $\alpha=3/5$.

The qualitative features of the phase transition are the same for unequal conformal subsystems: For $1 < r < r_c$, the transition is first order, at $r = r_c$ the phase transition is second order, and for $r>r_c$ it is a crossover. Furthermore, the critical value $r_c$ shows a rather weak dependence on $n_2/n_1$, it lies in the narrow interval $1.119\ldots < r_c < 1.25$, and the critical exponent $\alpha$ at the second-order phase transition point $r = r_c$ is always $2/3$  (for more details see Appendix \ref{App:PhaseTransition}).

\subsection{Massive subsystems}

\begin{figure}[tbp]
\centering 
\includegraphics[width=.55\textwidth]{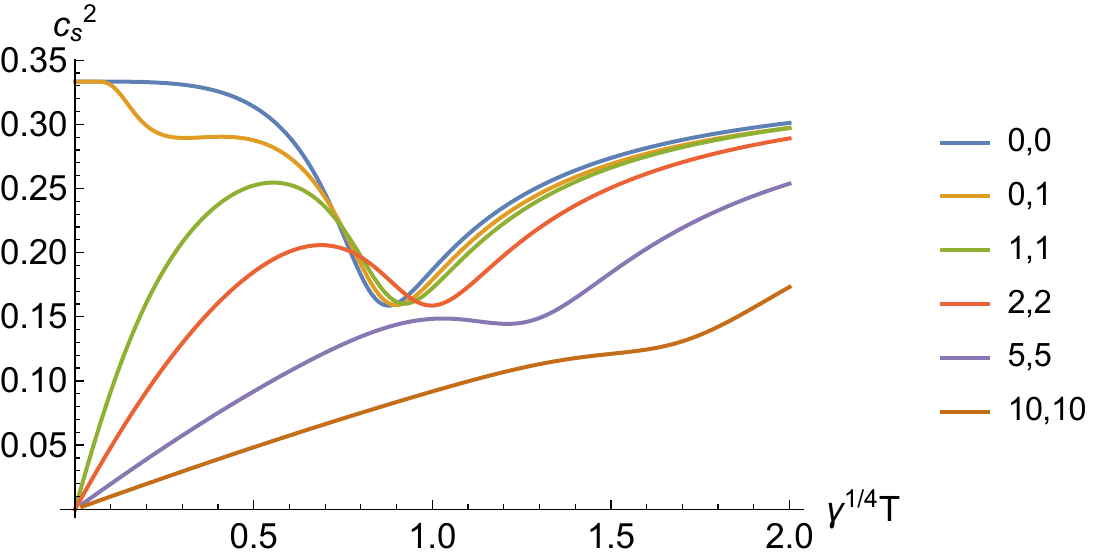}
\caption{\label{fig:nonconformalsubsystems} Speed of sound (squared) in two systems where one or both are replaced by a gas of
free massive bosons at $r=2$. If both systems are massive, the speed of sound starts from zero at zero temperature;
if one is still conformal, the lower end point remains at 1/3. 
The values given in the plot legend refer to the two masses in units of $\gamma^{-1/4}$.
(The massless case corresponds to $n_{1,2}=\pi^2/90$ in (\ref{bi-conformal}).)
}
\end{figure}

The simplest coupling rule
(\ref{coupling-rule}) with $r>1$ also makes sense for more general equations of state for the subsystems.
In Fig.~\ref{fig:nonconformalsubsystems} we display the results for the speed of sound (squared) that is
obtained by coupling two free Bose gases with various masses (for simplicity with $r=2$, where only a
crossover and no phase transition arises). When both subsystems have particles
with mass, the speed of sound starts from zero at $\gTf=0$, and approaches the conformal value
at large $\gTf$. (When one or both components contain massless particles, $c_s^2$ also starts from the
value $1/3$.)

The way approximate conformality is approached at high $\TT$ is again similar to the conformal case discussed above,
although we cannot demonstrate this analytically as in Appendix \ref{App:LowHigh}.
The high-temperature behavior (at $r>1$) is governed again by an asymptotically linear behavior of the metric
coefficients $\x,\xt,\y,\yt\sim \TT$. Such a behavior is at least consistent with the (simplest) coupling rules
(\ref{coupling-rule}): Once $\x,\xt,\y,\yt$ have grown sufficiently large, these equations are
homogeneous of degree two in the metric coefficients, provided the effective
temperatures $\To,\Tt$ become constant, which is the case when $\x,\xt\sim \TT$.

However, an important difference to the conformal case is that the trace-term $\Delta K\delta^\mu_\nu$
in the full \EM\ tensor is no longer subdominant, but in fact needed to cancel the contributions to
the trace of the full \EM\ tensor
at order $\TT^4$. This is a consequence of the form (\ref{Sdyn}) of the full entropy, $\ST=
\s(\To)\y^3+\st(\Tt)\yt^3\sim \TT^3$, together with 
thermodynamic consistency, $\ST=d\PT/d\TT$ (which is proved in Appendix \ref{App:ThdynConsistency}
for arbitrary equations of state of the subsystems).

We expect that it is equally possible to couple more involved equations of state 
than gases of free massive particles with the simplest coupling rule
and to obtain a UV-complete setup. However, we assume that the subsystems have self-interactions so that
they are able to thermalize on their own.

\section{Bi-hydrodynamics}\label{Sec:Bihydro}

In the following 
we investigate the linearized perturbations of the full hybrid system about thermal equilibrium
for given values of the hydrodynamic transport coefficients within the two subsystems, i.e.,
parameterising their \EM\ tensors to first order in the gradient expansion
according to
\be
t^\mn=(\e+\p)u^\mu u^\nu+\p g^\mn-2\eta_1 \sigma^\mn-\zeta_1 \theta P^\mn,
\ee
with $P^\mn=g^\mn+u^\mu u^\nu$, $\theta=\nabla_\mu u^\mu$, $\sigma^\mn=P^{\mu\alpha}P^{\mu\beta}\nabla_{(\alpha}u_{\beta)}
-\frac13 \theta P^\mn$, and similarly for the second subsystem with metric $\gt_\mn$.

Owing to the rotational symmetry of thermal equilibrium, the perturbations can be classified into three distinct sectors, which are called the shear, sound, and tensor channels. Each channel has distinct low energy characteristics. If we take the hydrodynamic limit in both sectors, only the shear and sound channels yield dynamic propagating modes with distinct forms of dispersion relations. The tensor channel in the bi-hydrodynamic limit consists only of a response local in space and time (without a pole) which is convenient for calculating the shear viscosity of the full hybrid system using the Kubo formula. For simplicity we will also analyze the case of conformal subsystems, and therefore we will set $\zeta_1 = \zeta_2 = 0$. Note $\zeta_{1,2}$ do not affect the shear channel in any case.

\subsection{Bi-hydro shear diffusion}\label{Sec:BiHydroShear}

In the shear sector, the velocity fields of both sectors point in the same direction but are orthogonal to the momentum (i.e., the direction of propagation)
of a perturbation. Without loss of generality we may assume that the momentum $\mathbf{k}$ is in the $z$-direction and the velocity fields are in the $x$-direction. The (normalized) velocity fields in both sectors including the infinitesimal linearized perturbations then assume the form:
 \begin{equation}\label{vel-shear}
\uo^\mu = \left(\frac{1}{\x},\wo e^{i (k z -\omega t)},0,0\right), \quad \ut^\mu = \left(\frac{1}{\xt},\wt e^{i (kz -\omega t)},0,0\right).
\end{equation}
The temperatures in both sectors remain unperturbed from their equilibrium values (in the shear channel).

Furthermore, we can consistently assume that the effective metrics are:
\begin{equation}\label{g-shear}
\g_{\mu\nu} = {\rm diag}(-\x^2, \y^2, \y^2, \y^2) + \delta \g_{\mu\nu},\quad \gt_{\mu\nu} = {\rm diag}(-\xt^2, \yt^2, \yt^2, \yt^2) + \delta \gt_{\mu\nu}
\end{equation}
with the non-vanishing components of $\delta g_{\mu\nu}$ and $\delta \gt_{\mu\nu}$ being:
\begin{eqnarray}\label{deltag-shear}
\delta \g_{01} = \bo e^{i(kz-\omega t)}, \quad \delta \g_{13} = \gammamet_{13} e^{i(kz-\omega t)}, \nonumber\\
\delta \gt_{01} = \bt e^{i(kz-\omega t)}, \quad \delta \gt_{13} = \gammamett_{13} e^{i(kz-\omega t)}.
\end{eqnarray}
Note that these metric perturbations 
preserve the norm of the velocity fields \eqref{vel-shear} at the linearized level.

It follows then that the hydrodynamic energy-momentum tensors of the two sectors including the linearized infinitesimal perturbations will assume the forms:
\begin{eqnarray}\label{deltat-shear}
 t^{\mu\nu} &=& {\rm diag}\left(\frac{\e (\To )}{\x^2},\frac{\p (\To )}{\y^2},\frac{\p (\To )}{\y^2},\frac{\p (\To )}{\y^2}\right) + \delta t^{\mu\nu}, \nonumber\\  \tit^{\mu\nu} &=& {\rm diag}\left(\frac{\et (\Tt )}{\xt^2},\frac{\pt (\Tt )}{\yt^2},\frac{\pt (\Tt )}{\yt^2},\frac{\pt (\Tt )}{\yt^2}\right) + \delta \tit^{\mu\nu} 
\end{eqnarray}
with the non-vanishing components of $\delta t^{\mu\nu}$ and $\delta \tit^{\mu\nu}$ being
\begin{eqnarray}\label{deltat-hydro-shear}
\delta t^{01} = - \frac{\p  \bo + (\p  +\e ) \wo  \x \y^2}{\x^2 \y^2} e^{i(kz-\omega t)}, 
\quad\delta t^{13} =\left(-\frac{\p \gammamet_{13}}{\y^4}-i k \frac{\eta_1 \wo}{\y^2}+ i \omega\frac{\eta_1\gammamet_{13}}{\x \y^4}\right)e^{i(kz-\omega t)},\nonumber\\
\delta \tit^{01} = - \frac{\pt  \bt +  (\pt  +\et )\wt \xt \yt^2}{\xt^2 \yt^2} e^{i(kz-\omega t)}, 
\quad\delta \tit^{13} =\left(-\frac{\pt \gammamett_{13}}{\yt^4}-i k \frac{\eta_2 \wt}{\yt^2}+ i \omega\frac{\eta_2\gammamett_{13}}{\xt \yt^4}\right)e^{i(kz-\omega t)}.\nonumber\\
\end{eqnarray}

The hydrodynamic equations (\ref{WI1}) and (\ref{WI2})
of the two sectors in the two (self-consistently perturbed) effective metrics are:
\begin{eqnarray}\label{hydro-shear}
\omega \frac{(\p  + \e )(\bo + \wo  \x \y^2)}{\x^2 \y^2} = - ik^2\frac{\eta_1 \wo }{\y^2}+ i \omega k\frac{\eta_1\gammamet_{13}}{\x \y^4}, \nonumber\\
\omega \frac{(\pt  + \et )(\bt+\wt \xt \yt^2)}{\xt^2 \yt^2} = - ik^2\frac{\eta_2 \wt}{\yt^2}+ i \omega k\frac{\eta_2\gammamett_{13}}{\xt \yt^4}.
\end{eqnarray}
These hydrodynamic equations automatically guarantee the conservation of the full energy-momentum tensor at the linearized level provided the metric perturbations $\bo$, $\bt$, $\gammamet_{13}$ and $\gammamett_{13}$ are solved self-consistently in terms of the physical variables $\wo$ and $\wt$ using the linearized version of the effective metric coupling equations. Once again we will assume that only the couplings $\ga$ and $\gap\equiv -r\ga$ are non-vanishing.

To proceed further, we will also assume that each fluid is conformal with equations of state given by \eqref{bi-conformal}. We will also parametrize:
\begin{equation}
\eta_1 = \frac{\kap}{\pi}  \no\To ^3, \quad \eta_2 =  \frac{\kapt}{\pi}  \nt\Tt ^3
\end{equation}
so that:
\begin{equation}
\frac{4\pi\eta_1}{\s }= \kap , \quad \frac{4\pi\eta_2}{\st } = \kapt .
\end{equation}

With a Minkowski background metric,
the linearized coupling equations determining $\bo$, $\bt$, $\gammamet_{13}$ and $\gammamett_{13}$ are simply
\begin{eqnarray}\label{coupling-shear-sector}
\delta \g_{03} = -\ga \delta \tit^{03}\xt\yt^3, \quad \delta \g_{13} = \ga \delta \tit^{13}\xt\yt^3, \nonumber\\
\delta \gt_{03} = -\ga \delta{t}^{03}{\x}{\y}^3, \quad \delta \gt_{13} = \ga \delta{t}^{13}{\x}{\y}^3.
\end{eqnarray}
With  \eqref{deltag-shear} and \eqref{deltat-hydro-shear} the solutions are:
\begin{eqnarray}
\bo &=& \frac{4 \ga \alft \Tt^4 \x \yt \left(\xt
   \yt^2\wt -\ga \alf \To^4  \y^3\wo\right)}{- \x \xt+\ga^2 \alf
   \alft  \y \yt \To^4 \Tt^4 }, \\\nonumber
\gammamet_{13} &=&\frac{i \ga k \alft \Tt^3 \y \left(\pi  \kapt \wt
   \xt \yt^2-\ga \kap \alf \To^3 \wo \x \y (\pi 
   \Tt \xt-i \kapt \omega )\right)}{\pi ^2 \left(\ga^2
   \alf \alft \To^4 \Tt^4 \x \xt-\y \yt\right)+\ga^2\alf \alft \To^3 \Tt^3(
   \kap \kapt  \omega ^2-i
   \pi  \omega  (\kapt \To
   \x+\kap \Tt \xt))}, 
\end{eqnarray}
and similarly for $\bt$ and $\gammamett_{13}$.
Inserting them into the linearized hydrodynamic equations \eqref{hydro-shear} yields equations for $\wo$ and $ \wt$
of the form
\begin{equation}\label{Q}
Q_{AB}(\omega, k) \w_B = 0,
\end{equation}
where $\w_A = (\wo,  \wt)$ and $Q_{AB}$ is a $2 \times 2$ matrix. The eigenmodes have dispersion relations $\omega(k)$ for which the determinant of $Q$ vanishes, i.e.
\begin{equation}\label{detQ}
{\rm det}Q(\omega(k), k) = 0,
\end{equation}
and the corresponding eigenvectors involve a momentum dependent combination of $\wo$ and $\wt$. It is to be noted that these modes are the intrinsic perturbations of the system which can exist without any extrinsic drive such as a perturbation to the fixed background metric $\gB_\mn$ where the full system lives.

Of particular interest are the shear-diffusion modes whose dispersion relations assume the characteristic form:
\begin{equation}
\omega_I = - i D_I k^2 + \mathcal{O}(k^3),
\end{equation}
where the index $I$ labels different solutions.

As discussed before, we can solve all equilibrium quantities as functions of $\TT $, $\ga$ and $\gap$ so that we can also express $D_I$ as functions of these variables. The perturbative expansions of the shear diffusion constants $D_I$ are given by:
\begin{eqnarray}\label{diff-pert}
D_a(\TT ) &=&\frac{\kap}{4 \pi  \TT}-\frac{\ga \kap
   \alft \TT^3}{\pi }+\frac{\ga^2 \kap \alft \TT^7 [\alft(\kap -\kapt)+\alf(9
   \kapt-5 \kap )]}{\pi
    (\kap-\kapt)}+\mathcal{O}(\ga^3
) ,\nonumber\\
D_b(\TT ) &=& \frac{\kapt}{4 \pi  \TT}-\frac{\ga
   \kapt \alf \TT^3}{\pi }+\frac{\ga^2 \kapt \alf \TT^7 [\alf(\kap
   -\kapt)-\alft(9 \kap -5 \kapt)]}{\pi  (\kap-\kapt)}+\mathcal{O}(\ga^3
).\quad
\end{eqnarray}
In the decoupling limit, $\gqT \rightarrow 0$ (with fixed $r$), these two shear diffusion modes clearly reduce to individual shear diffusion modes of
the subsystems 1 and 2; with nonzero coupling they instead involve both subsystems nontrivially. The propagating mode corresponding to the first diffusion constant $D_a$ involves velocity amplitudes with\footnote{Note that the combination of $\wo$ and $\wt$ in the propagating mode is $k-$independent. This is so because each element in the matrix $Q$ in \eqref{Q} is $\mathcal{O}(k^2)$ at the leading order on-shell, i.e.\ when $\omega = -i D_{a,b}k^2+ \cdots$.}
\begin{equation}
\wt = \left(\frac{4 \alf \kap}{(\kap -\kapt)}\ga \TT^4 + \mathcal{O}(\ga^2\TT^8)\right) \wo
\end{equation}
and therefore it is indeed localized mostly in the first subsystem when $\ga\TT^4$ is small. Similarly, the other propagating mode has
\begin{equation}
\wo = -\left(\frac{4 \alft \kapt}{(\kap -\kapt)}\ga \TT^4 + \mathcal{O}(\ga^2\TT^8)\right) \wt
\end{equation}
and thus is localized mostly in the second subsystem for small $\ga\TT^4$. For finite $\ga\TT^4$, both these modes receive significant contributions from both subsystems (see Fig.~\ref{fig:shearmix}).

Furthermore, the dependence on $\gap$ of the perturbative expansions \eqref{diff-pert} start only at third order in the perturbative expansion -- so this dependence is weak at small $\gTf$. We also note that the perturbation expansion in $\gTf$ evidently breaks down when $|\kap-\kapt| \lesssim \gTf $, 
irrespective of the values of $\alf $ and $\alft $. 

In the coincidence limit of $\kap=\kapt=\kappa$,
we instead obtain the following perturbative series
\begin{eqnarray}\label{diff-pert-2}
D_{a}(\mathcal{T}) & = & \frac{\kappa}{4\pi\TT},\\
D_{b}(\mathcal{T}) & = & \frac{\kappa}{4\pi\TT}-\frac{\ga\TT^{3}\kappa(\alf+\alft)}{\pi}+\frac{\ga^{2}\TT^{7}\ensuremath{\kappa}\left(\alf^{2}-10\alf\alft+\alft^{2}\right)}{\pi}+\mathcal{O}(\ga^{3}\TT^{11}),\nonumber
\end{eqnarray}
where one of the diffusion modes turns out to be independent of $\gTf$. The propagating mode corresponding to this $\gTf$-independent diffusion constant has
\begin{equation}
\wt =  \left( 1 + \frac{3}{2}(\alf -\alft) \gTf + \mathcal{O}(\ga^2,\gap^2)\right) \wo.
\end{equation}
When $\alf = \alft$, i.e.\ when the two subsystems are identical, then the propagating mode is exactly given by $\wt =  \wo$ (parallel and equal motion within the subsystems). In any case, this mode gets significant contributions from both subsystems \textit{even in the decoupling limit $\ga, \gap \rightarrow 0$}. The other propagating mode corresponding to the second diffusion constant $D_b$ in \eqref{diff-pert-2} is the following combination of $\wo$ and $\wt$ where
\begin{equation}
\wo = -\frac{\alft}{\alf} \left(1+\frac{9}{2}(\alf-\alft)\gTf+ \mathcal{O}(\ga^2, \gap^2)\right)\wt.
\end{equation}
When $\alf = \alft$, this mode is exactly given by $\wo = -\wt$ (anti-parallel and equal motion within the subsystems). This mode evidently gets significant contributions from
both subsystems 
{even in the decoupling limit}
$\gqT \rightarrow 0$ (as long as $|\kap-\kapt|\ll \gqT$). The nonperturbative dependence of $\wt/\wo$ on $\gqT$
is displayed in Fig.~\ref{fig:shearmix}.

\begin{figure}[tbp]
\centering 
\includegraphics[width=.5\textwidth]{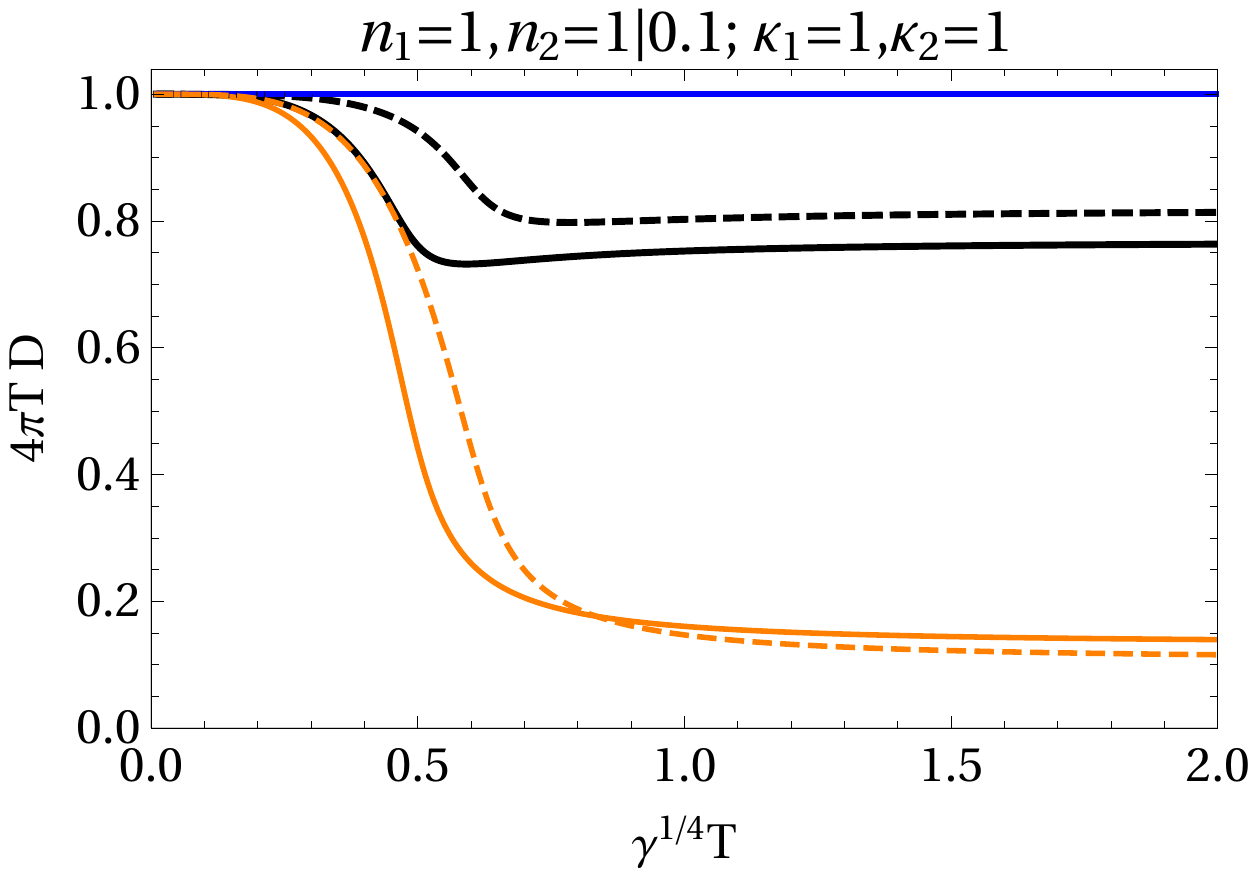}\includegraphics[width=.5\textwidth]{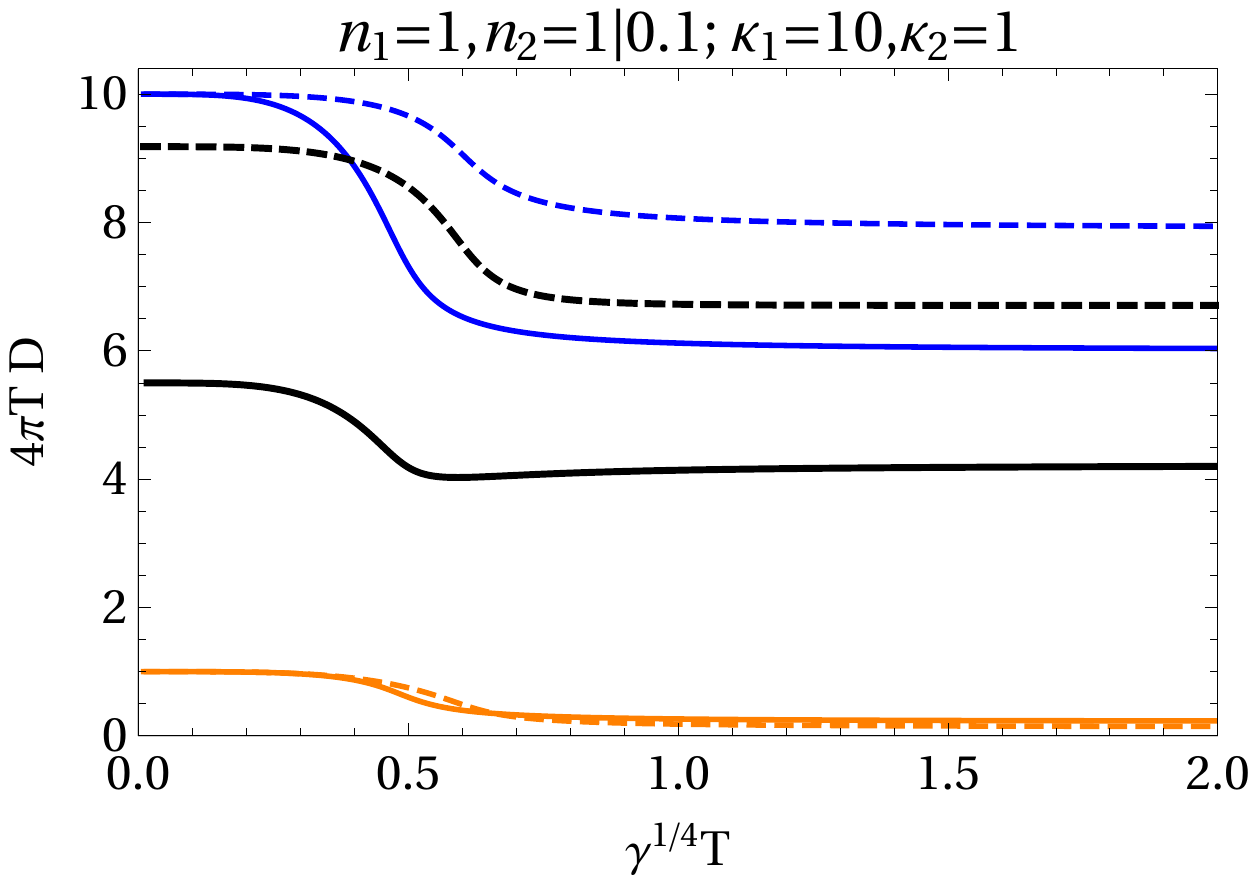}
\caption{\label{fig:TDs}Shear diffusion constants $D_{a,b}$ (blue and orange lines) corresponding to shear eigenmodes
in the hybrid fluid model for different parameters as a function of $\gqT$ compared to
the overall (Kubo) shear diffusion constant $\mathcal{D}$ (red lines) corresponding to the total shear viscosity $\eta/\ST\equiv T \mathcal{D}$.
Full and dashed lines correspond to equal numbers of degrees of freedom, $\no=\nt=1$, and unequal ones, $\no=1, \nt=1/10$, respectively.
The left panel has equal values of individual shear viscosities $\kappa_i=4\pi\eta_i/s_i=1$, the right panel has
$\kappa_1=10$ so that the first system corresponds to a more weakly coupled sector.
}
\end{figure}

\begin{figure}[tbp]
\centering 
\includegraphics[width=.5\textwidth]{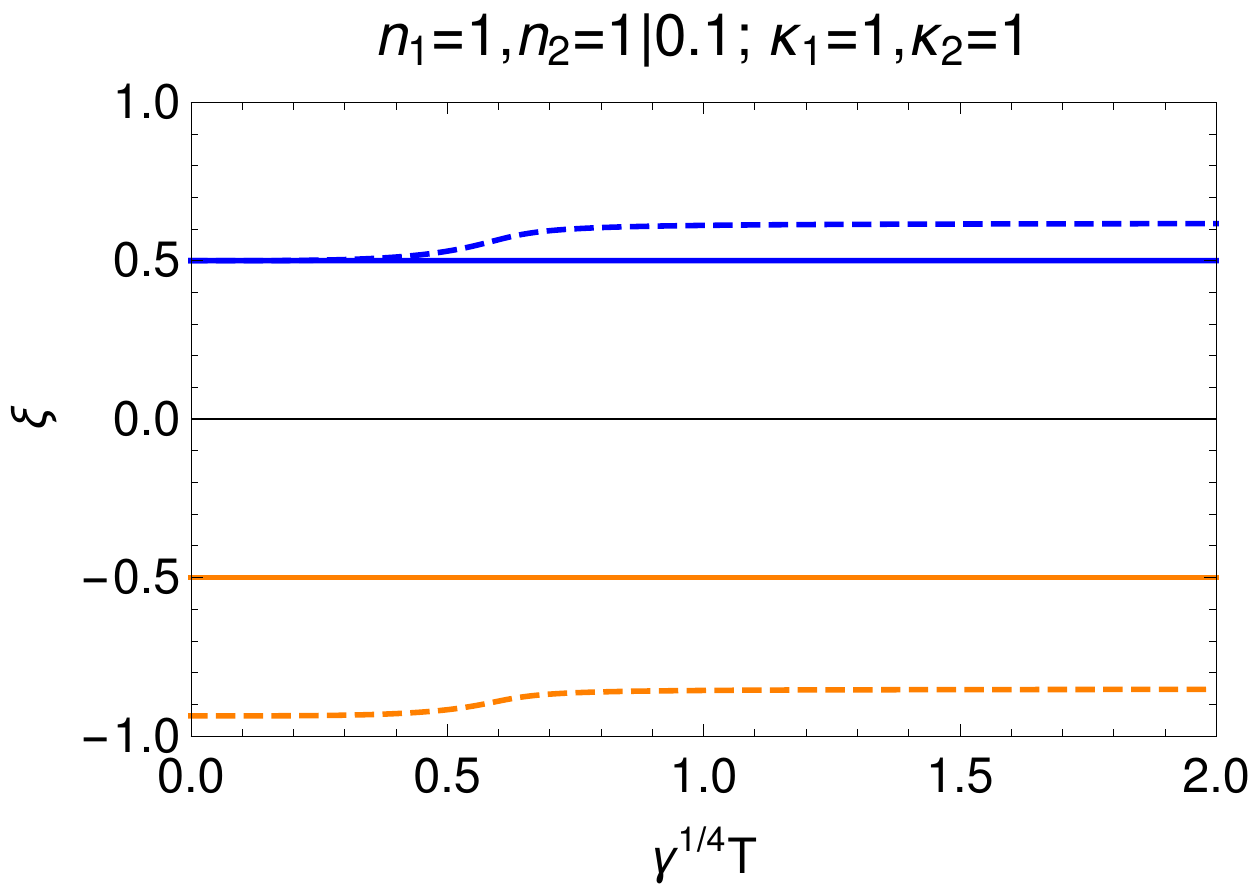}\includegraphics[width=.5\textwidth]{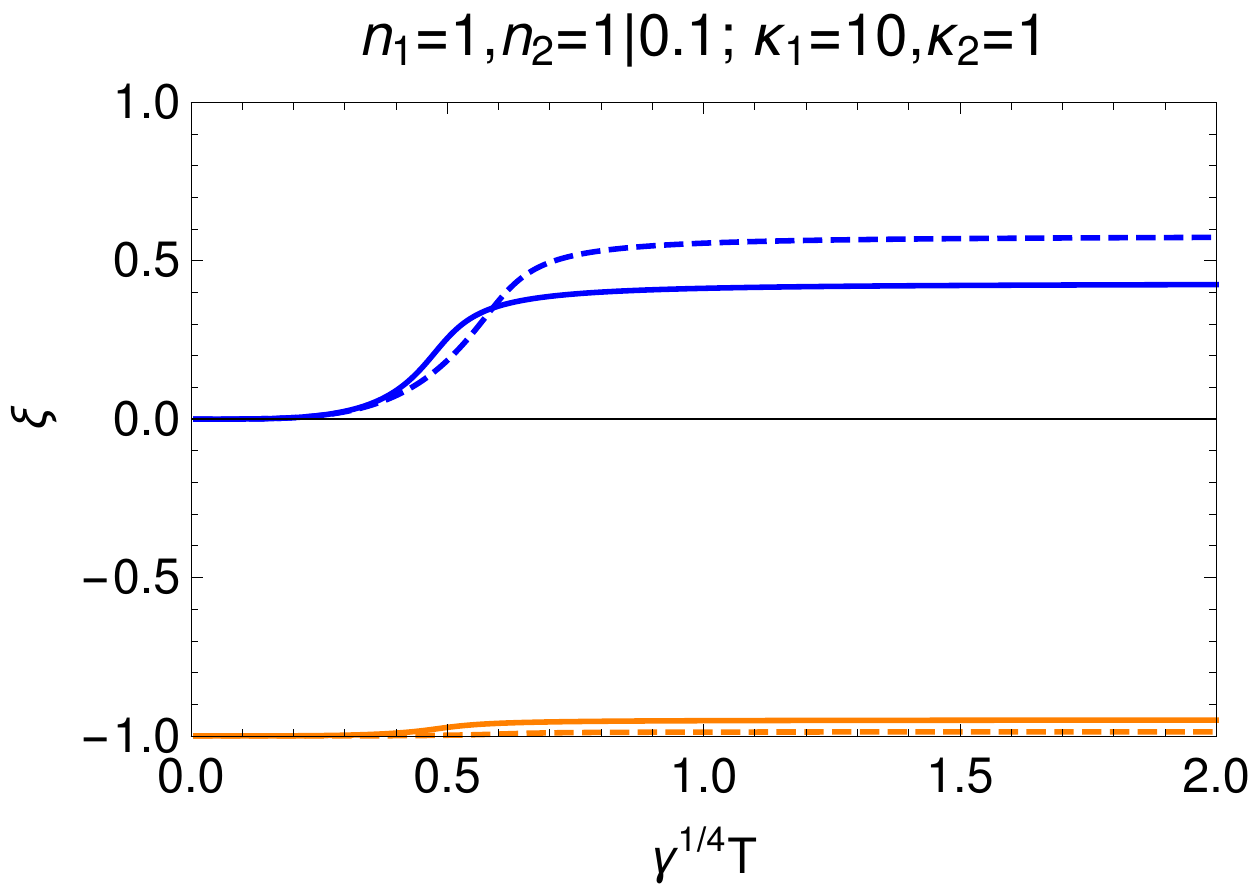}
\caption{\label{fig:shearmix}
The relation between the velocity amplitudes $\wo$ and $\wt$ of the shear eigenmodes displayed in Fig.~\ref{fig:TDs}
in the form $\xi\equiv \frac{2}{\pi}\arctan(\wt/\wo)$. A value of $\xi=0$ or $\xi=\pm1$ (with these two latter values to be identified)
means that the mode is carried only by subsystem 1 or 2, respectively;
$\xi=0.5$ or $\xi=-0.5$ corresponds to exactly equal amplitudes with equal or opposite phase.
}
\end{figure}

In order to define the shear viscosity of the full system, we can consider the tensor channel.  Consider an extrinsic \textit{homogeneous} perturbation such that the background metric in which the full system lives is perturbed by $h_{\mu\nu}(t)$ whose only non-vanishing component is $h_{13}(t)$. How does the full system respond? The coupling equations imply that the response involves homogeneous perturbations of the  effective metrics $\gammamet_{13}(t)$ and $\gammamett_{13}(t)$. Furthermore, the hydrodynamic equations of the individual systems in the individual effective metrics (cf. \eqref{hydro-shear}) imply that the velocity fields $\wo$ and $\wt$ also vanish for homogeneous $\gammamet_{13}$ and $\gammamett_{13}$ up to second order in the derivative expansion along with the perturbations of the temperatures in each sector.  Therefore, the linearized perturbations of the hydrodynamic energy-momentum tensors in the individual sectors assume the forms
\begin{eqnarray}\label{Kubo-ind}
\delta t^{13} &=& - \frac{\p }{\y^4} \gammamet_{13} - \frac{\eta_1}{\x \y^4} \dot\gammamet_{13} + \mathcal{O}(\exppar^2), \nonumber\\
\delta \tit^{13} &=& - \frac{\pt }{\yt^4} \tilde{\ga}_{13} - \frac{\eta_2}{\xt\yt^4} \dot{\tilde{\ga}}_{13} + \mathcal{O}(\exppar^2),
\end{eqnarray}
while the coupling rule (\ref{coupling-rule}) 
implies that the effective metric perturbations is 
determined by the extrinsic perturbation $h_{13}(t)$ 
according to the coupled linear equations
\begin{eqnarray}\label{couple-Kubo}
\gammamet_{13} &=& h_{13}\left(1- 2\ga \pt \xt \yt +\gap \left(- \frac{\et}{\xt^2}+ 3 \frac{\pt}{\yt^2}\right)\xt \yt^3\right) - \ga \pt  \frac{\xt}{\yt}\tilde{\ga}_{13} +\mathcal{O}(\exppar) , \nonumber\\
\tilde{\ga}_{13} &=& h_{13}\left(1- 2\ga \p \x \y +\gap \left(- \frac{\e}{\x^2}+ 3 \frac{\p}{\y^2}\right)\x \y^3\right) - \ga \p  \frac{{\x}}{{\y}}{\ga}_{13}+\mathcal{O}(\exppar).
\end{eqnarray}
Solving the above, we obtain
\begin{eqnarray}\label{g13s}
\gammamet_{13} &=& \frac{\left(1- 2\ga \pt \xt \yt +\gap \left(- \frac{\et}{\xt^2}+ 3 \frac{\pt}{\yt^2}\right)\xt \yt^3\right)- \ga \pt  \frac{\xt}{\yt}\left(1- 2\ga \p \x \y +\gap \left(- \frac{\e}{\x^2}+ 3 \frac{\p}{\y^2}\right)\x \y^3\right)}{1- \ga^2 \p  \pt  \frac{\x \xt}{\y \yt}}h_{13} \nonumber\\ &&+ \mathcal{O}(\exppar), \nonumber\\ 
\gammamett_{13} &=& \frac{\left(1- 2\ga \p \x \y +\gap \left(- \frac{\e}{\x^2}+ 3 \frac{\p}{\y^2}\right)\x \y^3\right)- \ga \p  \frac{{\x}}{{\y}}\left(1- 2\ga \pt \xt \yt +\gap \left(- \frac{\et}{\xt^2}+ 3 \frac{\pt}{\yt^2}\right)\xt \yt^3\right)}{1- \ga^2 \p  \pt  \frac{\x \xt}{\y \yt}}h_{13}\nonumber\\&& + \mathcal{O}(\exppar).
\end{eqnarray}
One can then readily compute the energy-momentum tensor of the full system including the linearized perturbation. We find that it assumes the standard hydrodynamic form with vanishing velocity and temperature perturbations.\footnote{Note it is a priori not obvious that even if the individual sector energy-momentum tensors are hydrodynamic, the full energy-momentum tensor also assumes a hydrodynamic form. This is particularly so because there are two independent entropy currents. Although in this specific example, the full energy-momentum tensor does indeed assume a hydrodynamic form, in the following subsection we will find counterexamples.} Explicitly,
\begin{equation}\label{Kubo}
\delta T^{13} = -\mathcal{P} h_{13}- \eta\dot{h}_{13} + \mathcal{O}(\exppar^2)
\end{equation}
where $\mathcal{P}$ is indeed the equilibrium pressure of the full system given by \eqref{EandP}. It also follows that the shear viscosity $\eta$ of the full system is given by
\begin{eqnarray}\label{eta-Kubo}
\eta = \frac{1}{1- \ga^2 \p  \pt  \frac{\x \xt}{\y \yt}}
&&\Biggl\{\eta_1 \y\biggl[\left(1- 2\ga \pt \xt \yt +\gap \left(- \frac{\et}{\xt^2}+ 3 \frac{\pt}{\yt^2}\right)\xt \yt^3\right)\nonumber\\
&&\qquad - \ga \pt  \frac{\xt}{\yt}\left(1- 2\ga \p \x \y +\gap \left(- \frac{\e}{\x^2}+ 3 \frac{\p}{\y^2}\right)\x \y^3\right)\biggr]\nonumber\\
&&
+\eta_2 \yt\biggl[\left(1- 2\ga \p \x \y +\gap \left(- \frac{\e}{\x^2}+ 3 \frac{\p}{\y^2}\right)\x \y^3\right)\nonumber\\
&&\qquad - \ga \p  \frac{{\x}}{{\y}}\left(1- 2\ga \pt \xt \yt +\gap \left(- \frac{\et}{\xt^2}+ 3 \frac{\pt}{\yt^2}\right)\xt \yt^3\right)\biggr]\Biggr\}.
\end{eqnarray}
It is to be noted that the bulk viscosity plays no role in the shear sector or in the response to a homogeneous $h_{13}(t)$ perturbation of the background metric. So even if the individual sectors have bulk viscosities all our results above remain valid. Given that $\ST$ of the full system is given by (\ref{Sdyn}) we readily obtain $\eta/\ST$. We may thus define the Kubo diffusion constant:
\begin{equation}
\mathcal{D} \equiv \frac{\eta}{\TT \ST} = \frac{\kap \alf +\kapt \alft  }{4\pi \TT(\alf + \alft)} + \mathcal{O}(\gamma).
\end{equation}

In Fig.~\ref{fig:TDs} the shear diffusion constants $D_I$ corresponding to shear eigenmodes are compared with the overall
diffusion constant $\mathcal{D}$ corresponding to the total shear viscosity $\eta/\ST\equiv T \mathcal{D}$ for various
parameters. The overall result obtained from the Kubo formula
is seen to be always in between the two $D_I$'s. 
The left panel shows the situation for two strongly coupled systems with $\eta_i/s_i=1/4\pi$, the right panel
the one for a more weakly coupled system $\mathfrak{S}_1$.
The dashed lines in both panels 
corresponds to the case that system $\mathfrak{S}_1$ contributes dominantly
to the pressure ($\no>\nt$). In this case the overall viscosity is closer to the viscosity of the dominant subsystem.

All results for the shear diffusion constants or specific viscosities decrease when the effective coupling $\gqT$ is increased from zero. In the case of the full viscosity there is
a slight nonmonotonic behavior in the crossover region between weak and strong coupling between the subsystems. At
large coupling all results appear to saturate at finite values.

Solving \eqref{detQ} one in fact obtains two additional eigenmodes which are spurious. First, these are non-hydrodynamic, meaning that $\omega$ is finite as $k$ vanishes. Second, when $k$ vanishes, these {eigenmodes} correspond to spontaneous fluctuation of the effective metric components $\gammamet_{13}$ and $\gammamett_{13}$ without involving any fluctuation of $\wo$, $\wt$ or any external background metric fluctuation (i.e., perturbation of $\gB_\mn$). Such a freaky fluctuation is possible because the perturbed energy-momentum tensors of each sector involves time-derivatives of the effective background metrics. Therefore, these make the coupling equations \eqref{coupling-shear-sector} dynamical in the sense that these are differential equations for $\gammamet_{13}$ and $\gammamett_{13}$.\footnote{One may see this from \eqref{deltat-hydro-shear} -- the first-order corrections omitted in these equations involves the time derivative of $\gammamet_{13}$ and $\gammamett_{13}$. Therefore even when $h_{13}$ is set to zero there exist solutions for $\gammamet_{13}$ and $\gammamett_{13}$! In this case, the two systems can just have fluctuating effective metrics without any extrinsic perturbation or change in the internal physical variables $\wo$, $\wt$, $\delta\To$ and $\delta\Tt$.} The spurious modes correspond to this spurious dynamics. The spurious modes are also badly behaved and are acausal (having positive imaginary parts in the dispersion relation) and this is related to the acausal behavior of first-order hydrodynamics. If we embed the hydrodynamics of each sector in kinetic theory/Israel-Stewart framework/holographic gravity, then these spurious modes disappear and are replaced by well-behaved relaxation modes. This will be one of the topics of the next section.

To summarize our findings for shear diffusion and specific viscosity:
\begin{enumerate}
 \item The full system has two shear diffusion modes with diffusion constants $D_{a,b}$ such that $\TT D_{a,b}$ decrease monotonically with increasing
 temperature $\TT$ before saturating at finite values at large $\TT$.
 \item The overall specific viscosity $\eta/\ST$ derived from the total conserved \EM\ tensor is in between
 the values of $\TT D_{a,b}$ with slight nonmonotonic behavior at the phase transition.
 \item When one of the systems has a dominant contribution to the total energy/pressure and a different specific viscosity, 
 the overall specific shear viscosity is closer to that of the dominant system.
\end{enumerate}

\subsection{Bi-hydro sounds and their attenuations}\label{Sec:BiHydroSound}
Owing to the rotational symmetry of the thermal equilibrium state, we can consistently assume that the velocity fluctuations in both sectors are longitudinal, i.e., pointing in the same direction as the momentum $\mathbf{k}$. This longitudinal alignment of the linearized velocity field defines the sound sector. Without loss of generality, we can take $\wo$, $\wt$ and $\mathbf{k}$ to be in the $z$-direction. The consistent forms of the effective metrics are:
\begin{equation}\label{g-sound}
\g_{\mu\nu} = {\rm diag}(-\x^2, \y^2, \y^2, \y^2) + \delta \g_{\mu\nu},\quad \gt_{\mu\nu} = {\rm diag}(-\xt^2, \yt^2, \yt^2, \yt^2) + \delta \gt_{\mu\nu}
\end{equation}
with the non-vanishing components of $\delta g_{\mu\nu}$ and $\delta \gt_{\mu\nu}$ being:
\begin{eqnarray}\label{deltag-sound}
&&\delta \g_{03} = \bo e^{i(kz-\omega t)}, \quad \delta \g_{00} = -2\x\,\delta\x\, e^{i(kz-\omega t)},\nonumber\\
&&\qquad\delta \g_{11} =\delta \g_{22} = (2\y\,\delta\y + \z)e^{i(kz-\omega t)},\quad
\delta \g_{33} = (2\y\,\delta\y -2 \z)e^{i(kz-\omega t)},\nonumber\\
&&\delta \gt_{03} = \bt e^{i(kz-\omega t)}, \quad \delta \gt_{00} = -2\xt\,\delta\xt\, e^{i(kz-\omega t)},\nonumber\\ 
&&\qquad\delta \gt_{11} =\delta \gt_{22} = (2\yt\,\delta\yt + \zt)e^{i(kz-\omega t)},\quad
\delta \gt_{33} = (2\yt\,\delta\yt -2 \zt)e^{i(kz-\omega t)}.
\end{eqnarray}
The four-velocity fields in the two sectors including the fluctuations thus are:
\begin{eqnarray}
\uo^\mu = \left(\frac{1}{\x}-\frac{1}{\x^2}\delta\x \, e^{i (k z -\omega t)} ,0,0,\wo e^{i (k z -\omega t)}\right), \nonumber\\
 \ut^\mu = \left(\frac{1}{\xt}-\frac{1}{\xt^2}\delta\xt \,e^{i (k z -\omega t)},0,0,\wt e^{i (kz -\omega t)}\right).
\end{eqnarray}
We may also anticipate that the temperatures also fluctuate from their equilibrium values so that we also have
\begin{equation}
\delta\To e^{i (k z -\omega t)}\quad{\rm and}\quad\delta\Tt e^{i (k z -\omega t)}.
\end{equation}
The non-vanishing components of the linearized perturbations of the individual hydrodynamic energy-momentum tensors then turn out to be:
\begin{eqnarray}
\delta t^{00} = \left(\frac{1}{\x^2} \frac{{\rm d}\e}{{\rm d}\To}\delta\To - 2 \frac{\e}{\x^3}\delta\x\right) e^{i(k z- \omega t)}, \quad \delta t^{03} = \left(\frac{\p}{\x^2 \y^2} \bo + \frac{\e + \p}{\x} \wo\right)e^{i(k z- \omega t)}, \nonumber\\
\delta t^{11} = \delta t^{22} = \left(\frac{1}{\y^2} \frac{{\rm d}\p}{{\rm d}\To}\delta\To - 2 \frac{\p}{\y^3}\delta\y - \frac{\p}{\y^4}\z +i \frac{2\eta_1}{3\y^2}k\wo +i\frac{\eta_1}{\x\y^4}\omega\z\right) e^{i(k z- \omega t)}, \nonumber\\
\delta t^{33} =  \left(\frac{1}{\y^2} \frac{{\rm d}\p}{{\rm d}\To}\delta\To - 2 \frac{\p}{\y^3}\delta\y +2 \frac{\p}{\y^4}\z -i \frac{4\eta_1}{3\y^2}k\wo - 2i\frac{\eta_1}{\x\y^4}\omega\z\right) e^{i(k z- \omega t)},
\end{eqnarray}
and similarly 
\begin{eqnarray}
\delta \tit^{00} = \left(\frac{1}{\xt^2} \frac{{\rm d}\et}{{\rm d}\Tt}\delta\Tt - 2 \frac{\et}{\xt^3}\delta\xt\right) e^{i(k z- \omega t)}, \quad \delta \tit^{03} = \left(\frac{\pt}{\xt^2 \yt^2} \bt + \frac{\et + \pt}{\xt} \wt\right)e^{i(k z- \omega t)}, \nonumber\\
\delta \tit^{11} = \delta \tit^{22} = \left(\frac{1}{\yt^2} \frac{{\rm d}\pt}{{\rm d}\Tt}\delta\Tt - 2 \frac{\pt}{\yt^3}\delta\yt - \frac{\pt}{\yt^4}\zt +i \frac{2\eta_2}{3\yt^2}k\wt +i\frac{\eta_2}{\xt\yt^4}\omega\zt\right) e^{i(k z- \omega t)}, \nonumber\\
\delta \tit^{33} =  \left(\frac{1}{\yt^2} \frac{{\rm d}\pt}{{\rm d}\Tt}\delta\Tt - 2 \frac{\pt}{\yt^3}\delta\yt +2 \frac{\pt}{\yt^4}\z -i \frac{4\eta_2}{3\yt^2}k\wt - 2i\frac{\eta_2}{\xt\yt^4}\omega\zt\right) e^{i(k z- \omega t)}.
\end{eqnarray}
The linearized coupling equations take the form:
\begin{eqnarray}\label{coupling-linear}
\delta \g_{\mu\nu} &=& \ga \left(\eta_{\mu\rho} \delta \tit^{\rho\sigma} \eta_{\sigma\nu}\sqrt{-\gt} +\frac{1}{2}\eta_{\mu\rho}  \tit^{{\rm (eq)}\rho\sigma} \eta_{\sigma\nu}\sqrt{-\gt} \gt^{\alpha\beta}\delta \gt_{\alpha\beta}\right) \nonumber\\ &&
+\gap  \left(\eta_{\rho\sigma} \delta \tit^{\rho\sigma} \eta_{\mu\nu}\sqrt{-\gt} +\frac{1}{2}\eta_{\rho\sigma}  \tit^{{\rm (eq)}\rho\sigma} \eta_{\mu\nu}\sqrt{-\gt} \gt^{\alpha\beta}\delta \gt_{\alpha\beta}\right),\nonumber\\
\delta \gt_{\mu\nu} &=& \ga \left(\eta_{\mu\rho} \delta t^{\rho\sigma} \eta_{\sigma\nu}\sqrt{-\g} +\frac{1}{2}\eta_{\mu\rho}  t^{{\rm (eq)}\rho\sigma} \eta_{\sigma\nu}\sqrt{-\g} \g^{\alpha\beta}\delta \g_{\alpha\beta}\right) \nonumber\\ &&
+\gap  \left(\eta_{\rho\sigma} \delta t^{\rho\sigma} \eta_{\mu\nu}\sqrt{-\g} +\frac{1}{2}\eta_{\rho\sigma}  t^{{\rm (eq)}\rho\sigma} \eta_{\mu\nu}\sqrt{-\g} \g^{\alpha\beta}\delta \g_{\alpha\beta}\right).
\end{eqnarray}
These should be utilized to eliminate $\delta\x$, $\delta\xt$, $\delta\y$, $\delta\yt$, $\z$, $\zt$, $\bo$ and $\bt$ in favour of the physical dynamical hydrodynamic variables $\delta\To$, $\delta\Tt$, $\wo$ and $\wt$. 

Assuming that the bulk viscosities of each individual system vanishes, the hydrodynamic equations of motion in the respective effective metrics take the form:
\begin{eqnarray}\label{sound-sector-fluc}
&&i k \x \wo - i \omega\, \left(\frac{\delta\s}{\s} + 3 \frac{\delta\y}{\y}\right) = 0,\nonumber\\
&&i k \xt \wt - i \omega\, \left(\frac{\delta\st}{\st} + 3 \frac{\delta\yt}{\yt}\right) = 0, \nonumber\\
&&i k\, \left(\frac{\delta\To}{\To} +  \frac{\delta\x}{\x}\right) - i\omega\left(\frac{\bo}{\x^2}+ \frac{\wo \y^2}{\x} \right)+ \frac{4}{3}k^2\frac{\eta_1}{\e + \p} \wo + 2\omega k\frac{\eta_1}{\e + \p}\frac{\z}{\x \y^2} = 0, \nonumber\\
&&i k\, \left(\frac{\delta\Tt}{\Tt} +  \frac{\delta\xt}{\xt}\right) - i\omega\left(\frac{\bt}{\xt^2}+ \frac{\wt \yt^2}{\xt} \right)+ \frac{4}{3}k^2\frac{\eta_2}{\et + \pt} \wt + 2\omega k\frac{\eta_2}{\et + \pt}\frac{\zt}{\xt \yt^2} = 0.
\end{eqnarray}
In order to find the eigenmodes, one can first solve for the effective metric fluctuations $\delta\x$, $\delta\y$, $\delta\xt$, $\delta\yt$, $\z$ and $\zt$ in terms of $\wo$, $\wt$, $\delta \To$ and $\delta \Tt$ using the linear algebraic equations \eqref{coupling-linear}. Substituting these forms above for the effective metric fluctuations, we obtain the four dynamical equations for the four variables $\wo$, $\wt$, $\delta \To$ and $\delta \Tt$ which yield a determinant. Finally, the dispersion relations of the eigenmodes are obtained by requiring that this determinant vanishes as in case of the shear sector.

Before considering the eigenmodes in detail, it is useful to examine the simple case of two identical perfect fluids, i.e., the case of $\alf = \alft$ and $\eta_1 = \eta_2 = 0$ (or rather we consider only the leading order in the derivative expansion). We want to prove that in this case one of the eigenmodes propagate exactly with the speed of sound of the full system provided the thermal equilibrium solution also yields identical effective metrics, i.e., $\x = \xt$ and $\y = \yt$. This result is valid even if the individual subsystems are not conformal. 

In the case of identical perfect fluid systems, we can also assume $\wo =\wt$ and $\delta\To = \delta\Tt$, and furthermore $\delta\x = \delta\xt$, $\delta\y= \delta\yt$, $\z = \zt = 0$ and $\bo = \bt$ so that the individual energy-momentum tensors and effective metrics are identical. Then this eigenmode can be obtained from
\begin{eqnarray}\label{sound-sector-fluc-2}
i k \,\x\, \wo - i \omega\, \left(\frac{\delta\s}{\s} + 3 \frac{\delta\y}{\y}\right) = 0,\nonumber\\
i k\, \left(\frac{\delta\To}{\To} +  \frac{\delta\x}{\x}\right) - i\omega\left(\frac{\bo}{\x^2}+ \frac{\wo \y^2}{\x} \right) = 0.
\end{eqnarray}
We also note that the full thermal equilibrium solution is parametrized by the temperature $\TT$. Therefore, a variation of $\TT$ which preserves its relationship with the (identical) individual system temperatures given by \eqref{eq-condition} will produce a solution corresponding to an infinitesimal change of the full system equilibrium temperature $\TT$. Thus we can obtain a solution with $\delta \x =\delta \xt = ({\rm d}\x(\TT)/ {\rm d}\TT)\delta\TT$,  $\delta \y =\delta \yt = ({\rm d}\y(\TT)/ {\rm d}\TT)\delta\TT$ and $\delta \To =\delta \Tt  = ({\rm d}\To(\TT)/ {\rm d}\TT)\delta\TT$ with
\begin{equation}
\delta\TT = \To(\TT)\delta\x+\x(\TT)\delta\To
\end{equation}
being satisfied. Furthermore, we can boost the full energy-momentum tensor. If the full system is boosted by an infinitesimal velocity 
$\vi$ in the $z$-direction (in background flat space), then the non-vanishing components of its energy-momentum tensor with an overall infinitesimal temperature fluctuation takes the linearized perfect fluid form:
\begin{equation}\label{full-em-fluc}
\ttot^{00} = \ET + \frac{{\rm d}\ET}{{\rm d}\TT} \delta \TT, \quad 
\ttot^{11} = 
\ttot^{22} = 
\ttot^{33}= \PT + \frac{{\rm d}\PT}{{\rm d}\TT} \delta \TT, 
\quad 
\ttot^{03} =(\ET + \PT) \vi.
\end{equation}
Of course if we make $\delta\TT$ space-time dependent we also need a spacetime dependent boost $\vi$ in order that we can ensure energy-momentum conservation. The conservation of the full energy-momentum tensor in flat space yields the linearized Euler equations:
\begin{equation} \label{Euler-2}
ik\vi - i\omega \frac{\delta\ST}{\ST} = 0, \quad  i k\frac{\delta\TT}{\TT} - i \omega \vi = 0.
\end{equation}
It is guaranteed that the diagonal components of the fluctuations can always be mapped to a change in $\delta\TT$ even if the systems are not identical. If we solve $\bo$ and  $\bt$ in terms of $\wo$ and $\wt$ using the off-diagonal $03$-component of the coupling equations, and then compute the off-diagonal $03$-component of the full energy-momentum tensor, we can also define the $\vi$ of the full system as an appropriate linear combination of $\wo$ and $\wt$ demanding the form \eqref{full-em-fluc} of the full energy-momentum tensor. This can always be done. In case of identical systems with identical energy-momentum tensors living in identical effective metrics, the procedure is simpler: eliminate $\bo$ in favour of $\wo$ from the coupling equation and obtain $\vi$ in terms of $\wo$ from the computed form of the full energy-momentum tensor.

To proceed further, we thus focus on the off-diagonal component $\ttot^{03}$. Specifically, we observe from \eqref{full-em-fluc} that
\begin{equation}\label{compare-1}
\delta\ttot^{03} =\delta\ttot^{0}_{\phantom{0}3} = (\ET+ \PT) \vi, \quad \delta\ttot_{0}^{\phantom{0}3} = -(\ET+ \PT) \vi.
\end{equation}
With our assumptions for the effective metric and the perfect fluid forms of the energy-momentum tensor, we should get
\begin{eqnarray}\label{compare-2}
&&\delta t^{03} =\delta \tit^{03} = \frac{\p}{\x^2 \y^2} \bo + \frac{\e + \p}{\x} \wo, \nonumber\\
&&\delta t_{0}^{\phantom{0}3} =\delta \tit_{0}^{\phantom{0}3} = - (\e + \p) \wo \x, \nonumber\\
&&\delta t^{0}_{\phantom{0}3} =\delta \tit^{0}_{\phantom{0}3} = (\e + \p)\left(\frac{\y^2}{\x}\wo + \frac{1}{\x^2}\bo\right).
\end{eqnarray}
From \eqref{full-em-fluc}, any consistent coupling equations should lead to
\begin{eqnarray}\label{above}
\delta\ttot_{0}^{\phantom{0}3} = 2\x\y^3\delta t_{0}^{\phantom{0}3}, \quad \delta\ttot^{0}_{\phantom{0}3} = 2\x\y^3\delta t^{0}_{\phantom{0}3}.
\end{eqnarray}
Furthermore thermodynamic identities for any consistent coupling ensure that $\ET + \PT = 2 \x\y^3(\e + \p)$. Therefore it follows from \eqref{compare-1}, \eqref{compare-2} and \eqref{above} that any consistent coupling equation should imply
\begin{equation}\label{v-iden}
\frac{\y^2}{\x}\wo + \frac{1}{\x^2}\beta = \wo \x = \vi.
\end{equation}

The coupling equations always ensure that conservation of the individual energy-mo\-men\-tum tensor in the individual effective metric leads to conservation of the full energy-momentum tensor in flat space. To show that  the eigenmode of the full system corresponds to the thermodynamic sound of the full system, we need to turn this {argument} around. We need to show that the Euler equations of the full energy-momentum tensor in flat space will lead to satisfying the individual Euler equations in individual effective metrics. Clearly, we will generically need identical systems with identical energy-momentum tensors living in identical effective metrics. Otherwise the number of conservation equations of the full system are outnumbered by the individual conservation equations. At the linearized level, we need to show that \eqref{Euler-2} implies \eqref{sound-sector-fluc-2}.

We note that thermodynamic variation ensures that $\delta \ST/\ST = 2 \delta s_1/ s_1 + 3 \delta \y/ \y$ since $\ST = 2 s_1 \y^3$ in the case of identical systems.  Similarly, $\delta\TT/\TT = \delta \To/\To + \delta\x/\x $ since $\TT = \To\x$. It is then easy to see that \eqref{Euler-2} implies \eqref{sound-sector-fluc-2} because of the two relations in \eqref{v-iden} which follows from consistent coupling equations. We then conclude that for any consistent coupling between two identical systems with identical effective metric solutions at equilibrium, the thermodynamic sound will correspond to one of the eigenmodes at the leading order in the derivative expansion. In this mode, the velocity fields in the two identical systems are parallel to each other so that $\wt = \wo$.

Even for identical perfect fluid systems there is another eigenmode where $\delta \To \neq \delta \Tt$ and $\wo \neq \wt$. In this mode, the velocity fields are anti-parallel to each other so that $\wt = -\wo$. Most importantly the thermodynamic relation $\delta\TT = \delta (\To \x) = \delta (\Tt \xt)$ is not satisfied by the fluctuations. This mode does not travel at the speed of thermodynamic sound.
When $\no\not=\nt$, it turns out that neither of the two eigenmodes does; in this case the thermodynamically defined
speed of sound is in between the velocities of the eigenmodes.

When the two systems are identical, and we consider the eigenmode which at leading order propagates at the speed of full system thermodynamic sound, we cannot map the first-order (identical) hydrodynamic fluctuations of the individual systems to that of a hydrodynamic form for the full system. To see this, we may repeat the steps of the above argument with $\z = \zt \neq 0$ and $\eta_1 = \eta_2 \neq 0$ and find that for generic $\eta_1$ the modified form of \eqref{v-iden} does not imply that we can obtain \eqref{sound-sector-fluc-2} with first-order corrections from the first-order correction of \eqref{Euler-2} (linearized Navier-Stokes equation in flat space).

\begin{figure}[tbp]
\centering 
\includegraphics[width=.5\textwidth]{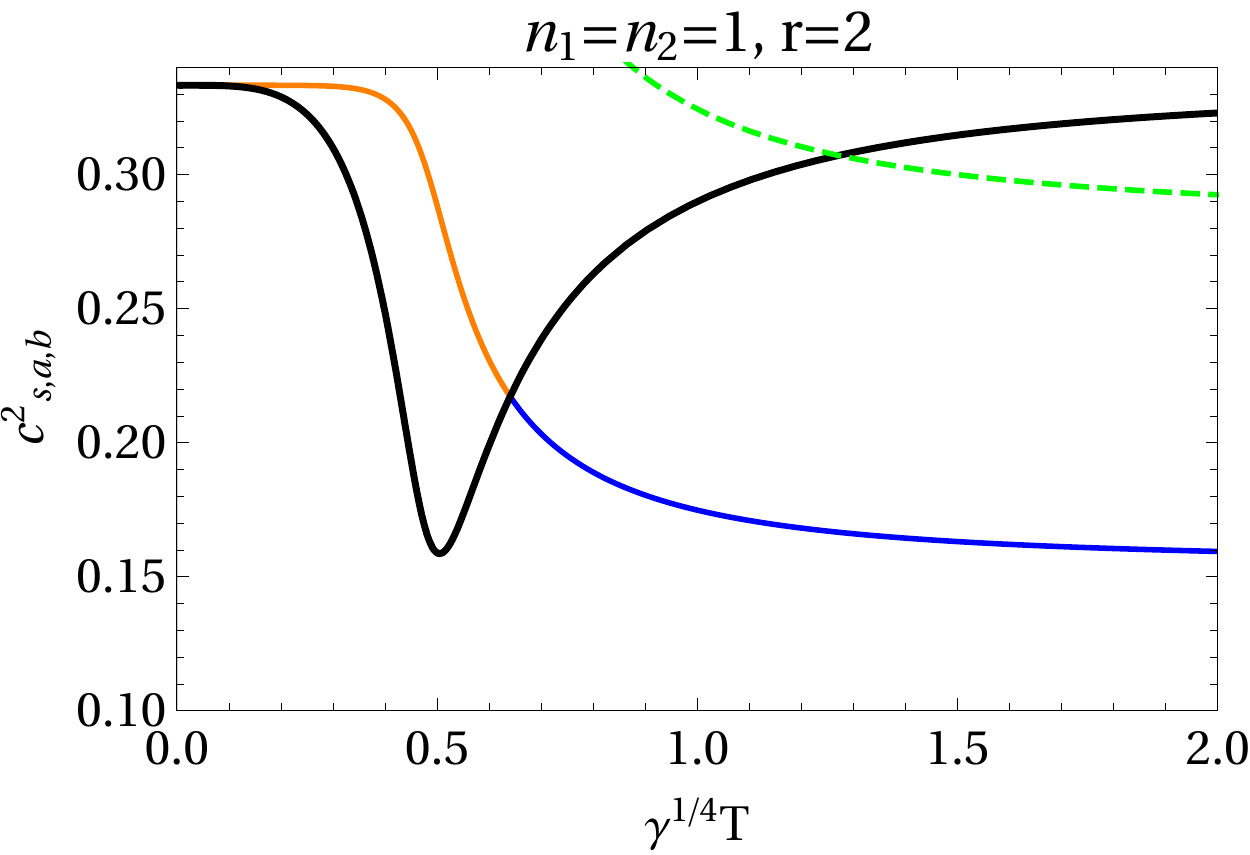}\includegraphics[width=.5\textwidth]{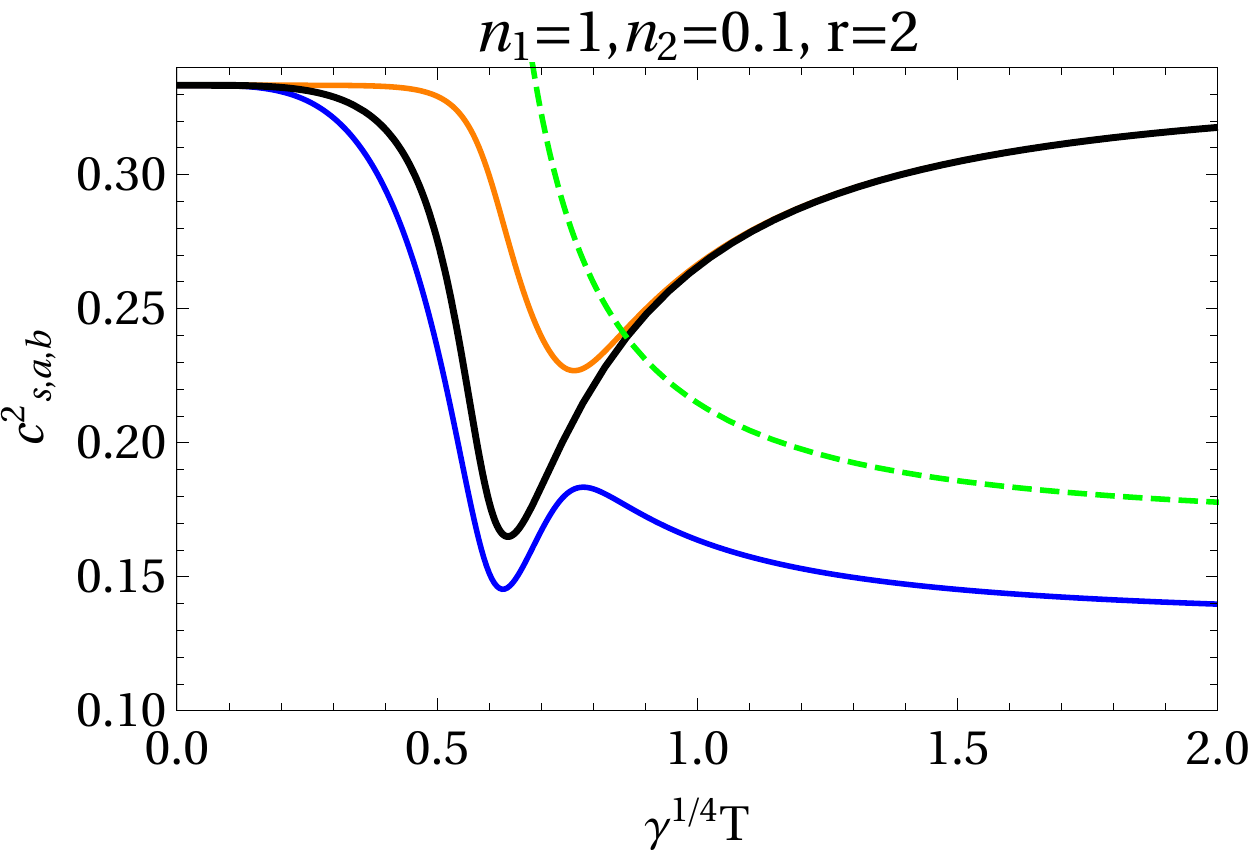}\\
\includegraphics[width=.5\textwidth]{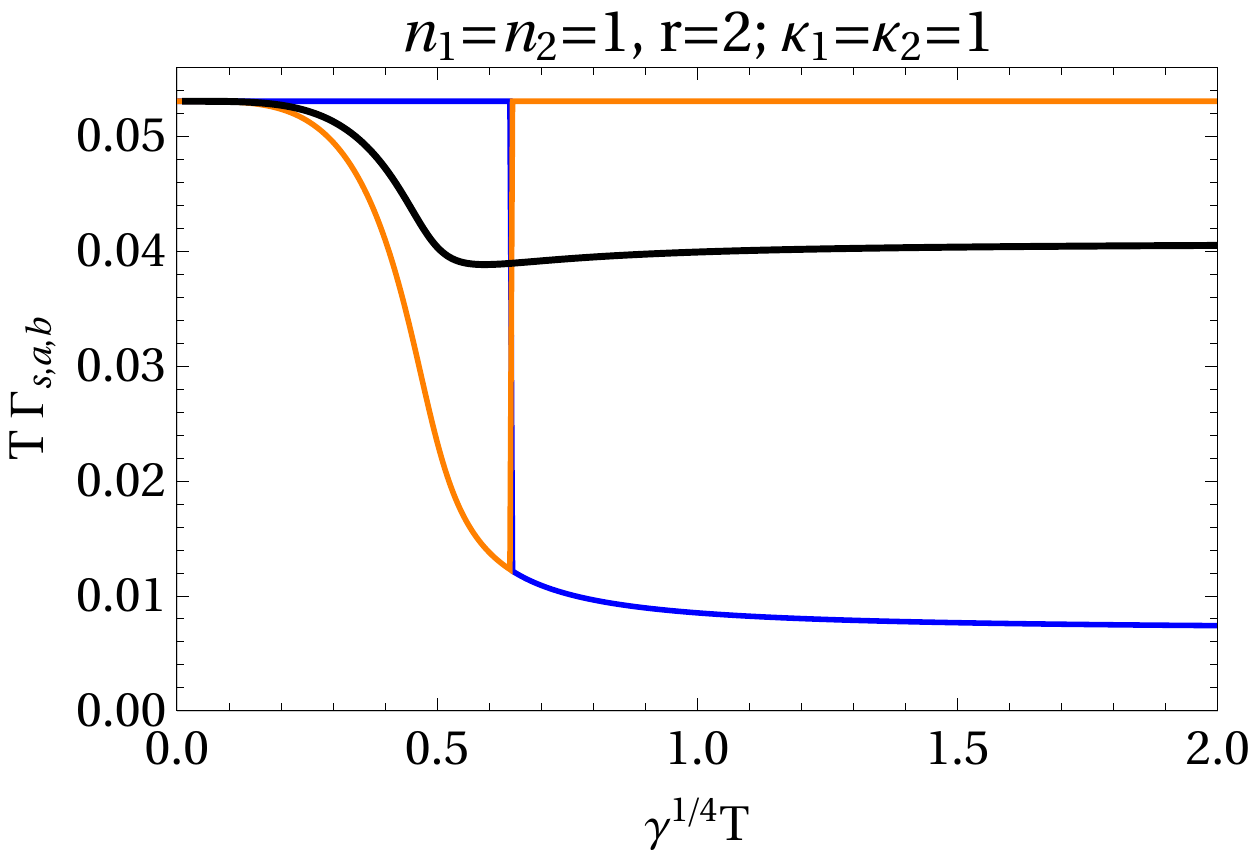}\includegraphics[width=.5\textwidth]{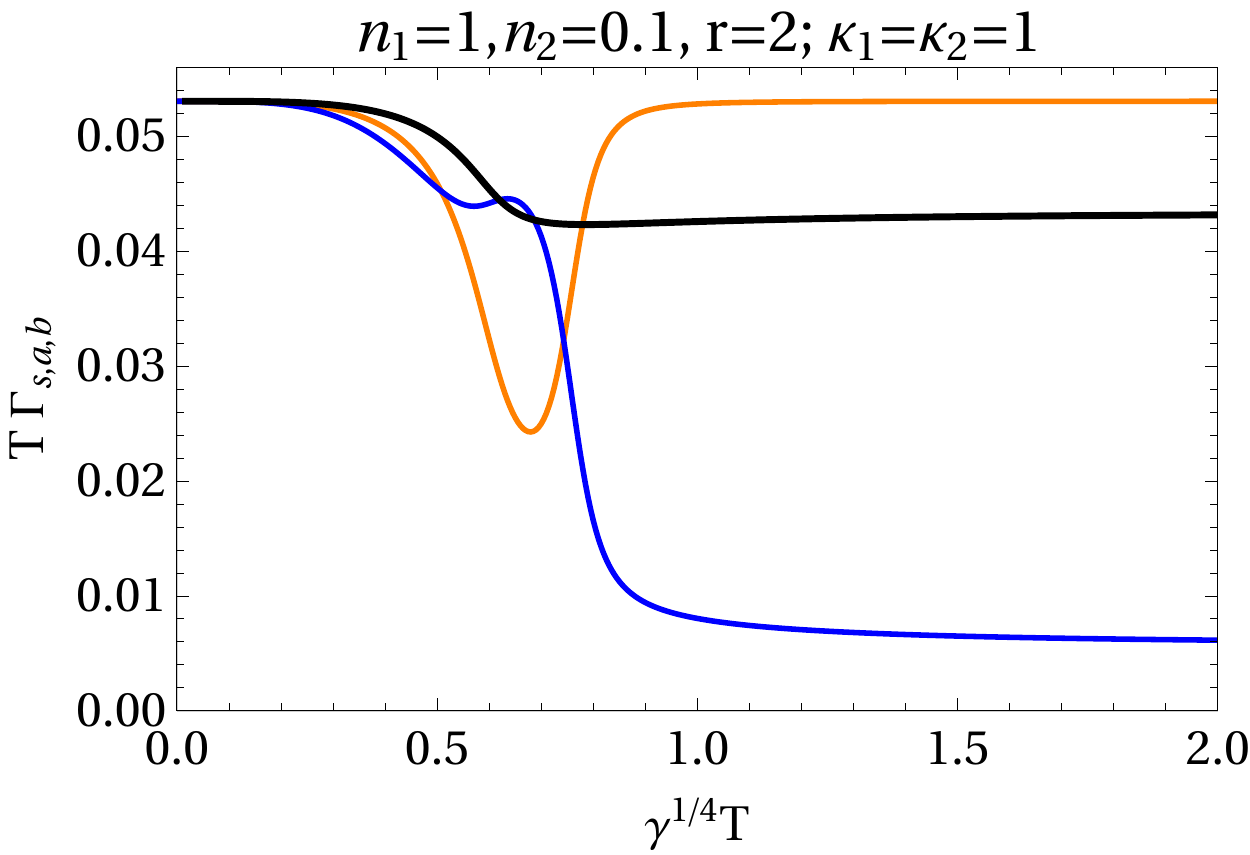}
\caption{\label{fig:cs-samek}Sound modes and their attenuation coefficients 
for equal and unequal conformal systems, same $\kappa=1$
(corresponding to $\eta_i/s_i=1/4\pi$), with the slower mode $a$ plotted in blue, and the faster mode $b$ plotted in orange.
The black line represents the
thermodynamic speed of sound and associated attenuation coefficient from the Kubo formula.
The green dashed line shows
the light-cone velocities squared of the two subsystems (in the case of $n_2=1/10$ only $\vt^2$ is in plot region).
In the case $n_1=n_2$ the lines for $c_{a,b}^2$ meet and could be continued smoothly by switching
the designation; however for any $n_1\not=n_2$ we have $c_b>c_a$ at nonzero $\gqT$. The discontinuous behavior
of the damping rates $\Gamma_{a,b}$ for $n_1=n_2$ is in fact the limit of smooth curves as $n_1\to n_2$ from
different starting values.}
\end{figure}

\begin{figure}[tbp]
\centering 
\includegraphics[width=.45\textwidth]{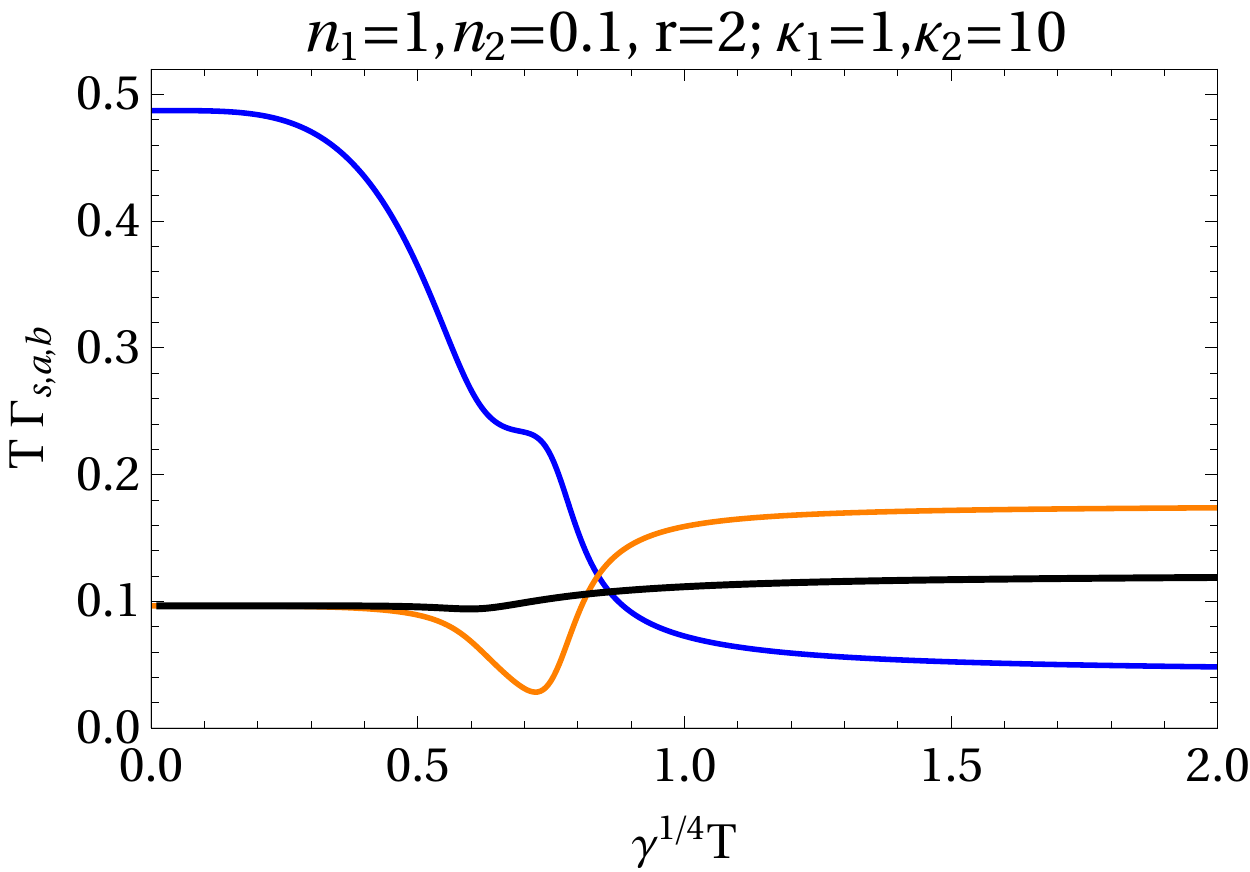}%
\includegraphics[width=.45\textwidth]{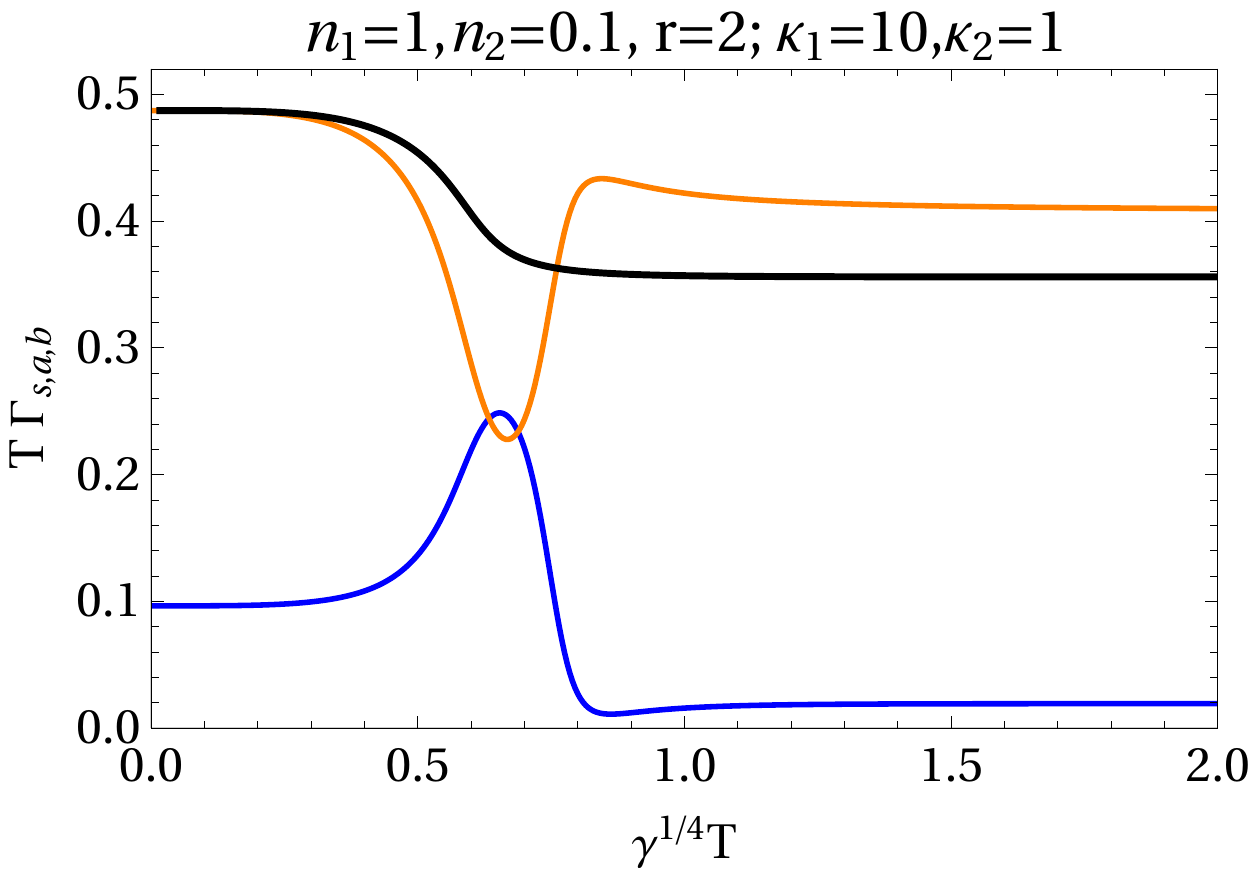}
\caption{\label{fig:cs-diffk}Attenuation coefficients $\Gamma_{a,b}$ of the 
sound eigenmodes (slower mode $a$ in blue, faster mode $b$ in orange) for 
unequal conformal systems with different $\kappa$. The black line gives the Kubo formula result
for sound attenuation.}
\end{figure}

Although in general different from the thermodynamically defined sound,
the dispersion relations of the eigenmodes have the same characteristic sound-like form,
\begin{equation}
\omega_{(a,b)} = \pm c_{(a, b)}k - i \Gamma_{(a,b)}k^2 + \mathcal{O}(k^3).
\end{equation}

The perturbative expansions of the results for the speed of sound modes
and their respective attenuation coefficients are given by
\begin{alignat}{1}
c_{a} & =\frac{1}{\sqrt{3}}\Big(1-2(\alf+\alft)\ga\TT^{4}-48\alf\alft\ga^{2}\TT^{8}\Big)+\mathcal{O}(\ga^{3}),\nonumber \\
\Gamma_{a} & =\frac{\kap\alf+\kapt\alft}{6\pi\TT(\alf+\alft)}-\frac{\alf\alft(9\kap\alf-\kapt\alf-\kap\alft+9\kapt\alft)}{3\pi(\alf+\alft)^{2}}\ga\TT^{3}+\mathcal{O}(\ga^{2}),\nonumber \\
c_{b} & =\frac{1}{\sqrt{3}}\Big(1-8\alf\alft\ga^{2}\TT^{8}\Big)+\mathcal{O}(\ga^{3}),\nonumber \\
\Gamma_{b} & =\frac{\kapt\alf+\kap\alft}{6\pi\TT(\alf+\alft)}-\frac{(\alf-\alft)\left(2\kapt\alf^{2}+7\alf\alft(\kap-\kapt)-2\kap\alft^{2}\right)}{3\pi(\alf+\alft)^{2}}\ga\TT^{3}+\mathcal{O}(\ga^{2}),
\end{alignat}
with a dependence on $r=-\gap/\ga$ showing up only in the higher-order terms.
For equal partial pressures, $\no=\nt$, the dependence of the sound attenuation coefficients on
$\kap$ and $\kapt$ simplifies. Both $\Gamma_{a}$ and $\Gamma_{b}$ are then proportional to $(\kap+\kapt)$ to all
orders in $\gqT$; the attenuation coefficient of the faster mode which then coincides with the thermodynamically
defined speed of sound moreover becomes independent of $\gqT$.

Mode $a$ has velocity and temperature fluctuation fields with perturbative expansions
\begin{eqnarray}
&&\wt = \frac{\alf}{\alft}\left(1 + \frac{21}{2}(\alft -\alf)\gTf + \mathcal{O}(\ga^2, \gap^2, k) \right)\wo,\nonumber\\ 
&&\delta \To =\pm \frac{\TT}{\sqrt{3}}\left(1 + 2 \alft\gTf + \mathcal{O}(\ga^2, \gap^2, k) \right) \wo, \nonumber\\
&&\delta \Tt = \pm \frac{\alf}{\alft}\frac{\TT}{\sqrt{3}}\left(1 + \frac{1}{2}\left(21 \alft - 17\alf\right) \gTf+ \mathcal{O}(\ga^2, \gap^2, k)\right) \wo.
\end{eqnarray}
Above, the $+$ sign refers to the case when the mode is propagating parallel to the momentum $\mathbf{k}$ and $-$ sign refers to the case of opposite propagation. Mode $b$ similarly is one in which
\begin{eqnarray}
&&\wt =-\left(1 - \frac{1}{2}(\alf -\alft)\gTf + \mathcal{O}(\ga^2, \gap^2, k) \right)\wo,\nonumber\\
&&\delta \To =\pm \frac{\TT}{\sqrt{3}}\left(1 + 2 \alft\gTf + \mathcal{O}(\ga^2, \gap^2, k) \right) \wo, \nonumber\\
&&\delta \Tt = \mp \frac{\TT}{\sqrt{3}}\left(1 + \frac{1}{2}\left( \alft + 3\alf\right) \gTf+ \mathcal{O}(\ga^2, \gap^2, k)\right) \wo.
\end{eqnarray}
For equal partial pressures, $\no=\nt$, mode $a$ and $b$ have $\wt=\wo$ and $\wt=-\wo$, respectively, to all orders.

\begin{figure}[tbp]
\centering 
\includegraphics[width=.5\textwidth]{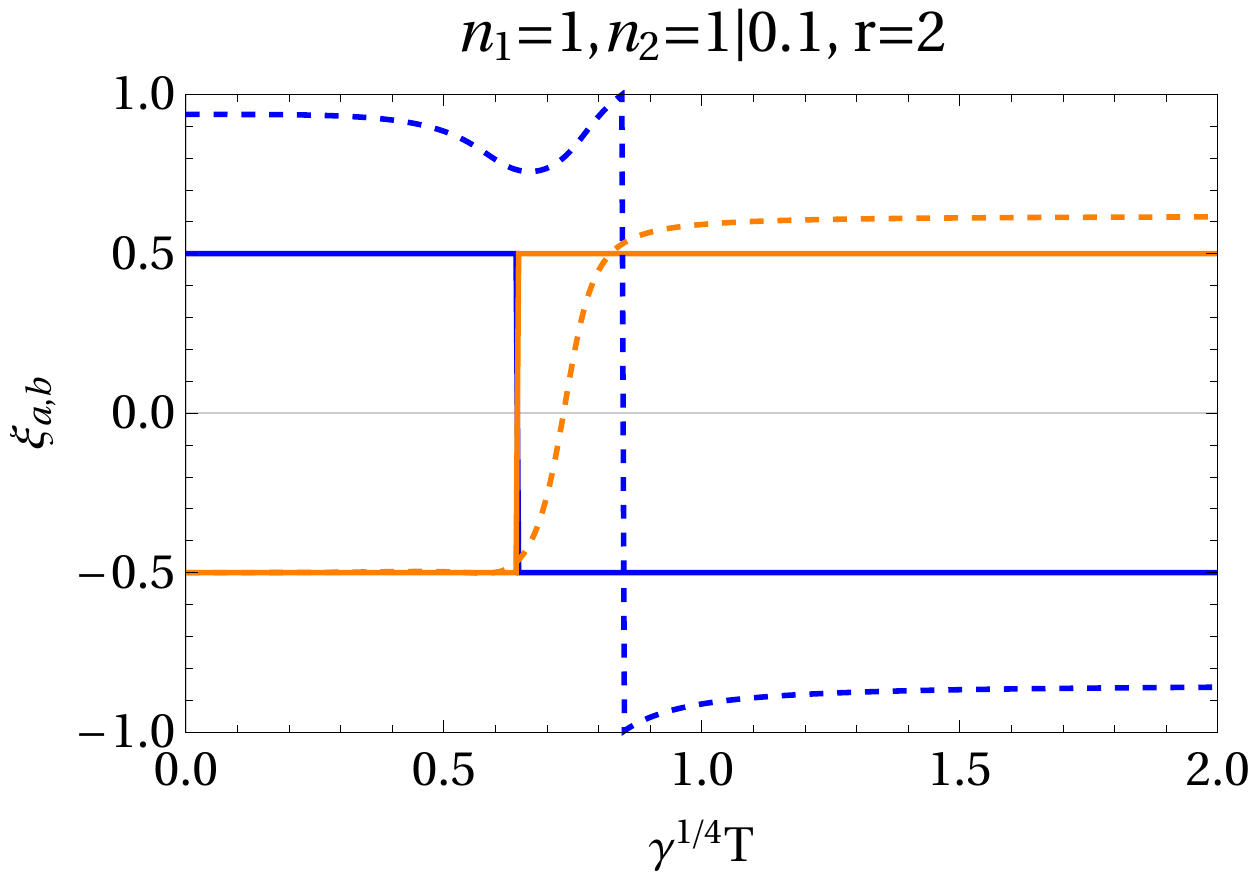}\includegraphics[width=.5\textwidth]{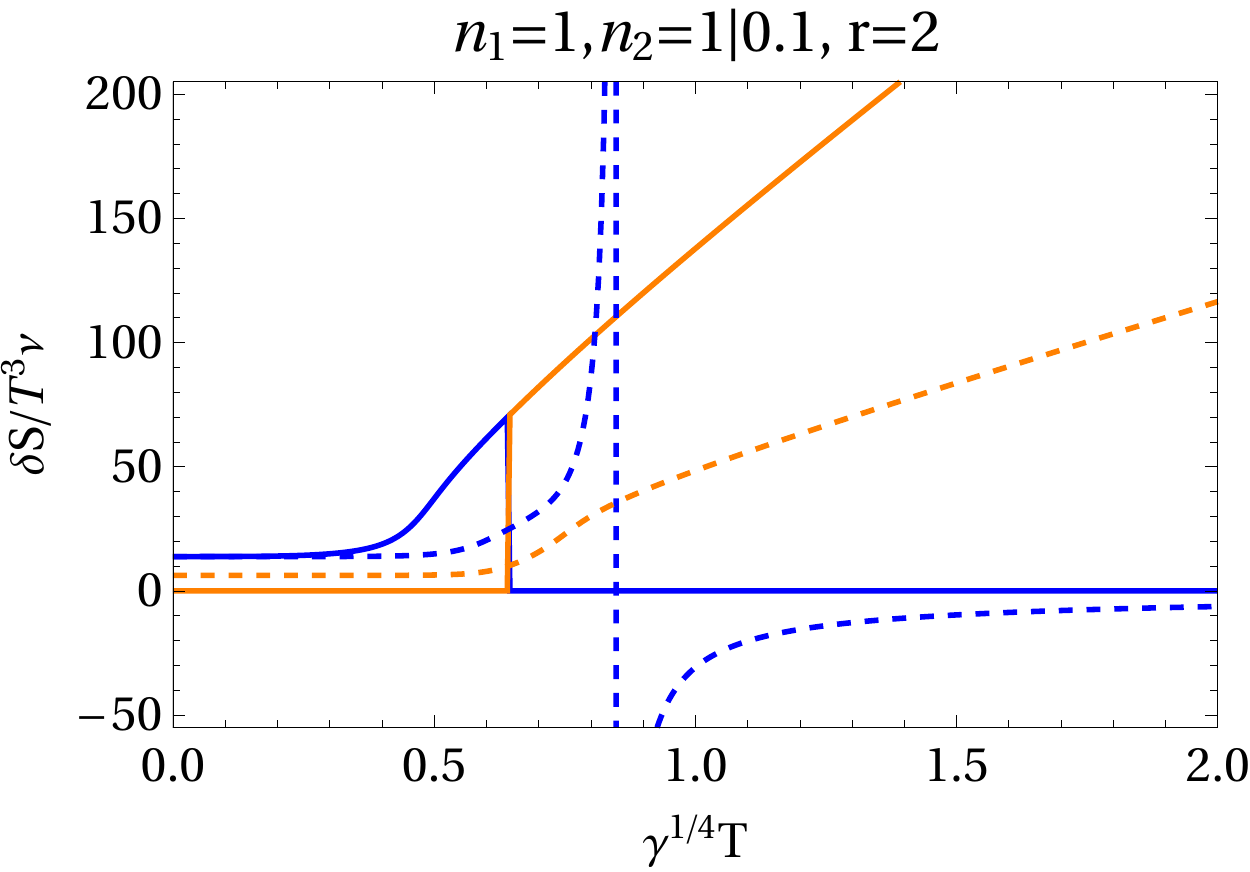}\\
\caption{\label{fig:soundeigenmodes}
Left panel: the relation between the velocity amplitudes $\wo$ and $\wt$ of the sound eigenmodes displayed in Fig.~\ref{fig:TDs}
in the form $\xi\equiv \frac{2}{\pi}\arctan(\wt/\wo)$; right panel: the corresponding fluctuation amplitude of the total entropy 
{density, $\delta\ST\equiv \delta\ST^{\mu=0}$}, divided by $\wo\TT^3$. Mode $a$ and $b$ are given in blue and orange, respectively, with full and dashed lines representing $\no=\nt=1$
and $\no=1, \nt=1/10$. The divergence of $\delta \ST/(\TT^3\wo)$ of mode $a$ at one value of $\gqT$ is due to a zero of $\wo$ (corresponding to $|\xi_a|=1$); here a velocity field is present only in subsystem 2. 
}
\end{figure}

It is instructive to compare the attenuation coefficients of these propagating modes with what would have been the hydrodynamic sound attenuation if one of these modes could have been interpreted as a sound channel hydrodynamic mode of the full system. First, assuming that each individual sector is conformal and hydrodynamic as we have done above, one can see from the expression of the conserved energy-momentum tensor of the full system that the trace of the full energy-momentum tensor does not contain any spatial or temporal derivative. This implies that the full system has vanishing bulk viscosity although it is not conformal but has a nonzero trace of the total \EM\ tensor. Second, with the bulk viscosity vanishing, the sound dispersion relation for a hydrodynamic system in flat space is given by
\begin{equation}
\omega = \pm c_s k - i \Gamma_s k^2 + \mathcal{O}(k^3), 
\end{equation}
with $c_s$ being the speed of thermodynamic sound and the attenuation coefficient
being $\Gamma_s = (2/3) (\eta/\TT \ST)$. With $\eta$ given by \eqref{eta-Kubo}, the latter would be the attenuation coefficient of one of the propagating modes if it could be interpreted as hydrodynamic motion in flat space. We find that none of the propagating modes attenuates in the hydrodynamic way even when one of them travels at the speed of thermodynamic sound as in the case of identical subsystems.

In Fig.~\ref{fig:cs-samek} and \ref{fig:cs-diffk} our nonperturbative results 
for the speeds and attenuations of the propagating modes of two strongly coupled fluids, and weakly plus strongly coupled fluids respectively have been plotted. Comparison has been made also with the hydrodynamic sound attenuation $\Gamma_s$ of the full system.
For equal partial pressures, $\no=\nt$, the results for $c_{a,b}$ coincide at $\gqT=0$ and one finite value of $\gqT$.
The crossing at the latter point is lifted for all $\no\not=\nt$ such that mode $b$ is always faster than mode $a$ for $\gqT>0$.
In the limit $\no\to\nt$, the results for $c_{a,b}$ develop a cusp, while $\Gamma_{a,b}$ even become discontinuous.
For exactly $\no=\nt$, ${\rm max}(\Gamma_a,\Gamma_b)$ is even a constant independent of $\gqT$ and the results
for the two modes could all be connected smoothly. We have however kept the mode labels corresponding to
taking the limit $\no\to\nt$.

Fig.~\ref{fig:soundeigenmodes} displays $\vt/v$ for the corresponding sound eigenmodes as well as the associated
{(adiabatic)} fluctuations of the total entropy density {$\ST^{\mu=0}$}. Again, the seemingly spurious discontinuities at $\no=\nt$ indeed arise when
taking the limit $\no\to\nt$ starting from $\no\not=\nt$.

In summary, our findings for the sound sector are:
\begin{enumerate}
\item The thermodynamic speed of sound of the full system $c_s$ is always between the velocities of the two sound modes $c_a$ and $c_b$, $c_a\le c_s\le c_b$, and coincides exactly with one of the modes for $\no=\nt$.
\item At temperatures above the crossover from weak to strong inter-system coupling, the velocity of the faster mode ($b$) quickly approaches the thermodynamically defined speed of sound.
\item Near the crossover temperature, the velocity fields of the two modes change their phase with $\v$ (or $\vt$) vanishing
at a certain value of $\gqT$ for mode $a$ (or $b$). Mode $a$ has out-of-phase oscillations for large $\gqT$ with decreasing total entropy fluctuations {$\delta\ST^{\mu=0}$} and speed slower than the thermodynamic speed of sound.

\item At high temperatures, $c_s$ and $c_b$ approach $1/\sqrt{3}$ due to emergent conformality.

\item While $c_s$ and $c_b$ can become larger than the effective lightcone speeds $\vo,\vt$, the velocity $c_a$ of the slower
mode remains smaller than $\vo,\vt$; this mode thus lies within both effective lightcones.
\item The value of the attenuation coefficient obtained from the Kubo formula is between that of the sound modes
for large $\gqT$.
\item While the dependence of the attenuation coefficients on $\gqT$ is in general complicated, at temperatures sufficiently above
the crossover region the slower {``non-acoustic''} sound mode is always the more weakly damped one.
\item {The coupling studied in our setup provides no pure damping modes, i.e. the imaginary part of the speed of sound vanishes as $k\rightarrow 0$. This reflects the fact, that this interaction is not sufficient to equilibrate the two subsystems, e.g. in a homogeneous configuration with subsystems at unequal temperatures.}
\end{enumerate}

\section{Coupling a kinetic sector to a strongly coupled fluid}\label{Sec:KinHydro}

In order to obtain a qualitative understanding of a coupled system of weakly interacting and strongly interacting degrees of freedom, we can study the consequences of mutual effective metric coupling of a gas of massless particles
(gluons) described by kinetic theory (as subsystem $\mathfrak{S}_1$)
and a strongly interacting holographic gauge theory described by dual gravitational perturbations of a black hole ($\mathfrak{S}_2$). Using the fluid/gravity correspondence, we may further simplify gravitational dynamics to that of a fluid with a low value of $\eta/s$ if we are interested in the long time dynamics\footnote{Here we are tacitly assuming that the strongly coupled sector thermalizes faster despite its coupling to the weakly coupled gluons. Our results here indicate that the correction to the relaxation dynamics of each sector can be very mild even though the hydrodynamic behavior is modified significantly by the democratic effective metric coupling.}. Due to the appearance of spurious modes and associated acausalities we need to embed first-order hydrodynamics in a more complete description. Therefore, we embed the strongly coupled fluid in an Israel-Stewart framework with an extremely small relaxation time. In the future we plan to do a more complete calculation by involving the relaxation dynamics of the strongly coupled sector as described holographically via quasi-normal mode perturbations of a black brane.

Following Refs.~\cite{Romatschke:2015gic,Kurkela:2017xis} and
ignoring for simplicity effects of quantum statistics,
the thermal equilibrium of the weakly coupled (and dilute) kinetic sector is described by a Maxwell-J\"uttner distribution
\begin{equation}
f_0(p^i) = n_0 e^{p^\mu u_\mu/{\To}},
\end{equation}
where $p^0$ is determined using the mass-shell condition, $p^\mu p^\nu \g_\mn = 0$ for massless gluons when
thermal corrections to the mass are neglected. 
Expressed in terms of $p \equiv \sqrt{{p^x}^2+ {p^y}^2+ {p^z}^2}$, 
we have $p^0 = p\y/\x$, $u_\mu = (-\x, 0,0,0)$, so that
the time-dilation factor $\x$
(but not the spatial dilation factor $\y$) drops out, giving
\begin{equation}
f_0(p^i) = n_0 e^{-{p\y}/{\To}}.
\end{equation}
The normalization constant $n_0$ can be fixed as follows. The energy-momentum tensor corresponding to a quasi-particle distribution $f$ is \cite{Andreasson:2011ng}
\begin{equation}\label{kinetic-emt}
t^{\mu\nu} = \sqrt{-g}\int_{-\infty}^\infty \frac{{\rm d}^3 p}{(2 \pi)^3}\frac{p^\mu p^\nu}{-p_0}  \,\,f(p^i, x^i, t)
\end{equation}
with $p_0=g_{0\mu}p^\mu$ satisfying the mass-shell condition. The equilibrium energy-momentum tensor then takes our previously assumed form:
\begin{equation}
t^{\mu\nu} = {\rm diag}\left(\frac{3 \alf \To^4}{\x^2},\frac{\alf \To^4}{\y^2},\frac{\alf \To^4}{\y^2},\frac{\alf \To^4}{\y^2}\right)
\end{equation}
where $\alf$ is our previously introduced (theory-dependent) parameter if 
\begin{equation}
n_0 = \alf \pi^2.
\end{equation}
We will therefore set $n_0$ to $\alf \pi^2$ so that we can directly use our previously obtained results to describe the equilibrium of the full system.

For convenience, we use spherical coordinates for the components of the momenta so that $p^x = p \sin\theta\cos\phi$, $p^y = p \sin\theta\sin\phi$ and $p^z = p \cos\theta$. A linearized fluctuation of the quasi-particle distribution about equilibrium can be written as:
\begin{equation}
f(p,\theta, \phi, x^i, t) = n_1 \pi^2 e^{-{p\y}/{\To}} +  \delta f(p,\theta, \phi, x^i, t).
\end{equation}
For computational purposes, it is useful to split the linear term $\delta f$ into two parts, each having a specific momentum $\mathbf{k}$ and a specific frequency $\omega$ component, according to
\begin{equation}\label{qp-linear}
\delta f(p,\theta, \phi, x^i,t) =\left(\delta f^{\rm (eq)}(p,\theta, \phi) + \Delta f(p,\theta, \phi)\right)e^{i(\mathbf{k}\cdot\mathbf{x} -\omega t)}.
\end{equation}
The term $\delta f^{(eq)}$ can be defined uniquely such that it produces a perturbation ${\delta t^{\mu\nu}}^{\rm (eq)}$ in the energy-momentum tensor that is of a perfect fluid form. Only the term $\Delta f$ will then contribute to the dissipation of energy and momentum. If $\delta\g$ is the (self-consistent) effective metric fluctuation in the kinetic theory, then in the relaxation time approximation $\delta f$ obeys the linearized Anderson-Wittig equation:
\begin{eqnarray}\label{AW}
\left(\partial_t + \frac{p^i}{p^0}\partial_i\right) \delta f - \delta\Gamma^i_{\beta\ga} \frac{p^\beta p^\ga}{p^0}\frac{\partial}{\partial p^i} f_0 = -\frac{\x}{\tauo} \Delta f
\end{eqnarray} 
with $\delta\Gamma^\mu_{\alpha\beta}$ being the linearized Levi-Civita connection obtained from $\delta\g$ and with $p^0$ also receiving a corresponding linear contribution so that the mass-shell condition is satisfied. Furthermore, in a conformal theory $\tauo$ should be proportional to $\To^{-1}$ and we may parametrize
\begin{equation}\label{tauo}
\tauo(\To) = \frac{5\kap}{4\pi \To}
\end{equation}
where $\kap$ is a constant which will be eventually identified with $4\pi\eta_1/\s$ as before. 

The relaxation time in the Israel-Stewart theory in which we are embedding the strongly coupled fluid is similarly set to 
\begin{equation}\label{taut}
\taut(\Tt) = \frac{5\lambda}{4\pi \Tt}.
\end{equation}
In order to isolate the strongly coupled fluid from the relaxation dynamics, we will take $\lambda$ very small so that $\taut$ is small. Unlike the kinetic sector where $\tauo$ determines the shear viscosity (this can be seen via consistent reduction to hydrodynamics), note that $\taut$ of the Israel-Stewart theory is an independent parameter which does not affect the shear viscosity but only second-order hydrodynamics. 

\subsection{Branch cut in response functions of the kinetic sector}

We can show that an infinite number of quasi-particle distribution fluctuations decouple from the strongly coupled sector in the sense that all perturbed observables will get contributions purely from the kinetic sector. For instance, fluctuations of the form
\begin{eqnarray}\label{df}
& \delta f  = F(p)G(\theta, \phi) e^{-i\omega t + i\mathbf{k}\cdot\mathbf{x}}, \quad {\rm with} \quad G(\theta,\phi) = H_1(\theta)\cos (n \phi)+H_2(\theta)\sin(n \phi) \nonumber\\& {\rm and} \quad n \geq 3
\end{eqnarray}
have vanishing fluctuations of the \EM\ tensor 
\begin{equation}
\delta t^{\mu\nu} \propto \int \frac{{\rm d}^3p}{p_0} \, p^\mu p^\nu \delta f(\mathbf{x}, \mathbf{p}, t) = 0.
\end{equation}
If also all perturbations in the strongly coupled sector are set to zero,
we can then self-consistently assume that
\begin{equation}\label{dgv}
\delta \g_\mn = \delta \gt_\mn = 0.
\end{equation}
In this case we have $\delta f = \Delta f$ and the linearized Anderson-Wittig equation \eqref{AW} reduces to
\begin{equation}
-i \left(\omega -\frac{\x}{\y}\mathbf{n}\cdot\mathbf{k}  + i \frac{\x}{\tauo}\right)\delta f  = 0, 
\end{equation}
where $n^i = p^i /p$ and $\tauo$ is the relaxation time in the kinetic sector. 
Choosing without loss of generality $\mathbf{k}$ along the $z$-direction we obtain
\begin{equation}\label{cut}
\omega = \frac{\x}{\y}k \cos\theta - i  \frac{\x}{\tauo(\To)}
=\frac{\x}{\y}k \cos\theta - i  \frac{1}{\tauo(\TT)},
\end{equation}
where we have used that $\To\x= \TT$ at equilibrium with $\TT$ being the physical temperature of the full system.
The above produces a cut in the response function that stretches in the lower half of the complex $\omega$ plane horizontally from $-(\x/\y)k- i /\tauo(\TT)$ to $(\x/\y)k- i /\tauo(\TT)$, 
see Fig.~\ref{fig:cut}. Physically, the factor of $\x/\y$ (the effective equilibrium lightcone velocity) reflects that the massless gluons propagate along this effective lightcone. The imaginary part turns out to receive no
correction when expressed in terms of the full system temperature $\TT$.

\begin{figure}[tbp]
\centering 
\begin{picture}(200,150)
\footnotesize
\put(0,100){\vector(1,0){200}}
\put(205,100){${\rm Re}\,\omega/\TT$}
\put(100,0){\vector(0,1){130}}
\put(90,140){${\rm Im}\,\omega/\TT$}
\dottedline{3}(15,100)(15,0)
\dottedline{3}(185,100)(185,0)
\put(0,105){$-k/\TT$}
\put(178,105){$k/\TT$}
\dottedline[$\scriptstyle\bullet$]{2}(50,40)(150,40)
\put(50,40){\circle*{4}}
\put(150,40){\circle*{4}}
\put(20,25){$-\frac{k}{\TT}\frac{a}{b}\!-\!i\frac{4\pi}{5\kappa_1}$}
\put(130,25){$\frac{k}{\TT}\frac{a}{b}\!-\!i\frac{4\pi}{5\kappa_1}$}
\put(97.5,29.5){\color[rgb]{0.5,0,.75}\normalsize\sf x}
\put(102,29){\color[rgb]{0.5,0,.75}$-i\frac{\Gamma_0}{\TT}$}
\end{picture}

\caption{\label{fig:cut}
Analytic structure of the response function in the kinetic sector.
The branch cut arising from (\ref{cut}) is given by the thick black line.
The pole corresponding to the pure damping mode (\ref{gamma0kinetic}), which lies on the
second Riemann sheet, is indicated by the cross in violet.}
\end{figure}
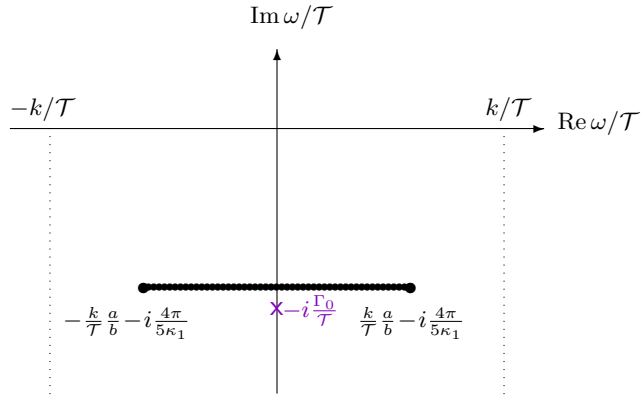

\subsection{Poles in response functions of the kinetic sector}

We now consider quasi-particle distribution fluctuations which cost energy and momentum. As before, we can split the propagating modes of the full theory into shear, sound, and tensor channels. We focus on the shear and sound channels for exactly the same reason as before -- the tensor channel has no hydrodynamic mode and to characterize it properly we require to embed the strongly coupled fluid into gravity which we have not done yet. 

As expected, we find that some of the propagating modes in both shear and sound channels are identical to the case of the conformal bi-hydrodynamic individual systems described before. This is simply because both the kinetic and Israel-Stewart sectors can be consistently truncated to conformal hydrodynamics individually. In particular, we will see that with the parametrization \eqref{tauo} of the kinetic relaxation time, we get exactly the same results as before with $\kap$ identified with $4\pi\eta_1/\s$. This reproduction of bi-hydrodynamics provides a consistency check of our calculations.

In addition to the  bi-hydrodynamic modes, there are two other non-hydrodynamic propagating modes in the full system in each shear and sound channel. These contribute poles in the response function. We find that one of these is continuously connected to the damping in the kinetic sector as we switch off the effective metric coupling. We will focus on this particularly because in case of the hydrodynamic sector, Israel-Stewart dynamics has been used simply as a tool for consistent embedding hydrodynamics and not for capturing actual relaxation dynamics. We will see that if the Israel-Stewart relaxation time is set to zero by taking $\lambda\rightarrow 0$ limit, the other damping mode has a smooth limit that captures the effective metric interactions of the kinetic sector with a strongly coupled fluid. Furthermore, if we take $\mathbf{k}\rightarrow 0$ limit, there is no way to distinguish the shear and sound channels owing to rotational symmetry of the equilibrium. The damping coefficient of the full system will then be the same in both shear and sound channels. This also provides a consistency check of our calculations.

Let us first focus on the shear channel. The effective metric fluctuations of the two sectors will be given by \eqref{g-shear} and \eqref{deltag-shear} as before. In the kinetic sector, the local mass-shell condition $p^\mu \g_{\mu\nu} p^\nu = 0$ will imply that at the linearized level:
\begin{eqnarray}\label{p0}
p^0(p,\theta, \phi, z, t) &=& \frac{p \y}{\x} +\delta p^0(p,\theta, \phi, z, t)  , \nonumber\\ \delta p^0 (p,\theta, \phi, z, t) &=& p\left( \frac{\bo}{\x^2} +\frac{\gammamet_{13}}{\x \y} \cos\theta\right)\sin\theta\cos\phi \, e^{i(kz -\omega t)}.
\end{eqnarray}
Assuming a self-consistent fluctuation of $\uo^\mu = (1/\x,\wo e^{i(kz -\omega t)},0,0) $ as before, we also obtain:
\begin{equation}
p_\mu \uo^\mu = - p \y + \frac{p}{\y}(\wo \y^3 - \gammamet_{13}\cos\theta)\sin\theta\cos\phi \, e^{i(kz -\omega t)}.
\end{equation}
There is no fluctuation in the temperature in the shear channel. The linearized fluctuation of the quasi-particle distribution function takes the form \eqref{qp-linear} with $\mathbf{k}$ in the $z$-direction and
\begin{eqnarray}
\delta f^{(eq)}(p,\theta, \phi) = e^{-\frac{p \y}{\To }}\frac{p}{\To \y}(\wo \y^3 - \gammamet_{13}\cos\theta)\sin\theta\cos\phi.
\end{eqnarray}
Above $\delta f^{(eq)}$ arises from the fluctuation in $p_\mu u^\mu$ since the local equilibrium distribution by definition takes the form $n_1 \pi^2 e^{-p_\mu u^\mu/T}$ and $\To$ does not fluctuate in the shear channel. Note that $\delta f^{(eq)}$ indeed reproduces the fluctuation in the energy-momentum tensor which takes a perfect fluid form.

The linearized Anderson-Witting equation \eqref{AW} can then be explicitly solved to obtain:
\begin{eqnarray}\label{Deltaf}
\Delta f(p, \theta, \phi) = f_0\frac{p \tauo\sin\theta\cos\phi\left(-(\bo+\wo \x \y^2)\y\omega+(k \wo \x^2 \y^2- \x \gammamet_{13}\omega)\cos\theta\right)}{\To  \x (-i \x \y- \y \tau_R \omega + k \x\tau_R\cos\theta)}.
\end{eqnarray}

In the kinetic sector, the energy-momentum tensor \eqref{kinetic-emt} after taking into account both effective metric and quasiparticle distribution fluctuations assume the linearized form
\begin{equation}\label{t-kinetic}
t^{\mu\nu} = {\rm diag}\left(\frac{\e }{\x^2},\frac{\p }{\y^2},\frac{\p }{\y^2},\frac{\p }{\y^2}\right) + \delta t^{\mu\nu}, \quad \e  = 3 \p   = 3 \alf \To ^4,
\end{equation}
with the non-vanishing components of $\delta t^{\mu\nu}$ being
\begin{eqnarray}\label{deltat-kinetic-shear}
\delta t^{01} = - \frac{\p  \bo + (\p  +\e ) \wo \x \y^2}{\x^2 \y^2} e^{i(kz-\omega t)}, 
\quad\delta t^{13} =\left(-\frac{\p \gammamet_{13}}{\y^4}+ \pio_{13}\right)e^{i(kz-\omega t)}
\end{eqnarray}
where
\begin{equation}\label{pi13-int}
\pio_{13} = \frac{1}{8\pi^3}\int_{-\infty}^\infty{\rm d} p\int_0^\pi {\rm d}\theta\int_0^{2\pi} {\rm d}\phi \,\,p^3 \y^2\cos\theta \sin^2{\theta}\cos\phi\,\, \Delta f(p,\theta, \phi).
\end{equation}
Comparing \eqref{deltat-kinetic-shear} with \eqref{deltat-hydro-shear} we see that the perfect fluid parts of the energy-momentum tensor perturbation that originates in the former case from the linearized perturbation in $p_0$, $p^0$ and $\delta f^{\rm (eq)}$ match perfectly. The dissipative contribution in \eqref{deltat-kinetic-shear} however originates from $\Delta f$ and is given by $\pio_{13}$. Using the solution \eqref{Deltaf} for $\Delta f$ in \eqref{pi13-int} we find that:
\begin{eqnarray}\label{eompi-shear}
\pio_{13} &=& \frac{2 \alf \To ^4}{k^5 \x^5 \y^2 \tauo^4} \Big( \gammamet_{13}\omega (i \x +\tauo \omega)+ k(-i \wo \x^2 \y^2 + \bo \tauo \omega)\Big)\\\nonumber&&
\Big(2 k^3 \x^3\tauo^3 + 3k \x \y^2 \tauo(\x -i \tauo\omega)^2
\\\nonumber&&+3(i \x \y+\y \tauo\omega)(-k^2 \x^2\tauo^2 +(i \x \y+ \y \tauo\omega)^2){\rm arctanh}\left(\frac{k \x \tauo}{i \x \y +\y \tauo\omega}\right)\Bigg).
\end{eqnarray}
In order to obtain the hydrodynamic limit we need to expand the right hand side above in $\tauo$ which yields
\begin{equation}\label{pi-hydro}
\pio_{13} = -i \frac{4 \alf \To ^4\tauo}{5 \x \y^4}\left(k\wo \x \y^2 -\gammamet_{13}\omega\right) + \mathcal{O}(\tauo^2).
\end{equation}
It is easy to note that the expansion in $\tauo$ is essentially the derivative expansion. Substituting the above form of $\pio_{13}$ in \eqref{deltat-kinetic-shear} and comparing again with the hydrodynamic form \eqref{deltat-hydro-shear}, we find a perfect match with
\begin{equation}
\eta_1 = \frac{4 \alf\To ^4\tauo}{5},
\end{equation}
i.e., $\eta_1/\s = \To \tauo/5$ and crucially $\kap = 4\pi \eta_1/\s$ as we have claimed. 

The energy-momentum conservation equation with $\delta t^{\mu\nu}$ given by \eqref{deltat-kinetic-shear} and the metric perturbation given by \eqref{deltag-shear} amounts to:
\begin{equation}\label{consv-shear}
(\e  + \p )(\bo+ \wo \x \y^2)\omega -k \x^2 \y^2\pio_{13} = 0.
\end{equation}
One can check that the above reduces to the standard hydrodynamic equation \eqref{hydro-shear} when $\pio_{13}$ is approximated by \eqref{pi-hydro}. We can regard \eqref{eompi-shear} and \eqref{consv-shear} as the dynamical equations for $\pio_{13}$ and $\wo$.

It is to be noted that one can explicitly check that the conservation equation \eqref{consv-shear} is equivalent to the linearized version of the matching condition $u_\mu (t^\mn - {t^\mn}^{\rm (eq)}) = 0$ which says that the projected energy-momentum tensor obtained from the full quasi-particle distribution $f$ should agree with that obtained from $f^{\rm (eq)}$. In fact, this matching condition is necessary to ensure energy-momentum conservation. At the level of linearized shear-sector fluctuation, the matching condition reduces to
\begin{equation}\label{matching}
\Delta t^{01} \equiv \frac{1}{8\pi^3}\int_{-\infty}^\infty{\rm d} p\int_0^\pi {\rm d}\theta\int_0^{2\pi} {\rm d}\phi \,\,\frac{p^3 \y^3}{\x} \sin^2{\theta}\cos\phi\,\, \Delta f(p,\theta, \phi) = 0.
\end{equation}
Explicitly, we can check that if we use $\Delta t^{01} = 0$ with the on-shell form of $\Delta f$ given by \eqref{Deltaf} and the equation of motion \eqref{eompi-shear} for $\pio_{13}$ to solve for the variables $\wo$ and $\pio_{13}$, we find that indeed \eqref{consv-shear} is satisfied leading to energy and momentum conservation.

Embedding the holographic conformal fluid (with $\et  = 3\pt  = 3\alft  \Tt ^4$ as before) in the Israel-Stewart framework  we obtain:
\begin{eqnarray}\label{deltat-is-shear}
\delta \tit^{01} = - \frac{\pt  \bt + (\pt  +\et )  (\wt \x \y)^2}{ \xt^2  \yt^2} e^{i(kz-\omega t)}, 
\quad\delta  \tit^{13} =\left(-\frac{\pt  \gammamett_{13}}{ \yt^4}+  \pit_{13}\right)e^{i(kz-\omega t)}.
\end{eqnarray}
The linearized Israel-Stewart equation of motion of $\pit_{13}$ is:
\begin{equation}\label{eompi-shear2}
-i   \yt  \taut   \pit_{13}\omega +   (\xt \yt)^4   \pit_{13}-i  \eta_2  \gammamett_{13}\omega - i k \eta_2   \wt \x \y^2 = 0.
\end{equation}
The conservation of energy-momentum tensor mirrors \eqref{consv-shear} of the kinetic sector and takes the form:
\begin{equation}\label{consv-shear2}
(\et  + \pt )(\bt+ \wt \x \y^2)\omega -k \xt^2\yt^2\pit_{13} = 0.
\end{equation}
The equations \eqref{eompi-shear2} and \eqref{consv-shear2} are the equations of motion for $\pit$ and $\wt$. Note that once again the hydrodynamic limit is reproduced by Taylor expansion in $\taut$ about $\taut = 0$. 

To ensure conformality, we once again parametrize:
\begin{equation}
\eta_2 =  \frac{\alft  \kapt}{\pi}  \Tt ^3
\end{equation}
as before. Furthermore, we will later take the limit $\lambda \rightarrow 0$ in which $\taut$ vanishes.

We now repeat the steps in the previous subsection. First, we use the coupling equations \eqref{coupling-shear-sector} to solve for $\bo$, $\bt$, $\gammamet_{13}$ and $\gammamett_{13}$ in terms of $\wo$, $\wt$, $\pio_{13}$ and $\pit_{13}$. Next, we substitute these solutions for $\bo$, $\bt$, $\gammamet_{13}$ and $\gammamett_{13}$ in the dynamical equations, namely \eqref{eompi-shear}, \eqref{consv-shear}, \eqref{eompi-shear2} and \eqref{consv-shear2} to obtain the $4 \times 4$ matrix equations:
\begin{equation}
Q_{AB}(\omega, k) \Lambda_B = 0
\end{equation}
where $\Lambda_B = (\wo, \pi_{13},\wt,\pit_{13})$. Finally, we obtain the eigenmodes $\omega(k)$ by solving ${\rm det}\, Q = 0$ at each $k$.

There are four propagating modes for each $k$ as discussed earlier. Two of these are exactly the bi-hydro shear-like eigenmodes obtained earlier with diffusion constants $\mathcal{D}_a$ and $\mathcal{D}_b$. We thus reproduce our previous results.

There are additionally two relaxation eigenmodes. 
One of these eigenmodes is related to the Israel-Stewart relaxational mode and its damping constant
becomes large for small $\lambda$ and therefore can be decoupled. The corresponding propagating mode in this limit is localized mostly in the Israel-Stewart sector and involves the following combination of $\pio_{13}$ and $\pit_{13}$ where
\begin{equation}
\pio_{13} = \left(\frac{4\alf }{5}\gTf + \mathcal{O}(\ga^2\TT^8)\right) \pit_{13}
\end{equation}
when $\ga\TT^4$ is small. 

The damping constant of the other relaxational mode remains finite.
It is of the form
\begin{equation}\label{gamma0kinetic}
\omega(k) = - i \left[\Gamma_0 + \mathcal{O}(k^2)\right]
\end{equation}
with perturbative expansion in the limit $\lambda \rightarrow 0$ according to
\begin{eqnarray}\label{Gamma-pert}
\Gamma_0 &=& \frac{4\pi\TT }{5 \kap } + \frac{16\pi\alf\alft (5\kap - 4\kapt )}{125\kap ^2}\ga^2\TT ^9  + \mathcal{O} (\ga^3).
\end{eqnarray}
This is interesting because the Anderson-Witting kinetic theory does not have on its own any non-hydrodynamic pole -- the mutual metric coupling evidently causes a pole to be generated from the cut discussed above (for $\gTf\to0$ it coincides with the cut).
This pole is farther from the real axis than the cut when $\kappa_1>\kappa_2$, i.e., when the kinetic
sector is more weakly coupled than the second sector described by pure hydrodynamics. The corresponding propagating mode involves the following combination of $\pio_{13}$ and $\pit_{13}$ where
\begin{equation}
\pit_{13} = \left(\frac{4\alft \kapt}{5\kap}\gTf + \mathcal{O}(\ga^2\TT^8)\right) \pio_{13}
\end{equation}
so that it is mostly localised in the kinetic sector as expected in the limit of small $\ga\TT^4$.

Interestingly, when $5\kap = 4 \kapt$ all corrections to $ \Gamma_0/\TT$ vanish so that it is exactly $4\pi/5\kap$  as the perturbation series \eqref{Gamma-pert} indicates. However, in this case the $\lambda\rightarrow 0$ limit becomes sick because the other mode becomes unstable. This is consistent with the expectation that the non-kinetic sector should have a lower $\eta/s$ as it is more strongly coupled.

Furthermore, the departure of $\Gamma_0/\TT$ from its decoupling limit value $4\pi/5\kap$ in the full calculation
are found to be very small for any value of $\gqT$ (see the left panel of Fig.~\ref{fig:G0T}). The damping
constant $\Gamma_I$ of the Israel-Stewart relaxational mode is evaluated in the right panel which
is indeed large for all $\gqT$ for the small value of $\lambda$ chosen. However, it turns out that one cannot
take the limit $\lambda \to 0$ for large $\gqT$, for $\Gamma_I$ diverges at a certain value of $\gqT$ beyond which
it turns negative, corresponding to an instability. One thus has to keep $\lambda$ finite in order to decouple this mode.

Repeating the same calculation in the sound channel, we find that we indeed reproduce the bi-hydro sound sector modes and the same damping coefficient $\Gamma_0$.

A remarkable outcome from our calculations is that 
non-hydrodynamic observables turn out to receive mild or no non-perturbative corrections even when the hydrodynamic sector receives large qualitative and quantitative modifications.

\begin{figure}[tbp]
\centering 
\includegraphics[width=.5\textwidth]{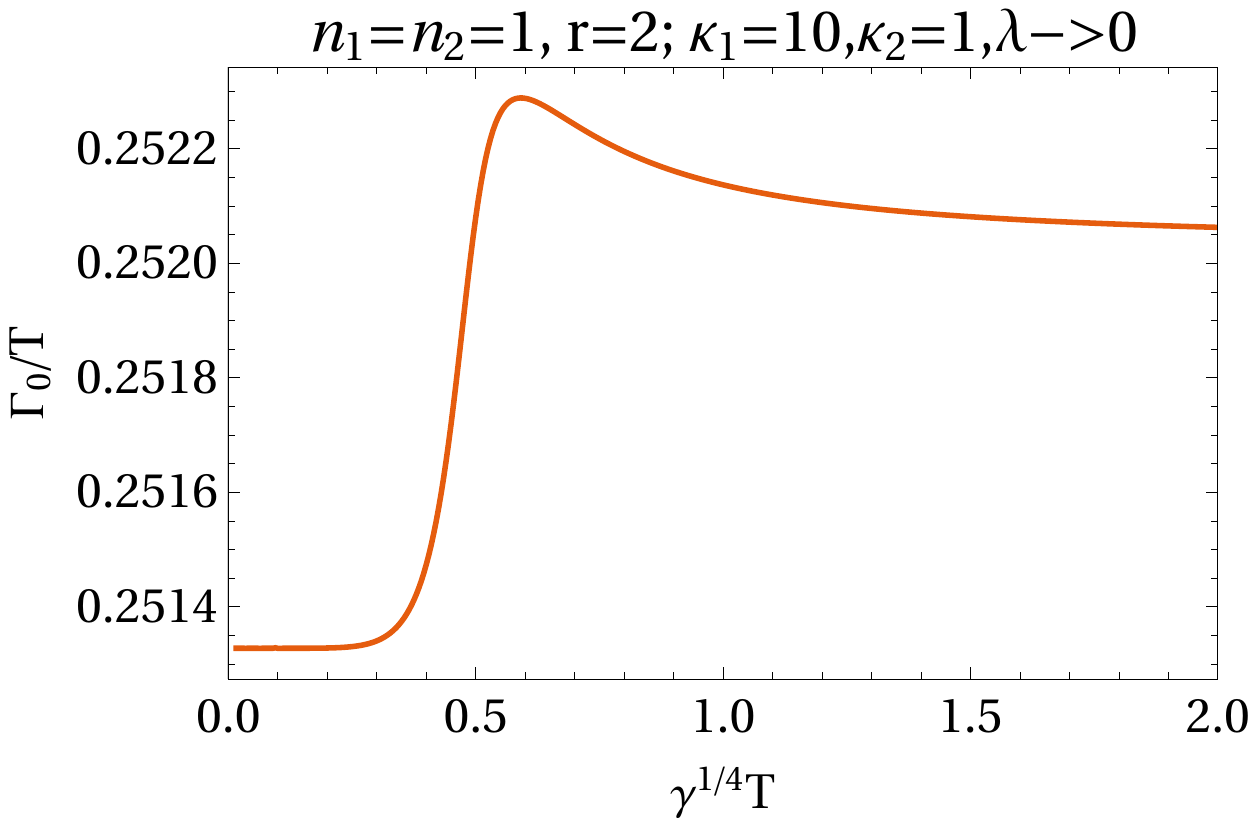}%
\includegraphics[width=.5\textwidth]{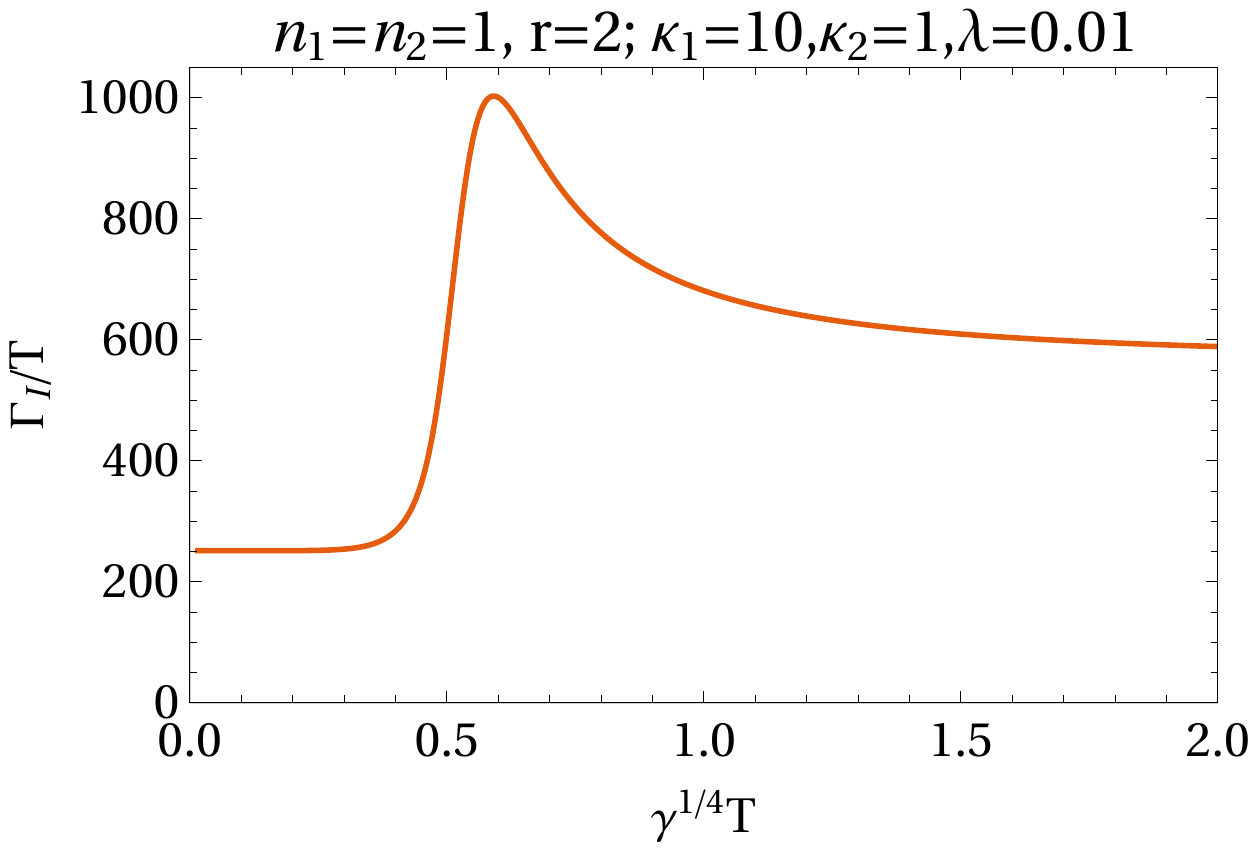}
\caption{\label{fig:G0T}
Pure damping modes (identical in shear sector and sound sector). Left panel: damping constant $\Gamma_0$ that remains
finite when $\lambda\to0$; right panel: damping constant $\Gamma_I$ of the Israel-Stewart relaxational mode which
is large for small $\lambda$ (however, $\lambda$ cannot be made arbitrarily small at large $\gqT$, see text).
}
\end{figure}

\section{Conclusions}\label{Sec:Outlook}

In the semi-holographic approach to the dynamics of quark-gluon plasma with its coexistence of strongly and weakly
interacting sectors it has been proposed to introduce a coupling of the respective marginal operators \cite{Iancu:2014ava,Mukhopadhyay:2015smb,Banerjee:2017ozx}.
In four dimensions, this always includes the energy-momentum tensors which can be coupled to the (effective)
metric of the complement subsystem. In this paper we have determined the most general ultralocal mutual effective metric coupling
which leads to a total energy-momentum tensor with respect to the flat Minkowski space of the complete system.
The effective metric tensors of the subsystems encode the interactions between them; in particular they
lead to state-dependent effective lightcone velocities within a subsystem that can be smaller than unity,
similar to thermal masses (but different in that the latter reduce the velocity of massless particles
depending on their energy).

We have then studied the consequences of mutual effective metric couplings in equilibrium and in a hydrodynamic limit
of near-equilibrium situations. Assuming full thermal equilibrium, we have found an interesting phase structure,
which can be separated by a first or second-order phase transition, or an analytic crossover, depending on the
coupling parameters. With only the two coupling constants of inverse dimension four turned on, we obtained two
distinct phases where the one at higher mutual coupling (or equivalently higher system temperature) has a
larger number of degrees of freedom per volume and eventually approaches conformality, which is curiously reminiscent of the deconfinement transition in QCD.

Studying the hydrodynamic behavior of such a two-fluid system, we found two modes in both the sound channel and in the shear
channel (with our detailed findings summarized already at the end of the respective subsections above). 
In the shear channel we found a decrease of shear diffusion constants as the mutual coupling is increased,
in accordance with the fact that shear diffusion is in general weaker at strong coupling.\footnote{However, since increasing
the mutual coupling is equivalent to higher system temperature, the behavior as a function of temperature here turns
out to be opposite to what is expected in QCD.} The overall shear viscosity, which is determined by the Kubo formula
involving the total energy momentum tensor, shows a similar behavior, numerically intermediate between the viscosity values
of the subsystems.

In the sound sector, we also found two modes, which correspond essentially to in-phase and out-of-phase density
perturbations of the two subsystems. One mode always has a velocity close (or equal) to the thermodynamically defined
speed of sound, while the other is slower and always below the effective lightcone velocity of the subsystems, and also
more weakly attenuated, suggestive of a quasiparticle nature. In the transition region, a role reversal takes place,
reminiscent of a similar phenomenon in other two-fluid systems \cite{Alford:2013koa}.
The damping of the two modes depends in a complicated manner on the shear viscosities in the individual subsystems and
their mutual interaction (or system temperature). In the long-wavelength limit, however, both attenuations vanish 
quadratically with the modulus of the wave vector, which means that completely homogeneous and isotropic density perturbations
do not equilibrate. This is similar to the behavior found in the semi-holographic toy model of Ref.~\cite{Mukhopadhyay:2015smb},
where the dual gravitational theory did not permit thermalization because in this limit there are no
propagating degrees of freedom (bulk gravitons). Instead one needs to turn on scalar degrees of freedom (the bulk dilaton) dual to
the Lagrangian density \cite{Ecker:2018ucc}. 

Finally, we also investigated the case where one subsystem is described by kinetic theory.
{In order to do so, we supplemented one of 
the subsystems with microscopic dynamics and chose to describe it in terms of a transport theory of 
a distribution function of particles $f(\vec x, \vec p,t)$.
We observe that because the metric coupling is mediated by local stretches of space-times of the individual
subsystems, all the modes with different $\vec p$ are affected uniformly. Because of this, the only effect the 
coupling between the two subsystems can have to the distribution 
function is to rescale the momentum variable $f(\vec x, \vec p,t) \rightarrow f(\vec x, \vec p' ,t)$ with $p'_i = \Lambda_{ij}(\vec x,t)p_j$.} In consequence the coupling cannot bring the distribution to the equilibrium form, and the non-hydrodynamic 
modes are unaffected by the coupling. Therefore, as in the toy example studied in \cite{Mukhopadhyay:2015smb} the thermalization of the full system must rely on thermalization of the individual subsectors, in the sense that coupling a non-dissipative subsystem to a dissipative one does not allow the non-dissipative subsystem to thermalize.
Considering the
damping of non-hydrodynamic relaxation modes we find that they 
receive typically small modifications, or none, through the mutual effective metric coupling.

Our results provide a glimpse of what can or cannot be expected from semi-holographic models
for equilibration and thermalization, rather independently of the internal dynamics of the perturbative
and nonperturbative sectors of the full system, when the semi-holographic coupling is exclusively through
the energy-momentum tensors of the subsectors. The semi-holographic model for the early stages of
heavy-ion collisions formulated in \cite{Iancu:2014ava,Mukhopadhyay:2015smb} is restricted to the phase where semi-hard gluons are
overoccupied so that they can be described by classical Yang-Mills equations while a thermal bath of
infrared degrees of freedom is building up. Some recent progress on the level of toy models has
been presented in \cite{Ecker:2018ucc} and is paving the way to more relevant qualitative studies. For later stages
of the formation and evolution of the quark-gluon plasma more refined models would be required which
involve a quantum kinetic theory for the hard degrees of freedom. Eventually, realistic studies
would have to turn to hard-thermal-loop resummed kinetic theory for the UV sector \cite{Arnold:2002zm} and a confining
model for the IR sector such as improved holographic QCD \cite{Gursoy:2010fj}. Before embarking on this, it will however
be interesting to study further the general qualitative features and consequences of a semi-holographic setup
by including couplings of operators other than the energy-momentum tensor, by studying more complicated
kinetic models, and by also considering the issue of fluctuations which may be crucial for full
thermalization but are beyond the large $N$ limit.

\section*{Acknowledgments}

We thank Y.\ Hidaka, E.\ Iancu, R.~D.\ Pisarski, A.\ Schmitt, and D.-L.\ Yang for useful discussions. 
A.\ Mukhopadhyay acknowledges support from a Lise Meitner fellowship
of the Austrian Science Fund (FWF), project no.\ M1893-N27, a research associateship at CERN,
and support from the Ramanujan Fellowship of
DST India and the new faculty initiation grant of IIT Madras.
F.\ Preis was supported by the FWF project P26328-N27,
and A.\ Soloviev was supported by the FWF doctoral program W1252.

\begin{appendix}
\section{General tensorial coupling rules}\label{App:GeneralCouplings}

In this appendix, we work out the most general ultralocal coupling rules
of the effective metric fields with the energy-momentum tensors of the complement subsystems.
The effective metrics cannot depend on the (covariant) derivatives of the
subsystem energy-momentum tensors because even the first derivatives of the
subsystem energy-momentum tensors can be discontinuous as for instance along a
phase boundary. The effective metrics, however, cannot have discontinuities, because in
that case one cannot formulate general covariant equations of motion as
covariant derivatives cannot be defined where the metric becomes discontinuous.
The energy-momentum tensor typically involves only first derivatives of the
fields and so does the action which determines the weights for field
configurations in the path integral. Therefore, field configurations where the
second derivatives of the fields and thus first derivatives of the
energy-momentum tensor are discontinuous are allowed to contribute to the path
integral. So one can generally argue that the effective metrics should be determined
only by polynomials of the subsystem energy-momentum tensors and not their
derivatives. 

With the short-hand notation
\be
\tu^\mu{}_\nu=\frac{\sqrt{-g}}{\sqrt{-\gB}}t^{\mu\rho}\gB_{\rho\nu},\quad
\tut^\mu{}_\nu=\frac{\sqrt{-\gt}}{\sqrt{-\gB}}\tit^{\mu\rho}\gB_{\rho\nu},
\quad t=\tu^\mu{}_\mu, \quad \tit=\tut^\mu{}_\mu
\ee
the generalization of (\ref{DeltaK})
to higher than quadratic order in the energy-momentum
tensors $t^\mn$ and $\tit^\mn$ 
can be written as
\be\label{DeltaKgeneral}
\Delta K = \sum_{m\ge0,j_i\ge0}\kappa_{mj_1 j_2 \ldots}(t \tit)^m \left(\tr{\tu\cdot\tut}\right)^{j_1} 
\left(\tr{(\tu\cdot\tut)^2}\right)^{j_2} \ldots,
\ee
where terms of order $2k$ have $m=0,\ldots k$ and $\sum j_i=k-m$.
Thus the number of terms in the interaction part of the total energy-momentum tensor at order $2k$,
denoted below by $|\kappa_k|$, is given by 
sums over the number-theoretic partition function
\be
|\kappa_k|=\sum_{m=0}^k p(k-m)
\ee
with $p(n)=1,1,2,3,5,7,11,15,22,30,42,\ldots$ for $n=0,1,2,3,4,5,6,7,8,9,10,\ldots$.

In order to have a conserved total energy-momentum tensor we need (for simplicity switching to a Minkowski background
metric $\gB_\mn$ for now)
\be\label{conservationtotalemt}
0=\partial_\mu K^\mu{}_\nu=\frac12 (\partial_\nu g_{\mu\sigma})\sqrt{-g}t^{\mu\sigma}
+\frac12 (\partial_\nu \gt_{\mu\sigma})\sqrt{-\gt}\tit^{\mu\sigma}+\partial_\nu \Delta K
\ee
where the Ward identities for the subsystems, (\ref{WI1}) and (\ref{WI2}), have been used.

The terms obtained by differentiating $\Delta K$ can be easily seen (using cyclicity of the trace)
to be matched by the ansatz
\ba
\g_\mn &=& \gB_\mn+\sum_{\ell\ge1,m\ge0,j_i\ge0} \gamma_{\ell|mj_1\ldots j_p} 
(\gB\cdot \tut \cdot (\tu\cdot \tut)^{\ell-1})_\mns (t \tit)^m \tr{\tu\cdot\tut}^{j_1} \tr{(\tu\cdot\tut)^2}^{j_2} \ldots
\nonumber\\
&+&  \gB_\mn \sum_{m\ge0,j_i\ge0}\gamma'_{1|mj_1\ldots j_p}t^m \tit^{m+1} \tr{\tu\cdot\tut}^{j_1} \tr{(\tu\cdot\tut)^2}^{j_2} \ldots 
\ea
and 
\ba
\gt_\mn &=& \gB_\mn+\sum_{\ell\ge1,m\ge0,j_i\ge0} \gamma_{\ell|mj_1\ldots j_p} 
(\gB\cdot \tu \cdot (\tut\cdot \tu)^{\ell-1})_\mns (t \tit)^m \tr{\tu\cdot\tut}^{j_1} \tr{(\tu\cdot\tut)^2}^{j_2} \ldots
\nonumber\\
&+&  \gB_\mn \sum_{m\ge0,j_i\ge0}\gamma'_{1|mj_1\ldots j_p}\tit^m t^{m+1} \tr{\tu\cdot\tut}^{j_1} \tr{(\tu\cdot\tut)^2}^{j_2} \ldots .
\ea
(Indices enclosed by round parentheses are to be symmetrized.)
To match
terms of order $2k$ in $\Delta K$ we have to take $\ell=1,\ldots k$, $m=0,\ldots k-\ell$, and $\sum j_i=k-\ell-m$.
The number of coefficients $\gamma$ and $\gamma'$ at this order (denoted by $|\gamma_k|$) is therefore
\be
|\gamma_k|=|\kappa_{k-1}| + \sum_{\ell=1}^k |\kappa_{k-\ell}|=2|\kappa_{k-1}|+|\kappa_{k-2}|+|\kappa_{k-3}|+\ldots
\ee
where it is convenient to define $|\kappa_{-n}|=0$ for $n=1,2,\ldots$. (Note that $|\kappa_0|=1$, corresponding
to the possibility of adding a cosmological constant, which we ignore because it does not lead to a gravitational
coupling of the two sectors.)

For the first few orders $2k$ the number of coefficients in the ans\"atze for $\Delta K$ and the two metric tensors 
read
\begin{center}
\begin{tabular}{l|rrrrrr}
 $k$ & 0 & 1 & 2 & 3 & 4 & 5 \\
 \hline
 $|\kappa_k|$ & 1 & 2 & 4 & 7 & 12 & 19 \\
 $|\gamma_k|$ & 0 & 2 & 5 & 11 & 21 & 38 \\
\end{tabular}
\end{center}

Plugging in the ans\"atze in (\ref{conservationtotalemt}) gives as many linear relations between the coefficients
as there are different terms produced by differentiating $\Delta K$. For each term in $\Delta K$ we get as many
different derivatives as there are different factors. We thus need to consider how many different parts a partition
of the number $k-m$ into $j_i$'s has. Define
\be\label{comb-id}
q(n)=\sum_{\text{partitions of $n$}} (\text{number of different parts of the partition})
\ee
From differentiating the trace terms with powers $j_i$ at order $2k$ we get
$q(k)+q(k-1)+\ldots+q(1)$ different terms, while from differentiating $(t \tit)^m$ with $m\ge1$ we get
$p(k-1)+p(k-2)+\ldots+p(0)=|\kappa_{k-1}|$ different terms (here $p(0)=1$ corresponds to the single term where $m=k$).
Now it turns out that\footnote{A proof of this non-obvious fact can be found
in M. D. Hirschhorn, ``THE NUMBER OF DIFFERENT PARTS IN THE PARTITIONS OF n'', Fibonacci Quaterly \textbf{52} (2014) 10--15
[http://web.maths.unsw.edu.au/\~{}mikeh/webpapers/paper192.pdf].}
$q(n)=p(n-1)+p(n-2)+\ldots+p(0)=|\kappa_{n-1}|$ and therefore
the number of relations is $2|\kappa_{k-1}|+|\kappa_{k-2}|+\ldots=|\gamma_k|$. Hence, there are always
as many relations as coefficients in the ansatz for the metric which can be used to determine them
in terms of the (free) coefficients $\kappa$ in $\Delta K$.

It is interesting to note that the most general form of the conserved energy-momentum tensor that we obtain here is that where the interaction term is the most arbitrary symmetric polynomial of $\tu^\mu{}_\nu$ and $\tut^\mu{}_\nu$, and (i) is thermodynamically consistent and (ii) such that the total entropy is the sum of the two subsystem entropies. We show in Appendix \ref{App:ThdynConsistency} that the latter requirements can indeed be satisfied if the interaction term is proportional to $\delta^\mu{}_\nu$ as we have here and is otherwise arbitrary. The combinatoric identity \eqref{comb-id} along with our general construction ensures that any such arbitrary interaction term in the full conserved energy-momentum tensor satisfying the above thermodynamic requirements can be interpreted as a democratic effective metric interaction, i.e., as an interaction that can be absorbed into an appropriate mutual modification of respective effective metrics. This is important because in absence of any symmetry argument to rule out a specific interaction term, it can be generically present, and therefore indeed we should be able to obtain it via an mutual effective metric coupling.

\subsection{Solutions to lowest orders}


When $\Delta K$ is a quadratic expression formed of the \EM\ tensors,
there are two coefficients each in $\Delta K$ and the metric ansatz, with two equations relating them:
\be
\Delta K=\kappa \, \tr{\tu\cdot\tut} + \kappa' t \tit,
\ee
where $\kappa\equiv \kappa_{01\dot0}$ and $\kappa'\equiv \kappa_{1\dot0}$ in terms of the
general multi-index coefficients introduced above, with $\dot0$ denoting an infinite string of zeros;
\be\label{coupling-rule-lowest}
g_\mn=\gB_\mn+\gamma (\gB\cdot \tut)_\mn+\gamma' \gB_\mn \tit, \quad
g_\mn=\gB_\mn+\gamma (\gB\cdot \tu)_\mn+\gamma' \gB_\mn t, 
\ee
where $\gamma\equiv \gamma_{1|\dot0}$ and $\gamma\equiv \gamma'_{1|\dot0}$.

Eq.~(\ref{conservationtotalemt}) yields
\be
\kappa=-\frac12 \gamma,\quad \kappa'=-\frac12 \gamma',
\ee
in accordance with (\ref{DeltaK}).


At the next higher order,
there are 4 coefficients in $\Delta K$,
\be
\kappa_{2\dot0},\quad
\kappa_{11\dot0},\quad
\kappa_{02\dot0},\quad
\kappa_{001\dot0},
\ee
and 5 coefficients in the metric tensors $g_\mn$ and $\gt_\mn$, constrained by 5 linear relations following from (\ref{conservationtotalemt}), which yield
\be
\gamma_{2|\dot0}=-\frac43 \kappa_{001\dot0},\quad
\gamma_{1|01\dot0}=-\frac43 \kappa_{02\dot0},\quad
\gamma'_{1|1\dot0}=-\frac43 \kappa_{2\dot0},\quad
\gamma_{1|1\dot0}=\gamma'_{1|01\dot0}=-\frac23\kappa_{11\dot0}.
\ee

\subsection{Action formulation}

When the subsystems can be described by an action principle,
one can formulate the democratic effective metric coupling of the two subsystems also through a joint action. Let the fundamental elementary fields of the first subsystem be denoted collectively as $\phi$ and those of the second subsystem as $\tilde{\phi}$. Then the dynamics of the full system with the lowest-order effective metric coupling (\ref{coupling-rule-lowest}) can be obtained from
\begin{eqnarray}\label{Alex-action}
S[\phi, \tilde{\phi}, \g_\mn, \gt_\mn, \gB_\mn] &=& \int{\rm d}^d x \sqrt{-\g} \mathcal{L}_1[\phi, \g_{\mu\nu}] +\int{\rm d}^d x \sqrt{-\gt} \mathcal{L}_2[\tilde{\phi}, \gt_{\mu\nu}] \\
\nonumber&&+  \frac{1}{2\ga}\int {\rm d}^d x \sqrt{-\gB}\left(\g_{\mu\alpha}-\gB_{\mu\alpha}\right)\gB{}^{\alpha\beta}\left(\gt_{\beta\nu}-\gB_{\beta\nu}\right)\gB{}^{\nu\mu} \\\nonumber
&&+\frac{1}{2\ga}\frac{\gap}{d\gap-\ga} \int {\rm d}^d x \sqrt{-\gB}\left( \g_{\mu\nu}\gB{}^{\mu\nu}-d\right)\left( \gt_{\alpha\beta}\gB{}^{\alpha\beta}-d\right).
\end{eqnarray}
We note that the above action is not simply a functional of $\phi$, $\tilde{\phi}$ and $\gB_\mn$ as usually is the case but also of the two effective metrics $\g_\mn$ and $\gt_\mn$. 
Thus $\g_\mn$ and $\gt_\mn$ appear as auxiliary fields and the interaction terms of the two subsystems represented by the last two lines of the above action merely implement the algebraic relations between the effective metrics and the subsystem energy-momentum tensors. On the other hand, if we vary with respect to $\phi$ and $\tilde\phi$ first, we evidently obtain the two subsystem dynamical equations in the respective effective metrics, since the last two lines are independent of $\phi$ and $\tilde\phi$. Since the individual subsystem actions are diffeomorphism invariant, these automatically imply that the subsystem Ward identities $\nablao_\mu t^\mu_{{\,}\nu} = 0$ and $\nablat_\mu \tit^\mu_{{\,}\nu} = 0$ hold on-shell. We can also explicitly check that when the full action is stationary for variation of $\phi$, $\tilde\phi$, $\g_\mn$ and $\gt_\mn$, then its variation with respect to the background metric $\gB_\mn$ yields the full energy-momentum tensor $\ttot^\mn$ as given in Section \ref{Sec:Setup}.\footnote{The existence of a full conserved energy-momentum tensor in any case follows from the individual subsystem Ward identities (which are easier to obtain from the above action as shown above). Typically a system has a unique energy-momentum tensor of a system up to possible local terms that are separately conserved by virtue of identities (i.e. without need of equations of motion). We can explicitly check that (\ref{Alex-action}) does not produce such additional terms.}

\section{Thermodynamic consistency}\label{App:ThdynConsistency}

\subsection{General proof}

We will show that any consistent effective metric coupling rule 
with a total conserved \EM\ tensor of the form (\ref{emt-full})
will imply that if the global equilibrium condition 
\begin{equation}
\To \x = \Tt\xt = \TT 
\end{equation}
is satisfied then thermodynamic consistency is also satisfied with total entropy $\ST(\TT) = \s(\To)\y^3 + \st(\Tt)\yt^3$. 
Note that this not only covers the simplest metric coupling rule (\ref{coupling-rule}) but the most general ansatz
for $\Delta K$ in (\ref{DeltaKgeneral}).

In order to prove this assertion, it is useful to consider the system in a static gravitational potential so that the background metric is:
\begin{equation}\label{bg-grav-pot}
\gB_\mn = {\rm diag}\left(-e^{-2\phi(\mathbf{x})}, 1,1,1\right),
\end{equation}
where $\phi(\mathbf{x})$ is static, i.e., not a function of time, but otherwise arbitrary. 
One can then make static ans\"atze for the effective metrics of the individual sectors, i.e. assume that
\begin{equation}\label{ind-em-x}
\g_{\mu\nu} = {\rm diag}(- \x(\mathbf{x})^2, \y(\mathbf{x})^2, \y(\mathbf{x})^2, \y(\mathbf{x})^2 ), \quad \gt_{\mu\nu} = {\rm diag}(-\xt(\mathbf{x})^2,\yt(\mathbf{x})^2,\yt(\mathbf{x})^2,\yt(\mathbf{x})^2 ).
\end{equation}
The effective metric coupling equations (constructed after removing anomalous terms in the individual energy-momentum tensors) with the assumption that each sector is in thermodynamic equilibrium at temperatures $\To$ and $\Tt$, respectively, i.e. with the forms
\begin{eqnarray}\label{ind-emt-x}
 t^{\mu\nu} &=& {\rm diag}\left(\frac{\e(\To(\mathbf{x}))}{\x(\mathbf{x})^2},\frac{\p(\To(\mathbf{x}))}{\y(\mathbf{x})^2},\frac{\p(\To(\mathbf{x}))}{\y(\mathbf{x})^2},\frac{\p(\To(\mathbf{x}))}{\y(\mathbf{x})^2}\right), \nonumber\\  \tit^{\mu\nu} &=& {\rm diag}\left(\frac{\et(\Tt(\mathbf{x}))}{\xt(\mathbf{x})^2},\frac{\pt(\Tt(\mathbf{x}))}{\yt(\mathbf{x})^2},\frac{\pt(\Tt(\mathbf{x}))}{\yt(\mathbf{x})^2},\frac{\pt(\Tt(\mathbf{x}))}{\yt(\mathbf{x})^2}\right),
\end{eqnarray}
involve no derivative of the metric so will still be algebraic (although the solutions will depend on the specific spatial point $\mathbf{x}$). 
However, the coupling equations 
need to be taken in their generalized form with nontrivial
background metric $\gB$. We will assume that we can obtain flat-space solutions by smoothly taking $\phi(\mathbf{x})\rightarrow 0$. 

To define an equilibrium solution, we need to relate $\To$ and $\Tt$ parametrizing each in terms of the full system temperature $\TT$. At global equilibrium, the inverse of the Euclidean time circle specifying the system temperature should be constant. The global equilibrium condition is simply then:
\begin{equation}\label{global-eq-grav-pot}
\To(\mathbf{x})\x (\mathbf{x})= \Tt (\mathbf{x})\xt(\mathbf{x}) = \TT (\mathbf{x})e^{-\phi(\mathbf{x})} = \TT_0,
\end{equation}
where $\TT_0$ is a constant, parametrizing the
global thermal equilibrium of the full system in the background metric \eqref{bg-grav-pot}. 

The first thing that we need to show is that the above is compatible with the conservation of the energy-momentum tensors. 
Using (\ref{WI1}),
we can check that the conservation of the individual thermal energy-momentum tensors \eqref{ind-emt-x} in the respective effective metrics \eqref{ind-em-x} imply simply that
\begin{eqnarray}\label{p-gradients}
\frac{\partial_i \p}{\e + \p}+  \frac{\partial_i \x}{\x} =0, \quad
\frac{\partial_i \pt}{\et + \pt}+  \frac{\partial_i \xt}{\xt} =0,
\end{eqnarray} 
respectively. Note that
$\y(\mathbf{x})$ and $\yt(\mathbf{x})$ do not feature directly in the above equations. Since ${\rm d}\p = \s {\rm d}\To$, $\e + \p = \To\s$, ${\rm d}\pt = \st {\rm d}\Tt$, $\et + \pt = \Tt\st$, 
the conservation equations (\ref{p-gradients}) are equivalent to
\begin{eqnarray}\label{consv-final}
\partial_i({\rm ln}(\To\x)) =0, \quad
\partial_i({\rm ln}(\Tt\xt)) =0,
\end{eqnarray} 
and thus 
implied by the global equilibrium condition \eqref{global-eq-grav-pot}.

By construction, the effective metric couplings ensure that the total \EM\ tensor, which can be parameterised as
\begin{equation}
\ttot^\mn = {\rm diag}\left(\ET(\TT(\mathbf{x}))e^{2\phi(\mathbf{x})},\PT(\TT(\mathbf{x})), \PT(\TT(\mathbf{x})), \PT(\TT(\mathbf{x})) \right),
\end{equation}
will be conserved in the background metric \eqref{bg-grav-pot}, since the individual thermal energy-momentum tensors are conserved with respect to the respective 
effective metrics. We therefore have 
\begin{equation}
\frac{\partial_i \PT}{\ET+ \PT} - \partial_i \phi = 0.
\end{equation}
Equations \eqref{consv-final} and \eqref{global-eq-grav-pot} imply that
\begin{equation}
\frac{\partial_i \TT}{\TT} - \partial_i \phi = 0,
\end{equation}
and therefore
\begin{equation}
\frac{\partial_i \TT}{\TT} = \frac{\partial_i \PT}{\ET+ \PT}.
\end{equation}
Identifying
$\ET+ \PT = \TT \ST$ 
this implies
\begin{equation}
\partial_i \PT = \ST \partial_i \TT.
\end{equation}
Since the above should hold for arbitrary smooth $\phi(\mathbf{x})$, we conclude that 
\begin{equation}
{\rm d}\PT = \ST {\rm d} \TT
\end{equation}
where the variation is taken by changing the constant parameter $\TT_0$. Together with $\ET+ \PT = \TT \ST$ the above implies
\begin{equation}
{\rm d}\ET = \TT {\rm d} \ST.
\end{equation}
This shows
that thermodynamic consistency follows from the conservation of the full energy-momentum tensor as ensured by our effective metric coupling.
In particular, assuming $\ET+ \PT = \TT \ST$ and the global equilibrium condition \eqref{global-eq-grav-pot}, we obtain ${\rm d}\ET = \TT {\rm d} \ST$ from the conservation of the full energy-momentum tensor. Clearly, we can take the limit $\phi(\mathbf{x})\rightarrow 0$ limit to obtain the
desired proof of thermodynamic consistency in flat space. 

We still need to show which form $\ST$ takes. 
Since the form of the full energy-momentum tensor with one contravariant and one covariant index is such that the explicit interaction terms 
involving $\Delta K$ are always diagonal, 
they come with opposite signs for $\ET$ and $\PT$. Therefore,
\begin{equation}
(\e + \p)\x\y^3 + (\et + \pt)\xt\yt^3 = (\ET + \PT)e^{-\phi(\mathbf{x})}.
\end{equation}
Thus from
$(\ET + \PT) = \TT \ST$ we get 
\begin{equation}
\To\s\x\y^3 + \Tt\st\xt\yt^3 = \TT \ST e^{-\phi(\mathbf{x})}.
\end{equation}
The relation between the temperatures \eqref{global-eq-grav-pot} 
reduces this to
\begin{equation}
\ST = \s\y^3 + \st\yt^3.
\end{equation}
The above form then holds for the general consistent effective metric coupling discussed in Appendix \ref{App:GeneralCouplings}.
Thus, we obtain a general proof of thermodynamic consistency with \eqref{global-eq-grav-pot} and the above form of the full entropy.

\subsection{Explicit check}

In the following, we explicitly verify thermodynamic consistency of the equilibrium solution
obtained in Sec.~\ref{Sec:Thermodynamics} for the simplest coupling rules (\ref{coupling-rule}).

With the results (\ref{EandP}), the thermodynamic relation
$\ET+\PT=\TT\ST$ is evidently fulfilled with
$\TT=\T\x=\Tt\xt$ and $\ST=\s\y^3+\st\yt^3$,
when $\epsilon_{1,2}+P_{1,2}=T_{1,2} s_{1,2}$. Here we shall check that then also
\be
\ST=\frac{d\PT}{d\TT}
\ee
holds provided the two subsystems satisfy
\be
\s=\frac{\e+\p}{\T}=\frac{d\p}{d\T},\quad
\st=\frac{\et+\pt}{\Tt}=\frac{d\pt}{d\Tt}.
\ee

We need to evaluate
\ba
\frac{d\PT}{d\TT}&=&\frac{d}{d\TT}\left[\p\x\y^3+\pt\xt\yt^3\right]-\frac{\gamma}2 \frac{d}{d\TT}\left[
\left\{\e\x^{-1}\y^3\right\}\left\{\et\xt^{-1}\yt^3\right\}
+3\left\{\p\x\y\right\}\left\{\pt\xt\yt\right\}\right]\nonumber\\
&&-\frac{\gamma'}2\frac{d}{d\TT}\left[
\left(-\e\x^{-2}+3\p\y^{-2}\right)\x\y^3
\left(-\et\xt^{-2}+3\pt\yt^{-2}\right)\xt\yt^3
\right].
\ea
Differentiating the equations for the metric factors allows us to substitute the derivatives of
the parts written within curly brackets as follows:
\ba
&&\gamma\frac{d}{d\TT}\left\{\e\x^{-1}\y^3\right\}=
\gamma'\frac{d}{d\TT}\left[\left(-\e\x^{-2}+3\p\y^{-2}\right)\x\y^3\right]-2\xt\xt',
\\
&&\gamma\frac{d}{d\TT}\left\{\p\x\y\right\}=
-\gamma'\frac{d}{d\TT}\left[\left(-\e\x^{-2}+3\p\y^{-2}\right)\x\y^3\right]+2\yt\yt',
\\
&&\gamma\frac{d}{d\TT}\left\{\et\xt^{-1}\yt^3\right\}=
\gamma'\frac{d}{d\TT}\left[\left(-\et\xt^{-2}+3\pt\yt^{-2}\right)\xt\yt^3\right]-2\x\x',
\\
&&\gamma\frac{d}{d\TT}\left\{\pt\xt\yt\right\}=
-\gamma'\frac{d}{d\TT}\left[\left(-\et\xt^{-2}+3\pt\yt^{-2}\right)\xt\yt^3\right]+2\y\y',
\ea
where a prime means differentiation w.r.t.\ $\TT$ (except in the case of $\gamma'$).
This leads to
\ba
\frac{d\PT}{d\TT}&=&\frac{d}{d\TT}\left[\p\x\y^3+\pt\xt\yt^3\right]
+\e\x'\y^3+\et\xt'\yt^3-3\p\x\y^2\y'-3\pt\xt\yt^2\yt'
\nonumber\\
&=&\p'\x\y^3+(\e+\p)\x'\y^3+\pt'\xt\yt^3+(\et+\pt)\xt'\yt^3
\nonumber\\
&=&\frac{d\p}{d\T} \frac{d\T}{d\TT}\x\y^3+\T \frac{d\p}{d\T}\x'\y^3+
\frac{d\pt}{d\Tt} \frac{d\Tt}{d\TT}\xt\yt^3+\Tt \frac{d\pt}{d\Tt}\xt'\yt^3
\nonumber\\
&=&\s\y^3\left( \frac{d\T}{d\TT}\x+\T\x'\right)+\st\yt^3\left( \frac{d\Tt}{d\TT}\xt+\Tt\xt'\right).
\ea
The two expressions within parentheses in the last line are both $d\TT/d\TT=1$, 
which completes the proof: ${d\PT}/{d\TT}=\ST$.

\section{Low and high temperature behavior of the equilibrium solution for conformal subsystems}\label{App:LowHigh}

In this appendix we collect some formulae that allows one to derive analytically the behavior of
the equilibrium solution for conformal subsystems (with arbitrary  $\no$, $\nt$) at
low and high temperatures, or, equivalently, small and large coupling $\ga$, in the case $r=-\gap/\ga>1$ such
that solutions exist for all values of the physical temperature $\TT$. {Moreover, we show how conformality
in the limit of large $\gTf$ comes about.}

For small $\gTf$, a power series expansion of the solutions to the set of equations (\ref{vb-equations}) can be easily
obtained.
The leading terms in the metric coefficients are
\ba
\x^2&=&1-{3 \nt}\gTf + (12r-27)\no\nt(\gTf)^2+O\left((\gTf)^3\right)\nonumber\\
\xt^2&=&   1-{3 \no}\gTf + (12r-27)\no\nt(\gTf)^2 +O\left((\gTf)^3\right)\nonumber\\
\y^2&=&   1+{\nt}\gTf +(12r+5)\no\nt(\gTf)^2 +O\left((\gTf)^3\right)\nonumber\\
\yt^2&=&1+{\no}\gTf +(12r+5)\no\nt(\gTf)^2+O\left((\gTf)^3\right)
\ea
and in the effective lightcone velocities:
\ba\label{v-perturbative}
\v=\x/\y&=&1-2 \nt\gTf -16\no\nt(\gTf)^2+O\left((\gTf)^3\right)\nonumber\\
\vt=\xt/\yt&=&1-2 \no\gTf -16\no\nt(\gTf)^2 +O\left((\gTf)^3\right)
\ea
(in the latter the dependence on $r$ first shows up at third order).

As discussed in Section \ref{sec:conformalsubsystems}, the lightcone velocities asymptote to finite
values $\v_\infty$, $\vt_\infty$ for large $\gTf$, provided $r>1$. These values are obtained by
solving the sixth-order algebraic equations (\ref{vasymp-unequal}), which reduces to a quadratic
equation with solution (\ref{vasymp-symmetric}) when $\no=\nt$.

The full, nonperturbative equation determining the lightcone velocities as a function of $\gTf$
is given by (\ref{gT4v-nonsymmetric1}) and (\ref{gT4v-nonsymmetric2})
which were obtained by solving 
first the quadratic equations for $\y^2$ and $\yt^2$ that are implied by (\ref{vb-equations}).
Using (\ref{gT4v-nonsymmetric1}) and (\ref{gT4v-nonsymmetric2}) in the relations for $\y^2$ and $\yt^2$
one finds
\ba\label{xy-asymptlinT}
\x^4&\equiv& \y^4 \v^4=\frac{3+\vt^2}{\v(1-\vt^2)}\n\gTf,\nonumber\\
\xt^4&\equiv& \yt^4 \vt^4=\frac{3+\v^2}{\vt(1-\v^2)}\nt\gTf.
\ea
Moreover, one can derive the simple identity
\be
\frac{\yt^2}{\y^2}=\frac{3+\v^2}{3+\vt^2}.
\ee

At small $\gTf$, all metric coefficients as well as $\v$ and $\vt$ tend to unity, with $1-\v^2$ and $1-\vt^2$
proportional to $\gTf$. As one can check easily, 
(\ref{xy-asymptlinT}) confirms the first-order coefficients in (\ref{v-perturbative}).

At large $\gTf$, where $\v$ and $\vt$ approach nonvanishing values $\v_\infty$ and $\vt_\infty$ below unity,
(\ref{xy-asymptlinT}) implies that the metric coefficients $\x,\xt,\y,\yt$ grow linearly with physical temperature $\TT$.
Since the effective temperatures of the subsystems are given by $\To=\TT/\x$ and $\Tt=\TT/\xt$,
this means that they saturate at finite values proportional to $\gamma^{-1/4}$,
\be
\gTf\to\infty \;\Rightarrow\quad
\To\to \left(\frac{3+\vt_\infty^2}{\v_\infty(1-\vt_\infty^2)}\no\ga\right)^{-1/4},\quad
\Tt\to \left(\frac{3+\v_\infty^2}{\vt_\infty(1-\v_\infty^2)}\nt\ga\right)^{-1/4}.
\ee

This behavior of the metric coefficients, together with saturation of $t^\mu_{\;\nu}$ and $\tit^\mu_{\;\nu}$,
implies that at large $\TT$ the coupling rules (\ref{coupling-rule}) become
\begin{eqnarray}
\g_{\mn}\approx\ga\,\gB_{\mu\rho}\tit^{\rho\sigma}\gB_{\sigma\nu}\frac{\sqrt{-\gt}}{\sqrt{-\gB}}+\gap\,\gB_{\rho\sigma}\tit^{\rho\sigma}\gB_{\mn}\frac{\sqrt{-\gt}}{\sqrt{-\gB}},\nonumber \\
\gt_{\mn}\approx\ga\,\gB_{\mu\rho}t^{\rho\sigma}\gB_{\sigma\nu}\frac{\sqrt{-g}}{\sqrt{-\gB}}+\gap\,\gB_{\rho\sigma}t^{\rho\sigma}\gB_{\mn}\frac{\sqrt{-g}}{\sqrt{-\gB}}.
\end{eqnarray}
Hence, for conformal subsystems
\be
t^{\mn}\left(\ga\,\gB_{\mu\rho}\tit^{\rho\sigma}\gB_{\sigma\nu}\frac{\sqrt{-\gt}}{\sqrt{-\gB}}+\gap\,\gB_{\rho\sigma}\tit^{\rho\sigma}\gB_{\mn}\frac{\sqrt{-\gt}}{\sqrt{-\gB}}\right)\approx t^{\mn} g_\mn=0,
\ee
or, equivalently,
\be
\tit^{\mn}\left(\ga\,\gB_{\mu\rho}t^{\rho\sigma}\gB_{\sigma\nu}\frac{\sqrt{-g}}{\sqrt{-\gB}}+\gap\,\gB_{\rho\sigma}t^{\rho\sigma}\gB_{\mn}\frac{\sqrt{-g}}{\sqrt{-\gB}}\right)\approx \tit^{\mn} g_\mn=0,
\ee
so that the 
pure trace terms in the full energy-momentum
tensor $\ttot_{\phantom{\mu}\nu}^{\mu}$ proportional to $\delta_{\phantom{\mu}\nu}^{\mu}$
become small compared $\TT^4$, $\Delta K/\TT^4\approx 0$. Hence, at large $\TT$,
\begin{equation}
\ttot_{\phantom{\mu}\mu}^{\mu}/\TT^4
\approx (t_{\phantom{\mu}\mu}^{\mu}\sqrt{-\g}+\tit_{\phantom{\mu}\mu}^{\mu}\sqrt{-\gt})/\TT^4=0.
\end{equation}
From the full solution we in fact find that $\ttot_{\phantom{\mu}\mu}^{\mu}/\TT^4 \sim 
\ga^{-1/2}\TT^{-2}$.

Note also that $\ttot^\mn$ should be interpreted as the non-anomalous part of the full energy-momentum tensor which is locally conserved by itself. (In flat background, the anomalous contribution vanishes.) Therefore, indeed in the limit $\gamma\TT^4 \rightarrow \infty$ the full system becomes conformal provided each system is conformal individually. As a corollary, the fluctuations of the full system in the limit $\TT \rightarrow \infty$ with $\ga$ and $\gap$ fixed will behave as that of a conformal system.

\section{A new kind of second-order phase transition and its critical exponent}\label{App:PhaseTransition}

In this section we analyze further the second-order phase transition that occurs at a particular
value of $r=-\gap/\ga$ and derive the value of the critical exponent $\alpha$ in the specific heat
of the full system,
\be\label{CVT}
\CVT=\TT \partial \ST/\partial \TT \sim |\TT-\TT_c|^{-\alpha}
\ee
when $\TT\to\TT_c$,
for the case of two conformal subsystems (\ref{bi-conformal})
where $\ST$ is given by (\ref{ST}).

Let us first study the simplest case of identical subsystems, $\no=\nt=n$,
where we can equate\footnote{It is a priori not excluded that there are solutions
$\v\not=\vt$ despite having set $\no=\nt$. Indeed, such a spontaneous symmetry breaking happens in the
region $\ga<0$ such that the broken phase has lower free energy. However, this region is unphysical
in that the effective lightcone velocity of the subsystems can be larger than the speed of light in the physical
Minkowski space. We have checked numerically that such symmetry breaking does not occur in the case $\ga>0$.}
$\v=\vt$.
In place of (\ref{gT4v-nonsymmetric1}) and (\ref{gT4v-nonsymmetric2}) we then have the simpler relation
\be\label{gT4v-symmetric}
n\gTf = \frac{v^5(1-v^2)(3+v^2)}{[3+v^4-3r(1-v^2)^2]^2}
\ee
which is plotted in Fig.~\ref{fig:gT4v} for the value of $r$ where the second-order phase transition occurs
and two values nearby in the crossover and in the first-order regime.
Note that in this plot only the part connected to $v=1$ (corresponding to $\gTf=0$) is physically realised;
increasing $\gTf$ from zero to infinity lowers $v$ to a finite limiting value given by the zero of
$[3+v^4-3r(1-v^2)^2]$ which we have given in (\ref{vasymp-symmetric}). 

\begin{figure}[tbp]
\centering 
\includegraphics[width=.6\textwidth]{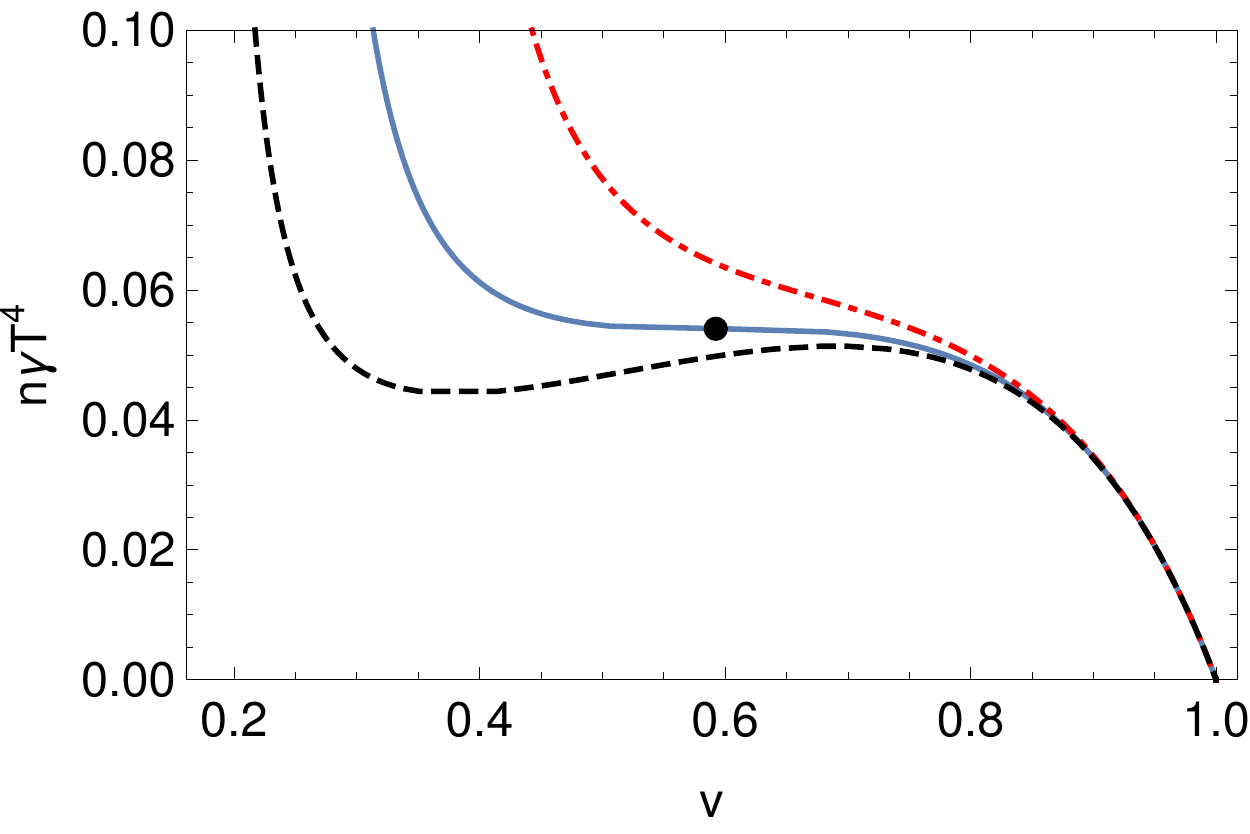}
\caption{\label{fig:gT4v} The relation between $\gTf$ and $v=\x/\y$ (for $\no=\nt$) 
at the critical
value $r=r_c$ (full line) with the critical point indicated by a black dot on top of it. The dotted and dashed
lines correspond to a crossover situation with $r=1.1 r_c$ and a first-order phase transition
with $r=0.95 r_c$, respectively.}
\end{figure}

For $r>r_c$, $\TT^4$ is a monotonic function of $v$ between $v_\infty$ and 1, whereas for $r<r_c$ it has multiple extrema
determined by the $d\TT^4/dv=0$. This equation can be written as
\be
\frac{v^8-2 v^6+36 v^4+42 v^2-45}{3 \left(v^2-1\right)^2 \left(v^4-10 v^2-15\right)}=r.
\ee
The maximum of $r(v)$ for $0<v<1$ determines the critical value $r_c$ beyond which no extrema of $\TT^4(v)$ occur. This is given by
\be
\frac{dr}{dv}=
\frac{4 v \left(v^2+3\right) \left(5 v^8+20 v^6-202 v^4-60 v^2+45\right)}{3
   \left(1-v^2\right)^3 \left(v^4-10 v^2-15\right)^2}=0.
\ee
The bi-quartic polynomial factor in the numerator has a single real root in the range $0<v<1$.
It can be given in closed form and reads
\be
v_c^2\Big|_{\no=\nt}=\frac{1}{5} \left(2 \sqrt{85+10 \sqrt{15}}-5-4 \sqrt{15}\right)\approx 0.35097,
\ee
and thus
\be\label{rcsymmetric}
r_c\Big|_{\no=\nt}=
\frac{1}{540} \left(195+43 \sqrt{15}+\sqrt{30 \left(4082-557 \sqrt{15}\right)}\right)
=1.114509\ldots,
\ee
which together inserted in (\ref{gT4v-symmetric}) yield $n\ga T_c^4\approx 0.0539768$.

The critical exponent $\alpha$ in the specific heat (\ref{CVT}) can now be inferred from
the simple relationship (\ref{ST}) between entropy and effective lightcone velocities. In the vicinity
of the critical point we have, for $\no=\nt$,
\be
|\ST-\ST_c|\sim 24n\TT_c^3 v_c^{-4}|v-v_c|.
\ee
As we have seen, the critical point is determined by the simultaneous vanishing of the first and second derivatives of
$\TT^4$ as given by (\ref{gT4v-symmetric}) with respect to $v$. Hence,
\be
|\TT^4-\TT^4_c|\sim 4\TT_c^3|\TT-\TT_c|\sim |v-v_c|^3
\ee
up to some constant prefactor, and thus
\be
|\ST-\ST_c|\sim |\TT-\TT_c|^{1/3}, \quad
\CVT\sim |\TT-\TT_c|^{-2/3}.
\ee

In the case of two conformal systems with $\no\not=\nt$, the critical parameters can no longer be obtained
in closed form. However, one can show that the critical exponent $\alpha$ is independent of $\nt/\no$
and only the values of $r_c$ and $\TT_c$ change.

In this case, one has to solve the two equations (\ref{gT4v-nonsymmetric1}) and (\ref{gT4v-nonsymmetric2})
numerically, which gives functions $\v=\v(\TT)$ and $\vt=\vt(\TT)$. For sufficiently large values of $r$,
both functions are single-valued; phase transitions occur when these functions develop infinite tangents.
Combining (\ref{gT4v-nonsymmetric1}) and (\ref{gT4v-nonsymmetric2}), one finds that 
\be
\frac{\nt}{\n} =\frac{\vt^5(1-\v^2)(3+\v^2)}{\v^5(1-\vt^2)(3+\vt^2)}\equiv \rho(\v,\vt)=const.
\ee
Because
\be
0=\frac{\partial \rho}{\partial\v} \frac{d\v}{d\TT}+\frac{\partial \rho}{\partial\vt} \frac{d\vt}{d\TT},
\ee
the zeros of $d\TT/d\v$ and $d\TT/d\vt$ have to occur simultaneously in general.
A critical endpoint with second-order phase transition appears when two zeros of $d\TT/d\v$ (or $d\TT/d\vt$)
merge as $r\to r_c$ from below, such that also $d^2\TT/d\v^2$ vanishes and a saddle point (in one dimension) arises.
In principle, such a saddle point could have the next two higher derivatives vanish, too,
which would change the critical exponent $\alpha$ to $-4/5$. However, with the one additional free parameter $n_2/n_1$
there is not enough freedom for a corresponding fine-tuning.

By exploring the solutions numerically, we have found that the situation analyzed above for $n_1=n_2$ is indeed generic.
Also for any $n_2/n_1\not=1$, there is a range $1<r< r_c$ where there is a first-order phase transition
at a finite value of $\gqT$, a second-order transition at $r=r_c$, and an analytic crossover for $r>r_c$.
In fact, $r_c$ depends rather weakly on $n_2/n_1$; it rises from its minimal value (\ref{rcsymmetric}) to 
$\approx 1.119$ for $n_2/n_1=0.1$ or 10, and can be shown always to
remain below $\frac54$. 

\end{appendix}

\bibliographystyle{JHEP} 

\bibliography{semiholo}

\end{document}